\documentclass[12pt,preprint]{aastex}

\usepackage{graphicx,natbib,multicol,multirow}

\newcommand{\degree}{\ensuremath{^\circ}}

\begin{document}

\title {New Debris Disks Around Young, Low Mass Stars Discovered With The \textit{Spitzer} Space Telescope}

\author{Peter Plavchan$^{1,2}$, M. W. Werner$^{2}$, C. H. Chen$^{3}$, K. R. Stapelfeldt$^{2}$, K. Y. L. Su$^{4}$,  J.R. Stauffer$^{5}$, and I. Song$^{5}$} 
\affil{$^{1}$ Infrared Processing and Analysis Center, California Institute of Technology, MC 100-22, 770 S. Wilson Ave, Pasadena, CA 91125; plavchan@ipac.caltech.edu\\
$^{2}$ Jet Propulsion Laboratory, California Institute of Technology, 4800 Oak Grove Drive, Pasadena, CA 91109 \\
$^{3}$ Space Telescope Science Institute, 3700 San Martin Drive, Baltimore, MD 21218\\
$^{4}$ Steward Observatory, University of Arizona, Tucson, AZ 85721\\
$^{5}$ \textit{Spitzer} Science Center, California Institute of Technology, MC 103-33, Pasadena, CA 91125\\
}

\begin{abstract}  
We present 24 $\mu$m and 70 $\mu$m MIPS (Multiband Imaging Photometer for \textit {Spitzer}) observations of 70 A through M-type dwarfs with estimated ages from 8 Myr to 1.1 Gyr, as part of a \textit {Spitzer} guaranteed time program, including a re-analysis of some previously published source photometry.   Our sample is selected from stars with common youth indicators such as lithium abundance, X-ray activity, chromospheric activity, and rapid rotation.  We compare our MIPS observations to empirically derived K$_{s}$-[24] colors as a function of the stellar effective temperature to identify 24 $\mu$m and 70$\mu$m excesses.  We place constraints or upper limits on dust temperatures and fractional infrared luminosities with a simple blackbody dust model.

We confirm the previously published 70 $\mu$m excesses for HD 92945, HD 112429, and AU Mic,  and provide updated flux density measurements for these sources.    We present the discovery of 70 $\mu$m excesses for five stars: HD 7590, HD 10008, HD 59967, HD 73350, and HD 135599.  HD 135599 is also a known \textit{Spitzer} IRS (Infra-Red Spectrograph) excess source, and we confirm the excess at 24 $\mu$m.  We also present the detection of 24 $\mu$m excesses for ten stars: HD 10008, GJ 3400A, HD 73350, HD 112429, HD 123998, HD 175742, AT Mic, BO Mic, HD 358623 and Gl 907.1.  We find that large 70 $\mu$m excesses are less common around stars with effective temperatures of less than 5000 K (3.7$^{+7.6}_{-1.1}$\%) than around stars with effective temperatures between 5000 K and 6000 K (21.4$^{+9.5}_{-5.7}$\%), despite the cooler stars having a younger median age in our sample (12 Myr vs. 340 Myr).  We find that the previously reported excess for TWA 13A at 70 $\mu$m is due to a nearby background galaxy, and the previously reported excess for HD 177724 is due to saturation of the near-infrared photometry used to predict the mid-infrared stellar flux contribution.   

In the Appendix, we present an updated analysis of dust grain removal time-scales due to grain-grain collisions and radiation pressure, Poynting-Robertson drag, stellar wind drag and planet-dust dynamical interaction.  We find that drag forces can be important for disk dynamics relative to grain-grain collisions for L$_{IR}$/L$_*$$<$10$^{-4}$, and that stellar wind drag is more important than P-R drag for K and M dwarfs, and possibly for young ($<$1 Gyr) G dwarfs as well.
\end{abstract}


\section{INTRODUCTION}

Nearly twenty-five years ago, the \textit{Infrared Astronomical Satellite} (IRAS) launched the study of infrared excesses around stars that we attribute to mature extra-solar planetary systems \citep[e.g.,][]{rhee07,zuckerman01a,fajardo00,mannings98,backman93,walker88,aumann84}.  Parent bodies in a planetary system -- $\sim$1 m and larger aggregates of rock and ice analogous to asteroids and Kuiper-belt objects in our own Solar System -- collide and produce dusty debris that is heated by incident stellar radiation. This optically thin ``debris disk'' re-emits the absorbed radiation at infrared wavelengths; it is detected in excess of the expected stellar radiation around $\sim$15\% of main sequence stars \citep[e.g.,][]{lagrange00}.  Ground-based infrared and sub-millimeter efforts \citep[e.g.,][]{lestrade06,liu04,weinberger04, song02}, the \textit{Infrared Space Observatory}  \citep[ISO; ][]{demuizon05,laureijs02,habing01,spangler01} and the \textit{Spitzer} Space Telescope \citep[]{werner04} have discovered $>$100 nearby stars with infrared excess.  These discoveries include several dozen debris disks around young stars less massive than the Sun \citep[e.g.,][]{rebull08,meyer07,smith06,low05,chen05a}.   Large samples of these systems observed with the \textit{Spitzer} Space Telescope are useful in determining overall trends such as disk frequency, disk infrared luminosity, and dust dynamics as a function of age or effective stellar temperature both in the field \citep[e.g.,][]{carpenter09,hillenbrand08,trilling08,wyatt08,gautier08, wyatt07,beichman06a,beichman06b,bryden06,su06,rieke05,chen05a} and in clusters and associations \citep[e.g.,][]{currie09,dahm09,cieza08,currie08b,hernandez08,cieza07,currie07,dahm07,gorlova07,hernandez07a,hernandez07b,carpenter06,chen05b,low05,stauffer05,gorlova04}.  High-contrast, high-resolution direct and coronagraphic imaging from a number of telescopes has spatially resolved $\sim$15 debris disks around solar-type stars in scattered and/or thermal emission that show disk structures and clearings indicative of possible unseen planets \citep[e.g., ][]{fitzgerald07,telesco05,krist05a,krist05b,marsh05,kalas04,stapelfeldt04,holland03,wahhaj03,heap00,schneider99,jayawardhana98,koerner98}.  The recent reported discoveries of Fomalhaut b \citep[]{kalas08,chiang09} and a possible planet orbiting Beta Pic \citep[]{lagrange09} support the hypothesis that planets can directly influence debris disk structure.  Spatially resolved systems provide further information about the dust dynamics and evolution, but discerning overall trends is limited by the small number of resolved systems.  

In parallel with infrared observations and high-contrast imaging of young stars, over 300 extrasolar planets have been discovered, primarily with the radial velocity technique, in the past decade, but also through direct imaging, transits, microlensing, and pulsar timing \citep[and references therein]{marois08,kalas08,butler06,marcy05,bond04,beaulieu06,wolszczan92}.   The synergy of these planet-finding methods with the study of debris disks can yield new insights into the overall circumstellar architecture of exo-planetary systems.  For example, radial velocity discovered extrasolar planets have revealed trends such as the Jovian planet frequency -- host star metallicity correlation \citep[]{gonzalez97}.  This correlation is not observed for debris disks \citep[]{beichman06a}. 

Fundamental questions remain about the dust dynamics, properties, and evolution of debris disks around young stars.  For example, the evolutionary time-scales for both primordial and debris disks appear to be dependent on spectral type.  Optically thick primordial disks are common ($\sim$50\%) at ages of $\sim$1 Myr around stars with spectral types of A through M \citep[]{haisch01, meyer97}.  These primordial disks provide the material to form planets and parent bodies that in turn can generate secondary debris disks.  In the 3--5 Myr Upper Sco association, \citet[]{carpenter06} observes that primordial disks continue to persist in an optically thick state around K and M dwarfs, but have already transitioned to optically thin disks around earlier type stars.   It might follow that debris disks should similarly persist for longer around K and M dwarfs due to lower stellar luminosities and masses (${\S}$A.2).  Debris disks can persist for stars older than $\sim$1 Gyr around A,F, and G-type stars \citep[]{trilling08,meyer07,su06,bryden06, rieke05}.    However, for K and M stars older than $\sim$50 Myr, there is a relative paucity of stars known with infrared excess associated with debris disks \citep[e.g.,][]{rhee07,plavchan05}.   Before Spitzer, selection effects from a flux-limited survey could partially account for this apparent difference in disk frequency and disk lifetime; debris disks around K and M dwarfs are thought to be relatively colder and fainter than around solar-type stars, due to the lower stellar luminosities \citep[]{rhee07,zuckerman01a}.  Indeed, two colder debris disks have been reported around M dwarfs with sub-millimeter observations \citep[]{lestrade06,liu04}.  More sensitive observations with the \textit{Spitzer} Space Telescope and ground-based efforts continue to identify a relative lack of bright 70 $\mu$m excesses (F$_{70}$/F$_*$$\gtrsim$2--10) around K and M stars older than $\sim$50 Myr \citep[]{trilling08,gautier08, beichman06a}.   This paper addresses and confirms this dearth of relatively bright excesses at 70 $\mu$m around stars cooler than our Sun, but find that small ($\sim$10--15\%) flux excesses at 24 $\mu$m are common at ages of $<$ 100 Myr.  \citet[]{forbrich08} also identifies 9 new candidate M dwarf debris disks with large 24 $\mu$m excesses in the $\sim$40 Myr open cluster NGC 2547, perhaps probing the epoch of icy planet formation for M dwarfs \citep[]{currie08b,kenyon08}.

Large ensemble infrared studies of young stars remain a valuable probe to address remaining questions about the dynamics, properties, and evolution of debris disks.  In this paper, we present the results of a \textit{Spitzer} guaranteed time program to search for debris disks.  We have obtained MIPS 24 $\mu$m and 70 $\mu$m observations of 70 young, nearby A-M stars with ages ranging from 8 Myr to $\sim$1 Gyr.  Preliminary results for a subset of these stars have been reported in \citet[]{rebull08,chen05a,low05}.  In ${\S}$2 we present our sample and selection criteria.  In ${\S}$3 we present the observations and our data analysis procedures.  In ${\S}$4, we present our analysis to evaluate the significance of our detections and non-detections, and place constraints on the disk properties of observed stars with and without infrared excess. In ${\S}$5 we present our results on stars with identified infrared excesses.    In ${\S}$6, we discuss individual stars with excess in our sample, and the transition from grain-grain collisions to drag-force dominated disks.  In ${\S}$7, we present our conclusions.  In the Appendix we present an updated analysis of the different time-scales for dust grain removal in debris disks from radiative forces, stellar wind forces, and grain-grain collisions.  We also consider the importance of planet-dust interaction relative to these processes.

\section{SAMPLE SELECTION}

Our sample of 70 stars were selected as being relatively young based upon a variety of youth indicators.  For F-M stars, selection criteria include high X-ray activity from \textit{ROSAT} detections, high lithium abundance, Calcium II H\&K line chromospheric activity, and rapid rotation \citep[]{gaidos98,gaidos00}.    For A stars, we select candidates based on inferred youth from Stroemgren photometry and distances inferred from Hipparcos parallaxes \citep[]{song01,song00,stauffer00}, with the exception of HD 123998 which was selected for its apparent IRAS excess \citep[confirmed in this work to be a false-positive, potentially due to IRAS beam-confusion; ][]{song00,song01}.   We select 27 stars with stellar effective temperatures less than 5000 K, 28 stars with stellar effective temperatures between 5000 K and 6000 K, and 15 stars with stellar effective temperatures greater than 6000 K.   We use these sub-groups within the sample throughout our analysis.  The choice of 5000 K and 6000 K to delineate the sub-groups is motivated by the different sample selection techniques applied in each temperature regime, the X-ray activity, and estimated ages (${\S}$4.4).  The sub-samples also highlight differences in our results between solar-type (5000-6000 K), low-mass ($<$ 5000 K) and high-mass radiative atmosphere stars ($>$ 6000 K).  Many of the stars in our sample are members or candidate members of young moving groups and associations.  In Table 1 we list our targets along with literature spectral types, distances, cluster and moving group memberships, and other youth indicators.

\section{OBSERVATIONS} 

We obtained 24 $\mu$m and 70 $\mu$m observations of our sample using the Multiband Imaging Photometer for Spitzer \citep[MIPS, ][]{rieke04} on the \textit{Spitzer} Space Telescope \citep[]{werner04} in photometry mode (default scale).   Each of our targets was observed in 2004 or 2005, using integration times of 48.2 s at 24 $\mu$m and 186--465 s at 70 $\mu$m.   The data were reduced using the MIPS Data Analysis Tool  \citep[DAT,][]{gordon05}, which performs the necessary steps need to produce a final mosaic of the target. We also applied additional reduction techniques to  remove instrumental artifacts and flux calibration as described by \citet[]{engelbracht07,gordon07} at 24 and 70 um, respectively.  We adopt flux density calibration factors at 24 $\mu$m and 70 $\mu$m of 0.0454 and 702 respectively to convert from instrumental units to MJy per steradian.  We also reduce and combine the ``bcd'' image data from the \textit{Spitzer} Science Center ``S13''/MOPEX/APEX pipeline.  The ``S13'' pipeline processes and removes image artifacts such as bad-pixels, cosmic rays, latent images, vertical array column ``jailbars'' from calibrating ``stim flashes'' for the 70 $\mu$m detector, and the ``droop effect'' at 24 $\mu$m \citep[]{gordon05}.

We extract photometry using Point-Spread Function (PSF) fits to the co-added image data with the DAT.  We examine the 70 $\mu$m images for any cirrus contamination.  We compare PSF-derived flux densities to aperture photometry derived with the DAT as in \citet[]{su06}, as well as to flux densities derived from the MOPEX/APEX pipeline using both aperture photometry and fitting to empirically derived Pixel-Response Functions \citep[PRFs,][]{makovoz05}.  In all four methods, we set a minimum detection threshold signal-to-noise ratio (S/N) of 3.  For both sets of comparison aperture photometry, we use a circular aperture with radii of 14.$^{\prime\prime}$94 (6 pixels) at 24 $\mu$m and 29.$^{\prime\prime}$5 (3 pixels) at 70 $\mu$m as in \citet[]{su06}, and use scalar aperture corrections of 1.143 and 1.694 respectively inferred from \textit{Spitzer} Tiny Tim models of the point-spread function \citep[]{krist02}.   We estimate background noise from the standard deviation of flux density in background annuli with inner and outer radii of 29$^{\prime\prime}$.55 and 42$^{\prime\prime}$.33 (12--17 pixels) and 39$^{\prime\prime}$.40 and 68$^{\prime\prime}$.95 (4--7 pixels) at 24 $\mu$m and 70 $\mu$m respectively.  This includes detector noise and noise due to cirrus structures present in the image.   If the observations were taken in regions with high cirrus, then the detection limits are dominated by cirrus noise.  Our sample cirrus levels span from very low ($<$ 5 MJy/sr at 70 $\mu$m) to high (e.g., $\sim$20 MJy/sr at 70 $\mu$m for HD 125158).  

In Table 2, we tabulate adopted 24 $\mu$m and 70 $\mu$m photometry, propagated uncertainties, S/N, expected photospheric flux densities (${\S}$4), and references in which \textit{Spitzer} photometry for these sources has been previously published.  All photometry in Table 2 assumes a $\lambda^{-2}$ flux density across the bandpass.  We find good agreement in the derived photometry for all four techniques at 24 $\mu$m, and for S/N $\raisebox{-4pt}{$\stackrel{>}{\sim}$}$ 5 at 70 $\mu$m.  At 70 $\mu$m, neither the DAT nor MOPEX/APEX demonstrate greater sensitivity at low S/N.  As noted in Table 2, there are seven instances where one or more methods do report a detection at 70 $\mu$m at S/N$>$3 whereas the other methods do not.   These results at 70 $\mu$m are below the adopted threshold for the detection of excess (${\S}$4.2) in all seven cases, and do not affect our results.  For the 24 $\mu$m excess source AT Mic, a partially resolved $\sim$3$^{\prime\prime}$ binary, we use a DAT aperture photometry rather than PSF-fitting at 24 $\mu$m to measure the combined flux density from both components. 

At both 24 and 70 $\mu$m we use the DAT PSF photometry when available due to the smaller flux density calibration uncertainties.  At 24 $\mu$m, we detect all of our sample targets.  At 70 $\mu$m, we detect 19 of our 70 sample targets with S/N$>$3 at the expected positions inferred from the 24 $\mu$m source position, unless otherwise noted in Table 2.  For non-detections at 70 $\mu$m, we calculate 3-$\sigma$ upper limits by measuring the standard deviation of flux density within a circular aperture with a radius of 29.5$^{\prime\prime}$ at the expected source location.   Our photometry are consistent with preliminary results in \citet[]{chen05a,low05}, but we obtain better overall precision and sensitivity.

\section{ANALYSIS}

In ${\S}$4.1, we use a $\chi^2$ minimization procedure to fit PHOENIX NextGen model photospheres \citep[]{hauschildt99a} to optical and near-infrared photometry to estimate stellar effective temperature and radii.  In ${\S}$4.2, we compare the K$_s$-[24] colors to the relation empirically derived by \citet[]{gautier08} from nearby ($<$5pc) disk-less main-sequence G, K and M dwarfs.  In ${\S}$4.3, we place constraints on the circumstellar material for detected and non-detected mid-infrared excesses.  In ${\S}$4.4 we estimate ages, masses, and calculate X-ray luminosities for our sample.  

\subsection{Modeling of Stellar Photospheres}

We obtain photo-electric optical UBVR$_c$I$_c$ photometry from the literature, with B and V magnitudes primarily from the Tycho catalog, transformed to the Johnson photometric system using the corrected relations in \citet[]{mamajek06,mamajek02}.  We obtain JHK$_s$ near-infrared photometry from the 2MASS All-Sky Catalog of Point Sources \citep[]{skrutskie06}.  In several cases, our sources saturate the 2MASS detectors, and for two of these sources  we obtain J and K band magnitudes from \citet[]{morel78}.  Due to saturation, we do not include 2MASS J-band photometry for HD 112429, HD 16555, HD 125158, and HD 144197, and H-band photometry for HD 112429.  We also do not include optical photometry for several sources when the literature magnitudes appear inconsistent with the magnitudes in other bands -- R$_c$-band magnitudes for HD 10008, TWA 18, and HK Aqr, and I-band magnitudes for SAO 145139, TWA 12, and HK Aqr.  We tabulate and cite all the photometry in Table 3.

PHOENIX NextGen models are used to fit the observed optical and near-IR photometry rather than a blackbody, since the stellar SEDs are very different from that of a blackbody at optical and near-IR wavelengths \citep[]{hauschildt99a,hauschildt99b,gray92,mullan89}.  We assume solar metallicity as is typical for main-sequence stars in the solar neighborhood \citep[e.g., ][]{boone06,drilling00}.  We assume log g=4.5 as is typical for main sequence stars, and A$_V$=0.  Our sample includes some K and M dwarfs that have not yet reached the main sequence and will have lower surface gravities, but this has a negligible effect on our derived stellar radii and effective temperatures.  To fit the NextGen spectra to the available photometry, we compute a $\chi^2$ minimization as a function of effective temperature and normalization.  We integrate model photospheres as a function of wavelength across effective bandpasses to compare to observed photometry for a given band.  We assign a weight of 1 to each band, and we normalize to K$_s$.  

For effective temperatures (T$_*$) greater than 3600 K, we interpolate (linearly in T$_*^4$) the NextGen model photospheres (provided in increments of 200 K) in increments of 20 K to improve the fits.  We find excellent agreement with effective temperatures inferred from literature spectral types \citep[]{luhman99,luhman98,hartigan94}.  For T$_*$ less than 3600 K, NextGen photospheres lack near-infrared opacity sources \citep[]{hauschildt99a}.  As a result, model photosphere fits for these M dwarfs tend towards lower effective temperatures (up to 200 K) than those implied from their previously identified spectral types.  The resulting fits in turn overpredict the observed 24 $\mu$m photometry by as much as 20\%.  For these sources, we fix the effective temperature in increments of 100 K to the literature spectral type.  This approach provided the most consistent results in predicted 24 $\mu$m photospheric flux density with the method in ${\S}$4.2 \citep[][]{gautier08}. We tabulate the derived stellar parameters from our model photosphere fits in Table 3.

\subsection{K$_s$-[24]  and [24]-[70] Colors}

To identify excesses at 24 and 70 $\mu$m, we use the effective temperatures derived in ${\S}$4.1 and the K$_s$-[24] color relation derived from nearby ($<$5pc) disk-less main sequence G, K and M dwarfs in \citet[]{gautier08} to predict the expected photopheric flux densities at 24 $\mu$m:

\begin{equation}
K_s-[24] =  \left\{ \begin{array}{lr}  -0.2915 + \left[1.267\times10^{-4} \:T_* + (2.389 \times 10^{-4} \:T_*)^2 + (2.2\times10^{-4} \:T_*)^3 \right]^{-1} & , T_*<5025 K\\
		 	0 & , T_*>5025 K
			\end{array} \right.
\end{equation} 
where [24] is the magnitude converted from the 24 $\mu$m photometry with a zero-point flux of 7.14 Jy, and $T_*$ is the effective stellar temperature in Kelvin.  At 70 $\mu$m we use a Rayleigh-Jeans extrapolation from the expected stellar flux at 24 $\mu$m.

We calculate the significance of the deviation of the measured photometry from expected photospheric flux densities, $\chi\equiv(F_\nu($observed$)-F_\nu($predicted$))/\sigma_\nu$ at 24 and 70 $\mu$m, where $\sigma_\nu$ includes the detection S/N and flux calibration uncertainties.  We list these values in Table 2 and plot them in Figure 1.  We identify sources with $\chi_{70}$$>$5 as 70 $\mu$m excesses, $\chi_{24}$$>$5 as 24 $\mu$m excesses.   In Figure 2 we plot F$_{70}$/F$_*$ as a function of effective temperature for our sample.  Figure 2 includes several strong detections of 70 $\mu$m disks reported here for the first time.  In Figure 3 we plot the observed K$_s$-[24] colors for our sample as a function of effective stellar temperature.  Figure 3 includes eleven detections of 24 $\mu$m disks reported here for the first time.  While some of the new 24 $\mu$m disks have been previously analyzed in \citet[]{chen05a}, the improved calibration uncertainties at 24 $\mu$m with the DAT pipeline have improved the sensitivity to small levels (e.g. $\sim$10--15\%) of flux density excess.
 
\subsection{Dust Disk Constraints}

We list the derived dust disk parameters and 70 $\mu$m color corrected photometry in Table 4.  To estimate the physical disk parameters  -- fractional infrared excess, dust temperature, dust mass, and orbital distance -- for the stars with and without excesses in our sample, we assume that the debris disks are optically thin.  After subtracting off the predicted photospheric flux density, we fit a single-temperature dust blackbody model to the 24 $\mu$m and 70 $\mu$m excess.  The derived fractional infrared excesses are consistent with our model assumption of an optically thin disk.  We estimate dust orbital radii assuming that the dust is in thermal equilibrium.  We color correct the 70 $\mu$m photometry based on the dust temperature, using the color corrections outlined in the \textit{Spitzer Observer's Manual}.  The color corrections are $<$1\% at 24 $\mu$m, and $\sim$10\% at 70 $\mu$m.  For the purposes of estimating upper limits to the observed fractional infrared excesses, we fit to 3-$\sigma$ upper limits for non-detections at 70 $\mu$m.   In Figure 4 we plot  the model photosphere and dust blackbody fits for stars with 70 $\mu$m detections.  For the sources with 24 $\mu$m excesses and 70 $\mu$m non-detections, we estimate the fractional infrared excess at 24 $\mu$m with the assumption that L$_{IR}$$\sim$$\nu F_\nu$(24 $\mu$m), after subtracting from $F_\nu$ the expected stellar flux density contribution.  We list this fractional infrared excess in Table 4 in addition to the upper limit derived from the 70 $\mu$m non-detection.

We estimate dust masses as in \citet[]{chen05a}, inferring the average grain size $\langle a\rangle$ using a collisional cascade grain size distribution power law exponent of -3.5, with a minimum grain size set by the radiative blowout radius a$_b$ ($\langle a\rangle$$=$1.32 a$_b$), and grain density of $\rho_g=2.5$ g/cm$^{-3}$ (Equation A22 in ${\S}$A.2.3).  We assume $\beta=1/2$ is the threshold criterion for radiative blowout, where $\beta$ is the ratio of radiation pressure to gravity for the purposes of calculating a$_b$ (Equation A18, ${\S}$A.2.2).  For the stars in our sample, only significant stellar winds of $\sim1000 \dot M_\odot$ will affect the blowout radius (${\S}$A.1,${\S}$A.2.2).    Such a large mass-loss rate is not supported by theoretical considerations and observations \citep[]{holzwarth07}, and hence only grain blowout by radiation pressure is relevant (${\S}$A.1).  We derive dust disk masses between 2.8$\times$10$^{-7}$ and 5.2$\times$10$^{-3}$ lunar masses for the disks in our sample.  We estimate minimum parent body masses required to produce this dust between 3$\times$10$^{-4}$ and 1.1 lunar masses (Equation A20).  The required minimum parent body masses are $\sim$10$^3$ times the derived dust disk mass.  These results imply both a broad dynamic range in mass of debris disks, and a diverse variety of asteroid or Kuiper Belt extra-solar analogs.

\subsection{X-Ray Luminosity, Age, Mass, and Stellar Wind Mass-Loss Rate Estimates}

We estimate properties of the host stars in our sample that are relevant to the investigation of debris disk evolution and spectral type dependence.  In Table 4, we summarize the derived ages, age determination methods, masses, L$_{X}$/L$_*$, and stellar wind mass-loss rates.  

\subsubsection{X-Ray Luminosity}

We derive X-ray apparent flux densities from \textit{ROSAT} PSPC detected count-rates using the conversion of \citet[]{schmitt95}:
\begin{equation}
\mbox{Apparent Flux}  = \mbox{\# cnts/s} \times (8.31 + 5.3\times\mbox{HR1}) \times 10^{-12} \mbox{ ergs/cm$^2$/s}
\end{equation}
where \# cnts/s is the \textit{ROSAT} count rate, and HR1 is the \textit{ROSAT} hardness ratio \#1.  We require that optical source positions fall within the 1-$\sigma$ \textit{ROSAT} position uncertainties to associate the \textit{ROSAT} detections with our sample stars.  We infer the fractional X-ray luminosity L$_{X}$/L$_*$ and the X-ray flux density F$_{X}$ at the stellar photosphere using the derived stellar properties (radius, luminosity) from synthetic spectral fitting in ${\S}$4.1.  

\subsubsection{Stellar Ages}

We use five main methods to estimate ages for the stars in our sample -- young moving group membership ($<$100 Myr), X-ray activity, Ca II H \& K activity, rotation periods, and fitting to theoretical isochrones.  First, we assign ages based on cluster and association memberships to AU Mic, AT Mic, and HD 358623 (Beta Pic moving group; 12 Myr), to TWA association sources (8 Myr), and AB Dor moving group members AB Dor, GJ 3400A and LO Peg (75$\pm$25 Myr).  For sources with measured X-ray activity, rotation periods or Ca II H \& K activity, we estimate ages using activity-age correlation relations derived in \citet[]{mamajek08}.  We use equations A3, 12--14, and 3 respectively, with appropriate limits (e.g., B--V$<$1.2, -5 $<$ Log(L$_X$/L$_*$) $<$ -4.3).  When two or three of these activity indexes are measured, we adopt the mean and standard deviation as the final age estimate and uncertainty.  When only a single indicator is available, we adopt an uncertainty of 20\% for CaII H\&K and X-ray activity.  For rotation-based age estimates, we propagate the uncertainties in the coefficients of the fit in \citet[]{mamajek08} to Equations 12--14, the relation first presented in \citet[]{barnes07}.  We obtain age estimates for 46 of our 70 target stars using these first four age determination methods. 

For eight of the remaining stars with T$_*<$ 6000 K and X-Ray activity implying ages $<$300 Myr, we fit the derived stellar radii and temperature in ${\S}$4.1 to \citet[]{seiss00} isochrones.  We adopt average ages and uncertainties from the range of isochrone grid points that are consistent to within 5\% of the derived stellar radius and temperature.  The 5\% criterion is chosen based on the estimated radius uncertainties in the fits in ${\S}$4.1, the S/N of Hipparcos measured parallax, and to recover reasonable age uncertainties ($\sim$50\%).  If a source is an unresolved binary, we adjust the derived radius appropriately before fitting to isochrones to avoid systematically under-estimating the age.  Three of the eight stars have measured 
Lithium abundances that support the estimated ages.  For stars that are implied to be young from measured activity indicators, the activity-based ages are consistent with isochrone fitting results.  For 14 stars with T$_*>$ 6000 K, we use the average age estimates of \citet[]{song00,song01,stauffer00} from fitting Stromgren photometry.   

We are unable to estimate ages for two target stars using the above five methods.  For the possible Pleiades moving group member RE 1816+541 lacking a trigonometric parallax, we assign an age of 115 Myr based upon the Pleiades age estimates from \citet[]{messina03,stauffer98}.  Finally, we obtain an age estimate of 1.1 Gyr from \citet[]{nordstrom04} for HD 17240 and adopt an age uncertainty of 50\%.  

In Figure 5, we plot the age distribution for our sample as a function of effective stellar temperature.  The K and M stars in our sample are primarily selected from young moving groups whereas the A-G type stars are primarily selected from other indicators.  As a result of these selection criteria, the stars in our sample cluster in age as a function of spectral type in Figure 5.    The T$_*$$<$ 5000 K low-mass K and M dwarfs are younger than the 5000--6000 K G dwarfs and the T$_*$ $>$ 6000 K stars, with median ages of 12.25 Myr, 340 Myr, and 200 Myr respectively.  This age bias limits comparisons across spectral types in our sample.

\subsubsection{Stellar Wind Mass-Loss Rates}

We wish to consider stellar wind drag from stellar wind protons as a disk dissipation mechanism (${\S}$A.1).  In order to evaluate this mechanism, we estimate the stellar wind mass-loss rates for stars in our sample.  If F$_X$$<$8$\times$10$^5$ ergs/s/cm$^2$ at the stellar surface (stars approximately older than 700 Myr), we can use the relation of \citet[]{wood02,wood05} to estimate stellar mass-loss rates, $\dot M_{sw}$:

\begin{equation}
\dot M_{sw} = \dot M_\odot \left(\frac{F_X }{ F_{X\odot}}\right)^{1.34} \left(\frac{R_*}{R_\odot}\right)^2
\end{equation}
where $\dot M_\odot=2\times10^{-14} M_\odot$/yr.  Seven stars with X-ray detections meet this criteria in our sample.   We caution against estimating mass-loss rates from X-Ray activity for stars younger than $\sim$700 Myr or with F$_X$$>$8$\times$10$^5$ ergs/s/cm$^2$ \citep[]{wood05}.   HD 135599 has an abnormally low X-ray luminosity given its estimated age of 214$\pm$17 Myr, and also possesses a strong 70 $\mu$m excess.

\subsubsection{Stellar Mass}

We estimate stellar masses for our sample using the estimated stellar ages, effective temperatures, and \citet[]{seiss00} isochrones.  We use the estimated stellar mass in the calculations of dust disk properties and time-scales.  For a given age and corresponding \citet[]{seiss00} isochrone, we linearly interpolate between the \citet[]{seiss00} stellar masses as a function of stellar effective temperature.

\section{RESULTS}

\subsection{Stars with Excess}

At 70 $\mu$m, we report the discovery of 70 $\mu$m excesses ($\chi_{70}$$>$5) for HD 7590, HD 10008, HD 59967, HD 73350, and HD 135599.  We confirm the previously detected MIPS excesses for HD 92945, HD 112429 and AU Mic \citep[]{low05,chen05a}.  We confirm that the \textit{Spitzer} Infra-Red Spectrograph (IRS) detected excess for HD 135599 extends to longer wavelengths \citep[]{lawler09}.  The model photospheres and empirical colors from ${\S}$4.1 and ${\S}$4.2 produce consistent values of $\chi_{70}$ at 70 $\mu$m.  We do not confirm the previously published 70 $\mu$m excess for TWA 13A, and instead attribute the 70 $\mu$m emission to a nearby background galaxy ($\sim$10$^{\prime\prime}$, Figure 6).   \citet[]{chen05a} previously reported an excess for HD 177724 at 24 $\mu$m and 70 $\mu$m that wasn't easily described by a simple dust blackbody model for the excess.  We instead attribute the previously reported excess to the large (0.28--0.36 mag) errors in the saturated J, H and K$_{s}$-band 2MASS photometry, and find that the MIPS photometry is consistent with the predicted photospheric flux density.  For all stars with excesses at 70 $\mu$m, we PSF subtract the detected sources to look for extended residuals.  AU Mic and HD 92945 show extended structures, and detailed analysis will be presented in \citet[]{bryden07}.  The 70 $\mu$m residuals for AU Mic and HD 92945 appear aligned with the position angles of the disks resolved in scattered optical and near-infrared light \citep[]{kalas04,metchev05,krist05a,krist05b}, albeit at low S/N.

At 24 $\mu$m, we report the discovery of eleven sources with 24 $\mu$m excesses  ($\chi_{24}$$>$5).  We identify flux density excesses for HD 10008 (16.6$\pm$2.0\%), GJ 3400A (13.9$\pm$3.3\%), HD 73350 (9.5$\pm$1.9\%), HD 112429 (11.1$\pm$1.9\%), HD 123998 (10.5$\pm$2.7\%), HD 135599 (10.7$\pm$2.8\%), HD 175742 (14.5$\pm$2.2\%), AT Mic (15.1$\pm$4.2\%), BO Mic (14.6$\pm$3.3\%), HD 358623 (11.4$\pm$2.3\%) and Gl 907.1 (9.1$\pm$2.3\%).  We attribute these small excesses to warmer circumstellar material than responsible for a 70 $\mu$m-only excess.  The above uncertainties in the flux density excess are slightly larger than implied by the $\chi_{24}$ values in Table 2 because we include the uncertainties in the K$_s$ band flux density as an estimate of the model photosphere flux density uncertainty at 24 $\mu$m.  Five of the 24 $\mu$m excess sources have effective temperatures between 3200 and 4800 K.  Thus this is one of the first reports of debris disks around cool, low mass stars in addition to \citet[]{forbrich08}.  

\subsubsection{Alternative Explanation for 24 $\mu$m Excess}

We investigate the possibility that the small excesses we detect at 24 $\mu$m are not due to re-processed radiation from dusty debris disks.    HD 10008, HD 73350, HD 112429 and HD 135599 also possess 70 $\mu$m excesses, supporting the assertion that our data and analysis are sensitive to $\sim$10\% excesses at 24 $\mu$m.  For the seven sources with 24 $\mu$m excesses and without 70 $\mu$m detections, the upper-limits at 70 $\mu$m are consistent with a debris disk SED.  Furthermore, excesses at 24 $\mu$m are more likely at younger ages.  The stars with 24 $\mu$m excesses we identify are generally much younger than our overall sample age distribution.

However, an unknown low-mass spectroscopic companion -- e.g., a M5 companion to a K0 star -- could produce an apparent $\sim$10\% flux excess at 24 $\mu$m.   The occurrence of spectroscopic binaries in the field with the mass ratio required to reproduce the 24 $\mu$m excess is much less than the 10\% (7/70) of stars for which we observe this excess \citep[]{hogeveen92}.   Nonetheless, we do not rule out low-mass companions for these sources.  BO Mic and GJ 3400A possess published IRAC photometry, but the [8]-[24] colors are inconclusive as to which scenario is favored \citep[]{chen05a}.  Positionally coincident background AGN and serendipitous asteroids could also account for the observed 70 and 24 $\mu$m excess, but are statistically unlikely for any given star ($<$1\%).  $>$0.1 mag stellar variability at K$_s$-band could also produce an apparent 24 $\mu$m excess for non-contemporaneous observations.

We note that the distribution of K$_s$-[24] photospheric colors for low-mass stars is less constrained than it is for warmer stars.  We compare the formal errors derived from the 24 $\mu$m S/N and K$_s$-band uncertainty, and compare it to the empirical K$_s$-[24] distribution after subtracting off the predicted photospheric color.  We derive an empirical distribution uncertainty of 0.03 mag from a Gaussian fit, which is consistent with our formal propagated errors.  9 of the 11 24 $\mu$m excess sources lie $>$3-$\sigma$ in excess above the fit to the empirical distribution.  The 10th, HD 112429, also possesses a 70 $\mu$m excess, and the 11th at 2.5-$\sigma$, is 12.5 Myr 4240 K GJ 907.1.  A 24 $\mu$m excess for GJ 907.1 is entirely plausible, but the excess for Gl 907.1 is the least robust detection of a 24$\mu$m excess that we include in our results.    For a sample size of 70 stars, we might expect $\sim$1 false-positive detection of excess from statistical fluctuations.

\subsection{Disk Fraction}

For our sample, we find an overall disk fraction of 21.4$^{+5.6}_{-4.1}$\% (15/70).  This disk fraction is smaller than the disk fraction for A stars \citep[33$\pm$5\%,][]{su06}, slightly higher than for young G stars \citep[14$^{+11}_{-7}$\%][]{carpenter09,hillenbrand08,meyer05}, and higher than the disk fraction for older, field G stars \citep[12--15\%,][]{trilling08,bryden06,lagrange00}.  We compute the 1-$\sigma$ uncertainties in the disk fraction using binomial statistics as in \citet[]{burgasser03}.    We list these values in Table 5 for a variety of sub-samples.

\subsubsection{Spectral Type}

At 70 $\mu$m, we can detect debris disks that are generally $>$100\% brighter than the expected photosphere (Figure 2).  For the 15 stars in our sample with T$_*$$>$6000 K, we find a disk fraction of 13.3$^{+13.3}_{-4.6}$\%.  This is smaller than the disk fraction derived from analogous studies of A-type stars \citep[33$\pm$5\%,][]{su06}, but the results are consistent to within 1-$\sigma$ given the small number statistics.  In our sample 70 $\mu$m excesses are $\sim$6 times less common around stars with effective temperatures of less than 5000 K (3.7$^{+7.6}_{-1.1}$\%) than around stars with effective temperatures between 5000 K and 6000 K (21.4$^{+9.5}_{-5.7}$\%).  These two disk fractions differ at the $\sim$2-$\sigma$ level.    We observe this discrepancy even though the cooler stars have a younger median age in our sample (12.25 Myr vs. 340 Myr).  At younger ages, debris disks might be expected to be brighter by factors of a few.  This paucity of late-type star debris disks at 70 $\mu$m has been found by, and is consistent with, several other more detailed studies \citep[and references therein]{hillenbrand08,gautier08,trilling08,beichman06a}.  The interpretation of our result is complicated and limited by several factors and selection effects in our sample that we consider in more detail in ${\S}$A.3.  

At 24 $\mu$m, we can detect much smaller ($\sim$10--15\%) excesses relative to the photosphere (Figure 3).  When we include the detections of excesses at 24 $\mu$m, the disk fraction dependence on spectral type disappears.  We find an overall disk excess fraction of 22.2$^{+9.8}_{-5.9}$\% for stars with effective temperatures of less than 5000 K, and 25.0$^{+9.6}_{-6.3}$\% for stars with effective temperatures between 5000 K and 6000 K.   This result suggests that debris disks are just as common around cool stars as they are around hotter stars, but the dust disks around hotter stars may generally be more massive and brighter relative to the stellar photosphere.  This result is consistent with the frequency of $\sim$40 Myr M dwarfs with 24 $\mu$m excesses identified in \citet[]{forbrich08}.

\subsubsection{Binarity}

Our sample includes 24 known binary systems (Table 1), and constitutes a sample that has been fairly well studied for AO, wide separation and spectroscopic companions.  The binary completeness of our sample, however, is undoubtedly $<$100\%.  The incompleteness of the known binaries in our sample limit our interpretation, but we can make a tentative comparison.  We do not identify a statistically significant difference between the disk fractions for binaries and single stars.  We derive a single star disk fraction of 22.7$^{+3.0}_{-5.0}$\% (10/44), and a binary disk fraction of 19.2$^{+9.8}_{-5.4}$\% (5/26).  \citet[]{trilling07} carried out an investigation of debris disks around A and F-type binaries, and concluded that binaries have a higher disk fraction on average than single stars.  We do not confirm this result when investigating lower mass stars, with the aforementioned caveats.  \citet[]{trilling07} identified that tight (separations $<$3 AU) and wide separation ($>$50 AU) binaries have a higher disk fraction than intermediate separation binaries (3--50 AU).   Our sample of binaries is evenly split between intermediate separation (12) and wide separation binaries (14), and we find 3 and 2 debris disks in each category respectively.  

\subsection{Age}

In Figure 7, we plot the fractional infrared excess as a function of stellar age in our sample.   We over-plot a t$_*^{-1}$ power law consistent with collisional evolution of debris disks \citep[${\S}$A.2.1, e.g.,][]{wyatt08,wyatt07,rieke05}. The data are not consistent with this power-law evolution.  However, the youngest stars in our sample (e.g. $<$100 Myr) are dominated by the coolest stars (T$_*$ $<$ 5000 K) and vice-versa.  Consequently, this bias masks any true underlying trend in L$_{IR}$/L$_*$ vs. age.  We  caution against drawing any significant conclusions from this figure.  The dependence of debris disk frequency and brightness at 24 and 70 $\mu$m has been investigated in other more detailed studies \citep[e.g.,][and references therein]{carpenter09,trilling08,su06}.   \citet[]{carpenter09} and \citet[]{trilling08} both note that the 70 $\mu$m age evolution for the disk frequency and disk brightness for solar-type stars does not decrease with age as strongly as is observed for A-type stars \citep[]{su06,rieke05}.

Five of the seven 24 $\mu$m excess sources without large 70 $\mu$m excesses are younger than $\sim$50 Myr.  This age range is consistent with the epoch of terrestrial and icy planet formation \citep[e.g., ][]{kenyon08,currie08b}, and is much younger than the median age of our sample (200 Myr).    The average age of stars with T$_*$ $<$ 5000 K is 21 Myr and 60 Myr for stars with and without detected infrared excess respectively (at either 24 or 70 $\mu$m).  For comparison, the averages are 312 Myr and 395 Myr respectively for 5000 K $<$ T$_*$ $<$ 6000 K stars.  The stars with excesses tend towards younger ages, and this is consistent with a collisional-cascade production of dust from parent bodies \citep[and references therein]{wyatt08}.

\subsection{Stellar Rotation Period}

In Figure 8, we plot the fractional infrared excess as a function of stellar rotation period for our sample for all stars with T$_*$$<$6000 K.   We use measured stellar rotation periods when available in Table 1.  If not, we estimate the rotation period from measured $v \sin i$ rotational velocities in Table 1, estimated radii in Table 3, and we assume an average rotation axis inclination with respect to our line of site of 30$\degree$.  For HD 92945, we assume the star matches the inclination of the debris disk, estimated to be 25$\degree$ from edge on \citep[]{krist05b}.  

We identify a weak correlation between rotational periods and observed fractional infrared excesses for stars with excess.  We derive a Spearman rank correlation coefficient $\rho_s$$=$0.35.  Our sample spans an age range older than the sample in \citet[]{rebull08}, who report a similar weak correlation between rotation and excess for stars in the 12 Myr Beta Pic moving group.  Our observations support the premise presented in \citet[]{rebull08} and \citet[]{currie08a} that massive primordial disks evolve into massive debris disks.  Massive primordial disks regulate the initial angular momentum evolution of stars before the primordial disks dissipate, and the persisting excess--period correlation for debris disks is an echo of the primordial disk rotation correlation.  A correlation between rotation and excess was also presented in \citet[]{stauffer07}. \citet[]{stauffer05} identified 3 debris disks in the $\sim$100 Myr Pleiades, and did not identify any anti-correlation between excess and rotational velocity.

\subsection{Activity Indicators}

In our sample, stars with 24 $\mu$m excess appear well-mixed with non-excess stars for X-ray saturated stars with $\log$(L$_X$/L$_*$)$>$-4.  Rotation periods between $\sim$5.5--7 days are favored for the detection of 70 $\mu$m excess around stars with  $\log$(L$_X$/L$_*$)$<$-4.  We do not identify any significant correlations of fractional infrared excess with either the lithium abundance or the activity index R$^\prime_{HK}$.  We do however note in ${\S}$6.1.3 and Figure 9 that the bright 70 $\mu$m excess stars are outliers when computing the ratio of the rotation period to both $\log$(L$_X$/L$_*$) and R$^\prime_{HK}$.  Given our small sample size and broad age and spectral type ranges, this trend is intriguing but inconclusive and possibly spurious.

For the seven stars with F$_X$$<$8$\times$10$^5$ ergs/s/cm$^2$, we can use the X-ray luminosity as a proxy for stellar wind mass loss rate \citep[${\S}$A.2, ][]{plavchan05,wood05}.  We find that the smallest mass-loss rate in this sub-sample is seen for the only star with an excess -- HD 135599.  The other 6 stars possess excess upper limits considerably smaller than the excess detected for HD 135599.  At a fixed age, such an anti-correlation is expected from the steady-state model of \citet[]{plavchan05} and has been tentatively observed in the younger Sco-Cen association \citep[]{chen05b}.   Since we only have seven stars in this sub-sample and only one detected excess, we cannot draw any conclusions except to note that these observations do not contradict the steady-state model of \citet[]{plavchan05}.  

\section{DISCUSSION}

In ${\S}$6.1, we present notes about the stars in our sample with the brightest excesses.   In ${\S}$6.2, we discuss the impact of various grain removal mechanisms on debris disk evolution and for the debris disks in our sample.  These mechanisms are presented in detail in the Appendix.

\subsection{Individual Star Notes}

\subsubsection{AU Mic}

Since the discovery of a resolved disk around AU Mic in \citet[]{kalas04}, much work has been done to characterize and model AU Mic's debris disk \citep[]{fitzgerald07,strubbe06,quillen07,krist05a,krist05b,metchev05}.  Our updated 70 $\mu$m flux density measurement should not significantly alter interpretations.  We update the application of the stellar wind drag steady-state model to AU Mic in \citet[]{plavchan05}.  The discovery in \citet[]{wood05} that X-ray activity does not correlate with stellar wind mass-loss rates for F$_X$$>$8$\times$10$^5$ ergs/s/cm$^2$ invalidates the application in \citet[]{plavchan05} of a steady-state stellar wind drag model to AU Mic; AU Mic possesses an X-ray surface flux density of $\sim$1.8$\times$10$^7$ ergs/s/cm$^2$.  We estimate the grain-grain collision time-scale within the ``birth ring'' at 43AU \citep[the location of the dust-producing ring of parent bodies,][]{strubbe06} to be $\sim$10$^4$ times shorter than the stellar wind drag time-scale for $\dot M_{sw}=10 \dot M_{\odot}$, a dust disk mass of 10$^{-2}$M$_\oplus$, and a disk width of 4.3 AU \citep[Equation A23, ${\S}$A.2,][]{fitzgerald07,metchev05,liu04,augereau06}.  The evolution of the AU Mic debris disk is currently dominated by grain-grain collisions and the presence of any approximately Jovian mass planets (${\S}$A.2).

\subsubsection{HD 135599}
For this K0 star ($\sim$1/3L$_\odot$), the inferred mass-loss rate implies that the drag time-scale from stellar wind is $\sim$10 times shorter than P-R drag (Equations A8--10, ${\S}$A.1,${\S}$A.2),. The stellar wind drag time-scale is $\sim$0.4 times the grain-grain collision time-scale (Equations A14,A23) for the dust radius (12 AU) and dust mass (6.6$\times10^{-5}$ M$_{\mbox{moon}}$) estimated in Table 4 (${\S}$4.3).    This implies that stellar wind drag is dynamically important in the evolution of the dust in this disk, and makes HD 135599 a canonical example to study the transition between disk evolution dominated by grain-grain collisions and drag-force dominated disk evolution (${\S}$6.2 \& ${\S}$A2.2.3).  The \textit{Spitzer} IRS spectra of HD 135599 are consistent with the ``power-law'' excess predicted for a drag-force dominated debris disk \citep[e.g. F$_\nu$(dust)$\propto$$\nu^\alpha$, ${\S}$6.2 and ${\S}$A.2 in this paper,][]{lawler09}.  A more accurate dust mass measurement from sub-mm observations and high-contrast imaging to obtain radial surface brightness profiles is warranted to clarify this interpretation further.  

A power-law extrapolation of the IRS excess for HD 135599 predicts a flux at 70 $\mu$m that is smaller than what is observed \citep[]{lawler09}.  This implies that the entire disk for HD 135599 is not described by a single dominant dust removal mechanism. Due to the different power-law dependencies on orbital distance for drag forces and grain-grain collisions, and because the collisional time-scale depends on the local dust disk density, the evolution of dust at different orbital radii can be dictated by different dust removal mechanisms (${\S}$A.2).  For HD 135599, it is plausible that there is a ring of parent bodies producing dust through mutual collisions where the dust collision time-scales are short relative to the stellar wind drag time-scale, but not significantly shorter.  This would produce the ``excess excess'' seen at 70 $\mu$m.  Then, a fraction of the dust produced in the ring spirals inward under the action of stellar wind drag into smaller orbital radii where the collisional time-scales become longer than the drag force time-scales.  This would then explain the observed IRS power-law excess.

\subsubsection{HD 7590, HD 73350, HD 92945, and HD 135599:  An Activity--Excess Anti-Correlation?}

We consider the large fractional infrared excesses (F$_{70}$/F$_*$$>$10) observed for HD 7590, HD 73350, HD 92945, and HD 135599 (the ``bright disk stars'').   For HD 7590, HD 73350 and HD 135599, the stellar ages inferred from X-ray activity and R$^\prime_{HK}$ tend towards older ages, whereas ages inferred from rotation periods tend towards younger ages\footnote{For HD 7590, we estimate ages of 496, 427, and 463 Myr from X-ray, rotation period and R$^\prime_{HK}$ indicators respectively; for HD 73350 we estimate ages of 613, 358, and 569 Myr from X-ray, rotation period and R$^\prime_{HK}$ indicators respectively; for HD 92945, we estimate ages of 311 and  278 Myr from X-ray and R$^\prime_{HK}$ activity respectively; for HD 135599, we estimate an age of 214 Myr from its rotation period and association membership, while we ignore the X-ray estimated age of 1.1 Gyr.}.  This trend is not present for stars with no excess, and follows directly by comparing the ratio of the three activity indexes.  Within our sample, HD 7590, HD 73350, and HD 135599 are outliers when computing the ratio of the rotation period to both Log(R$^\prime_{HK}$) and Log(L$_X$/L$_*$) (Figure 9).   Given the small size of our sample, this result is inconclusive and may be spurious.  A more rigorous treatment with a larger debris disk sample of solar-type stars with measured rotation periods and activity indexes is warranted.

If HD 7590, HD 73350 and HD 135599 are younger than we estimate from X-ray and R$^\prime_{HK}$ activity, then the activity for these stars are abnormally low.  This presents the possibility that stars with large excesses are more likely to be less active at a given age.  This anti-correlation between activity and excess is consistent with the possibility that stellar wind drag is the dominant grain removal mechanism (${\S}$A.1).  At a fixed age, stellar wind drag predicts that less active stars will have brighter disks due to the longer wind drag time-scales \citep[${\S}$A.2, ][]{chen05b}.  However, at these ages of $<$700 Myr, X-ray activity is not a reliable proxy for stellar wind mass-loss rate as discovered in \citep[]{wood05}.  Alternatively, this anti-correlation is also consistent with massive debris disks evolving from massive primordial disks, which in turn regulated the host star angular momentum \citep[]{currie08a}.  

A second hypothesis is that the ``bright disk stars'' are undergoing an epoch of enhanced planetesimal collisions akin to the Late Heavy Bombardment within our own Solar System when it was $\sim$700 Myr old \citep[]{kleine02,chambers04,gomes05,strom05}.  HD 10008, HD 59967, and HD 123998 also fall within the estimated age range of 200--500 Myr.

\subsection{The Transition Between Grain-Grain Collisions and Drag Forces in Disk Dynamical Evolution -- Observational Predictions}

Dust is produced in debris disks via a collisional cascade of dust grains that result from the collision of parent bodies.  The collisions of parent bodies decay with time \citep[e.g., ][ and references therein]{wyatt08,kenyon05}.  The evolution of the observable debris disk depends on the how the dust is removed from the system, primarily thought to occur through drag forces causing dust to spiral into the star (drag force dominated), radiative blowout as the end stage of a collisional cascade of dust grains (collision dominated), and/or planet-dust interactions (${\S}$A.2).  Drag-force dominated disk evolution predicts a constant vertical optical depth as a function of orbital radius.  This produces a power-law infrared spectrum,  where F$_\nu$(dust)$\propto$$\nu^\alpha$ over a wavelength regime consistent with the orbital distances of the in-spiraling dust in thermal equilibrium, and $\alpha$ varies from $\sim$ -1-- -3 depending on the wavelength dependence of the grain emissivity (${\S}$A.2.2).  This is in contrast to a dust excess consistent with a single-temperature blackbody, such as produced by a single ring of dust (${\S}$A.2.1).  Collision-dominated disks are necessarily relatively bright (log(L$_{IR}$/L$_*$)$\sim$-3-- -4).  Resolved collision-dominated debris disks are typically consistent with the ring geometry \citep[e.g.,][]{kalas08,fitzgerald07,telesco05,krist05a,krist05b,kalas04,stapelfeldt04}, and the IRS excess spectra consistent with a single-temperature blackbody \citep[e.g.,][]{morales09,chen06}.

To evaluate the importance of different grain removal mechanisms in our observed debris disks, we calculate the ratio of the drag time-scales to the grain-grain collision time-scales using equation A23 (${\S}$A.2.3).  This ratio is approximately equal to or less than 1 at the minimum dust orbital radii and dust disk masses inferred for HD 10008, GJ 3400A, HD 59967, HD 123998, HD 175742, AT Mic, BO Mic, HD 358623 and Gl 907.1 (hereafter the ``faint disk stars'').  This ratio suggests that dust could be removed under that action of drag forces before they are blown out from radiation pressure and collisional grinding.   The fractional infrared excesses range from 6--80$\times10^{-6}$ for the ``faint disk stars''.  These ``faint disk stars'' possess flux excesses that are smaller than have been typically investigated previously with Spitzer at 70 $\mu$m \citep[e.g. log(L$_{IR}$/L$*$)$\gtrsim$-4,][]{carpenter09,hillenbrand08,trilling08,bryden06,beichman06a}.  Consequently, with this sample we are probing a dust dynamics regime previously unexplored.  From the ratio of collision and P-R/wind drag time-scales, we infer that the transition between collisional-cascade disk evolution to P-R drag disk evolution could occur at a fractional infrared luminosity as high as $L_{IR}/L_*\sim8\times10^{-5}$ (${\S}$A.2.3).  This result is in disagreement with the conclusions of \citet[]{wyatt05,dominik03} which predict a transition at $L_{IR}/L_*\sim4\times10^{-7}$, more than two orders of magnitude fainter.  

The \textit{Spitzer} IRS power-law spectrum for HD 135599 is a canonical example for the transition of the dominant dust dynamical mechanisms in a debris disk, with $L_{IR}/L_*=1.0\times10^{-4}$ (${\S}$6.1.2).  A detailed IRS study of debris disks with a range of fractional infrared excess would further test this hypothesis, and this has been carried out in \citet[]{lawler09} and \citet[]{morales09}.  \citet[]{lawler09} and \citet[]{morales09} find IRS excesses that are consistent with a power-law for excesses between $L_{IR}/L_*\sim3\times10^{-6}$ and $L_{IR}/L_*\sim3\times10^{-5}$.  This fractional infrared excess range is coincident with the fractional infrared excesses of the ``faint disk stars.''  As noted for HD 135599 in ${\S}$6.1.2, the transition from grain collision-dominated disk evolution to drag force-dominated disk evolution is not a sharp transition.  Most disks in transition between collision time-scales and drag time-scales for dust removal will have different evolutionary dynamics at different orbital radii.  Rings of parent bodies will produce dust locally that will initially be subject to grain-grain collisions, but drag forces can act to fill in the interior parts of the disk \citep[e.g., ][]{strubbe06}.  

For HD 7590, HD 73350, HD 92945, HD 112429, HD 135599, and AU Mic, stellar winds of $\sim$3--600 times solar are necessary for drag force time-scales to overcome grain-grain collisions in dictating the disk evolution.  While this is feasible for the ages of these stars, an important observational test is infrared spectra.  \textit{Spitzer} IRS observations of AU Mic and HD 92945 do not show the power law behavior expected if drag forces are the dominant grain removal mechanism \citep[]{chen08,chen06}.  This implies the stellar wind mass-loss rates are less than $\sim$50 $\dot M_\odot$, and the evolution of these disks are dictated by the grain-grain collision time-scales.

Finally, we note that planets of sufficient mass can truncate the inward spiral of dust grains under the action of drag forces before the dust reaches the sublimation radius (Equation A30 and ${\S}$A.2.4) .  The presence of a power-law excess can be used to place an upper limit to the mass of any planet orbiting within the orbital radii of observed dust.

\section{CONCLUSIONS}

Infrared photometry provides basic constraints on fractional flux excess, dust temperature, and dust orbital radii in debris disks associated with young planetary systems.   We identify new 70 $\mu$m excesses for HD 7590, HD 10008, HD 59967, HD 73350, and HD 135599, and constrain their dust disk properties.  We confirm the excesses at 70 $\mu$m previously identified for HD 92945, HD 112429, and AU Mic.  We present the discovery of 24 $\mu$m excesses for HD 10008, GJ 3400A, HD 73350, HD 112429, HD 123998, HD 175742, AT Mic, BO Mic, HD 358623 and Gl 907.1, and confirm the 24 $\mu$m excess for HD 135599.

Within our 70 star sample selected for youth indicators, we find that large 70 $\mu$m excesses (L$_d$/L$_*$$>$10$^{-4}$) associated with debris disks are more common around $\sim$200--500 Myr solar-type stars with T$_*$ $>$ 5000 K than for $\sim$8--50 Myr stars with T$_*$ $<$ 5000 K.  We find that small 24 $\mu$m excesses (L$_d$/L$_*$$<$10$^{-4}$) associated with debris disks are equally common among young stars of all spectral types.  The tentative correlations we present with activity indicators are speculative given the small and non-uniform sample, but suggest a more detailed investigation is warranted. 

From arguments presented in the Appendix and in ${\S}$6.2, we predict that drag-forces are relevant compared to grain-grain collisions for the dynamical evolution of disks around G-M dwarfs with L$_{IR}/$L$_*$$<$10$^{-4}$.  We find that stellar wind drag is more important than P-R drag for removing dust around K and M dwarfs with these fractional infrared excesses, and possibly for G dwarfs as well.

\subsection{Acknowledgements}

This work is based [in part] on observations made with the \textit{Spitzer} Space Telescope, which is operated by the Jet Propulsion Laboratory, California Institute of Technology under a contract with NASA. Support for this work was provided by NASA.

This publication makes use of data products from the Two Micron All Sky Survey, which is a joint project of the University of Massachusetts and the Infrared Processing and Analysis Center/California Institute of Technology, funded by the National Aeronautics and Space Administration and the National Science Foundation.  

This research has made use of the NASA/ IPAC Infrared Science Archive, which is operated by the Jet Propulsion Laboratory, California Institute of Technology, under contract with the National Aeronautics and Space Administration.

This research has made use of the SIMBAD database, operated at CDS, Strasbourg, France.  

This research has made use of the NASA Star and Exoplanet Database (NStED), operated by the Infrared Processing and Analysis Center, California Institute of Technology, under a contract with NASA.

Thanks to Geoff Bryden, Farisa Morales, and Thayne Currie for their conversations and comments.   We thank the anonymous referee for a thorough referee report.

\appendix
\section{Appendix}

In ${\S}$A.1, we present a brief overview of stellar wind drag.  In ${\S}$A.2 we summarize and update equations for the grain lifetimes due to grain-grain collisions, stellar wind forces, radiative forces, and the presence of planets.   This section can serve as a reference to quickly evaluate the relative importance of these processes in debris disk evolution and dynamics.   These processes are discussed in more detail in \citet[]{hillenbrand08,krivov07a,krivov07b,krivov06,strubbe06,minato06,najita05,plavchan05,chen05a,wyatt05,wyatt03,dominik03, jura95, gustafson94, backman93}.  In ${\S}$A.3, we present five factors that affect the interpretation of the debris disk fraction as a function of spectral type in ${\S}$5.2.1.

\subsection{Stellar Wind Drag Primer}

\citet[]{plavchan05} and \citet[]{minato06} propose that stellar wind drag acting on dust grains in a debris disk could account for the relative paucity of late-type stars older than $\sim$50 Myr with detectable infrared excess.  For our Sun, protons from the solar wind produce a radial outward pressure, $\sim$3$\times$10$^3$ times smaller in magnitude than the analogous radiation pressure \citep[]{gustafson94}.     Radiation pressure also gives rise to the Poynting-Robertson drag force that is reduced in magnitude from the radial component by a factor of $v_{orb}/c$, the ratio of the dust grain orbital velocity to the speed of light.  Stellar wind pressure also has an analogous drag term, reduced in magnitude from the radial component by a factor of  $v_{orb}/v_{sw}$, where $v_{sw}$ is the stellar wind velocity.  This factor is the aberration angle of the protons seen by a dust grain as it orbits a star.  A dust grain preferentially forward scatters stellar wind protons in the direction it is traveling, causing it to lose angular momentum and slowly spiral inwards towards the star.  Since stellar wind velocities are on the order of a few hundreds of km/s, it follows that $v_{orb}/v_{sw}$$\sim$10$^3$ $v_{orb}/c$. While stellar wind radial pressure is effectively negligible compared to radiation pressure for dust dynamics around solar-type stars, the drag terms are instead comparable in magnitude.   

For the Sun, the strength of stellar wind drag is $\sim$1/3$^{rd}$ the strength of Poynting-Robertson drag \citep[]{gustafson94}.   From ${\S}$A.2, we find that stellar wind drag will be more important than Poynting-Robertson drag for stars with mass-loss rates comparable to the Sun and stellar luminosities less than 1/3 L$_\odot$ -- e.g. K and M dwarfs -- and for stars with enhanced mass-loss rates -- e.g. young stars.  As an example, for a $\sim$1/3 L$_\odot$ K dwarf, a mass-loss rate of 5 time solar would lead to a stellar wind drag force that is 5 times stronger than P-R drag.  This in turn shortens dust lifetimes by a factor of 5, and lowers the observable fractional infrared excess compared to a disk with only P-R drag considered.  This follows directly from Equation A10 and \citet[]{plavchan05}.

Quantitatively evaluating the relevance of stellar wind drag in low-mass star debris disk evolution is complicated by the uncertainty in the strength and evolution of stellar winds from low mass stars.   Recent observations confirm that stellar winds from M dwarfs in M dwarf -- white dwarf binaries are $\sim$2--10 times the mass-loss rate from the Sun at ages of several Gyr \citep[]{schmidt07,debes06}.  Additionally, \citet[]{wood05} observe that stellar wind mass-loss rates correlate with X-ray activity, implying that young stars also have higher mass-loss rates and hence shorter dust grain lifetimes under the action of stellar wind drag.  However, for stars younger than $\sim$500-700 Myr, the correlation in \citet[]{wood05} breaks down, and the X-ray luminosity cannot be used as a direct proxy for measuring the stellar wind mass-loss rate.    Young late-type stars exhibit enhanced X-ray activity, but this is not a direct indication of a correspondingly large radial stellar wind mass-loss rate.  Measuring stellar wind mass-loss rates and their evolution from young ($\sim$5--500 Myr), low mass stars remains an open problem.  

We can use the relation of \citet[]{wood05} to infer that M dwarfs older than $\sim$700 Myr, such as the M dwarfs observed in \citet[]{gautier08}, possess mass loss rates per units surface area that are comparable to solar-type stars.   For a dust disk comparable in mass to the zodiacal dust cloud in our Solar System, the stellar wind drag time-scale will be shorter than both the P-R drag and grain-grain collision time-scales, and the disk evolution will be dominated by stellar wind drag.  This can account for the lack of older M dwarfs with observed debris disks \citep[]{gautier08,minato06,plavchan05}.  We explore the role of stellar wind drag relative to other grain removal processes in more detail ${\S}$A.2.

\subsection{Dust Grain Removal time-scales}

\subsubsection{Grain-Grain Collisions}

The average time for a grain to collide with another grain is estimated with the expression in \citet[]{minato06,najita05} as:

\begin{equation}
\tau_{\mbox{coll}} = 2\times10^3\:\mbox{yr} \left(\frac{r}{50\mbox{AU}}\right)^{7/2}\left(\frac{a}{\mu\mbox{m}}\right)\left(\frac{\rho_g}{\mbox{g cm}^{-3}}\right)\left(\frac{\pi a^2}{S_z}\right)\left(\frac{M_*}{M_\odot}\right)^{-1/2}\left(\frac{M_d}{10^{-3}M_\oplus}\right)^{-1}
\end{equation}
where $a$ is the grain radius, $r$ is the dust orbital distance, $\rho_g$ is the dust grain density, $M_*$ is the stellar mass, $M_d$ is the dust mass, $M_\odot$ is the solar mass, $M_\oplus$ is the Earth's mass, and $(\pi a^2/S_z)=1/Q_{\mbox{coll}}$, the ratio of the effective grain-grain collision cross-section to the geometric cross-section ($Q_{\mbox{coll}}\sim1$).  

Equation A1 relies on two important assumptions.   First, the grains are assumed to be sufficiently inclined in their orbits to see the full vertical optical depth of the disk at the orbital radius of the grain (the grain traverses the full optical depth of the disk twice as it orbits the parent star).  For optically thin debris disks, the average grain inclination should approximate the ratio of the scale height of the dust disk to the orbital distance, $h/r$.

For the second assumption, the dust is assumed to be evenly distributed over a disk surface area $\pi$D$_{OUT}^2$.  This second assumption is only valid when the entire dust mass $M_d$ is interior to the orbital distance under consideration, ie $r=D_{OUT}$.  In this sense, $\tau_{\mbox{coll}}$ in Equation A1 represents the time-scale for all grains at the disk outer radius $D_{OUT}$ of size $a$ and density $\rho_g$  in an entire disk of mass $M_d$ to undergo a collision.

With approximately a dozen bright, spatially resolved disks, the observed dust geometries are primarily narrow rings  \citep[]{kalas08,fitzgerald07,krist05a,krist05b,telesco05,marsh05,stapelfeldt04,kalas04,holland03,wahhaj03,heap00,schneider99,jayawardhana98}. Here we update Equation A1 and do away with the second assumption to avoid mis-application of collisional time-scales as a function of orbital radius, and to make more realistic estimates.

We define the radial surface density profile $\Sigma_{r}(r)$ such that:

\begin{equation}
M_d = \int_{D_{IN}}^{D_{OUT}} \Sigma_{r}(r)\times r \;dr 
\end{equation}
where $D_{IN}$ and $D_{OUT}$ are the inner and outer disk radii respectively, and we have assumed azimuthal symmetry.  Note we have incorporated the constant from the azimuthal integration into $\Sigma_{r}(r)$.  Then the general form for Equation A1 becomes:

\begin{equation}
\tau_{\mbox{coll}} = 2\times10^3\:\mbox{yr} \left(\frac{r}{50\mbox{AU}}\right)^{3/2}\left(\frac{a}{\mu\mbox{m}}\right)\left(\frac{\rho_g}{\mbox{g cm}^{-3}}\right)\left(\frac{\pi a^2}{S_z}\right)\left(\frac{M_*}{M_\odot}\right)^{-1/2}\left(\frac{2}{\Sigma_{r}(r)} \times \frac{10^{-3} M_{\oplus}}{(50\mbox{AU})^2}\right)
\end{equation}

From Equation A2, if we set $D_{IN}=0$, and $\Sigma_{r}(r)=\Sigma'_{0}$, we get that $\Sigma_r(r)=\Sigma'_0= 2 M_D / D_{OUT}^2$, and we recover Equation A1.

We consider a generic radial surface density profile of:
\begin{equation}
\Sigma_{r}(r) = \Sigma_0 r^\alpha
\end{equation}
with inner and and outer disk radii $D_{IN}$ and $D_{OUT}$, which can be set appropriately depending on $\alpha$.  From Equation A2, we find 
\begin{equation}
\Sigma_0= \left\{ \begin{array}{lr}   M_D(\alpha+2) / (D_{OUT}^{\alpha+2} -  D_{IN}^{\alpha+2})& , \alpha\neq-2\\
		 	M_D/\ln(D_{OUT}/D_{IN}) & \alpha=-2 
			\end{array} \right.
\end{equation} 
and equation A3 becomes:
\begin{equation}
\tau_{\mbox{coll}} = 2\times10^3\:\mbox{yr} \:\mbox{\LARGE $\gamma$} \left(\frac{r}{50\mbox{AU}}\right)^{3/2-\alpha}\left(\frac{a}{\mu\mbox{m}}\right)\left(\frac{\rho_g}{\mbox{g cm}^{-3}}\right)\left(\frac{\pi a^2}{S_z}\right)\left(\frac{M_*}{M_\odot}\right)^{-1/2}\left(\frac{M_d}{10^{-3}M_\oplus}\right)^{-1}
\end{equation}
where:
\begin{equation}
\mbox{\LARGE $\gamma$} = \left\{ \begin{array}{lr}   \left(\frac{2}{\alpha+2}\right)\left[\left(\frac{D_{OUT}}{50\:AU}\right)^{\alpha+2}-\left(\frac{D_{IN}}{50\:AU}\right)^{\alpha+2}\right] & , \alpha\neq-2\\
		 	2\ln(D_{OUT}/D_{IN}) & , \alpha=-2 
			\end{array} \right.
\end{equation}

The gamma factor in Equations A6 and A7, encapsulating the disk geometry, is important for estimating grain lifetimes.  For example, consider a ring of dust at radius D$_{ring}$ and width $\Delta$D$_{ring}$=0.1D$_{ring}$.  The collision time-scale estimate in Equation A1 will differ from the actual collision time-scale by a factor of $\gamma\sim$2$\Delta D_{ring}$=0.2 for $\alpha$=-2 or 0, and overestimate the collision frequency for any of the dust that is scattered or dragged out of the ring.  As a second example, we consider $\alpha=-1.5$ with $D_{IN}$=0 AU and $D_{OUT}$=200AU.  The collision time as a function of orbital distance will scale as $r^3$, and the collision time-scales will be $\gamma=$8 times longer.

\subsubsection{Radiative and Stellar Wind Drag}

A drag force removes angular-momentum from an orbiting dust grain, causing it to spiral into the star.  The time for a dust grain to spiral into a star under the action of P-R drag is estimated in \citet[]{minato06} as:

\begin{equation}
\tau_{PR} = 2\times10^6\:\mbox{yr}\left(\frac{r}{50\mbox{AU}}\right)^{2}\left(\frac{a}{\mu\mbox{m}}\right)\left(\frac{\rho_g}{\mbox{g cm}^{-3}}\right)\left(\frac{\pi a^2}{C_{ph}}\right)\left(\frac{L_*}{L_\odot}\right)^{-1}
\end{equation}
where ($\pi a^2 / C_{ph}$) = 1/Q$_{rad}$, where $Q_{rad}$ is the radiative coupling coefficient, the ratio of the effective to geometric cross-sections.  Similarly, the time for a dust grain to spiral into a star under the action of stellar wind drag is estimated in \citet[]{minato06} as:
\begin{equation}
\tau_{SW} = 6\times10^6\:\mbox{yr}\left(\frac{r}{50\mbox{AU}}\right)^{2}\left(\frac{a}{\mu\mbox{m}}\right)\left(\frac{\rho_g}{\mbox{g cm}^{-3}}\right)\left(\frac{\pi a^2}{C_{sw}}\right)\left(\frac{\dot M_*}{\dot M_\odot}\right)^{-1}
\end{equation}
where ($\pi a^2 / C_{sw}$) = 1/Q$_{SW}$, where $Q_{SW}$ is the stellar wind coupling coefficient, the ratio of the effective to geometric cross-sections.  While \citet[]{plavchan05} assumed $Q_{SW}=1$, this assumption is confirmed with modeling of dust grains in \citet[]{minato06}.  

From Equations A8 and A9 it follows:
\begin{equation}
\frac{\tau_{SW}}{\tau_{PR}} = 3\frac{Q_{rad}}{Q_{SW}}\frac{L_*}{L_\odot}\frac{\dot M_\odot}{\dot M_*}
\end{equation}
as in \citet[]{plavchan05} with $Q_{rad}/Q_{SW}=1$ assumed.  We note that Equation 3 in \citet[]{plavchan05} is more fundamental than Equation A10 since the solar wind mass-loss rate is variable, but this is a good approximation.  Using the X-ray mass-loss rate relation in \citet[]{wood05}, we can re-express A10 as:
\begin{equation}
\frac{\tau_{SW}}{\tau_{PR}} = 3\frac{Q_{rad}}{Q_{SW}}\frac{L_*}{L_\odot}\left(\frac{L_{X,\odot}}{L_{X,*}}\right)^{1.34}\left(\frac{R_{*}}{R_\odot}\right)^{0.68}
\end{equation}
For the Sun, $L_X/L_\odot\sim5\times10^{-7}$.  When $\tau_{SW}/\tau_{PR}<1$, P-R drag time-scales are longer than stellar wind drag time-scales, and stellar wind drag is relevant.   Using A11 and the approximation that 1.34=4/3, stellar wind drag dominates P-R drag when:
\begin{eqnarray}
\frac{L_X}{L_*}  & > 3^{0.75}\left(\frac{L_{X,\odot}}{L_\odot}\right)\left(\frac{T_\odot}{T_*}\right) \nonumber\\
			   &> 1.13\times10^{-6}\left(\frac{T_\odot}{T_*}\right)
\end{eqnarray}
Equations A11--12 are subject to the condition that F$_X$$<$8$\times$10$^5$ ergs/s/cm$^2$, which can be re-written as:
\begin{equation}
\frac{L_X}{L_*}<1.26\times10^{-5}\left(\frac{T_\odot}{T_*}\right)^4
\end{equation}
For ``saturated'' stars with F$_X$$>$8$\times$10$^5$ ergs/s/cm$^2$, stellar wind mass-loss can still dominate P-R drag, but the mass-loss rates can't be estimated accurately from the X-ray activity.

Stellar wind drag and P-R drag are additive, and can be combined as done in 
\citet[]{strubbe06,fitzgerald07}:
\begin{equation}
\tau_{drag} = 2\times10^6\:\mbox{yr}\left(\frac{r}{50\mbox{AU}}\right)^{2}\left(\frac{a}{\mu\mbox{m}}\right)\left(\frac{\rho_g}{\mbox{g cm}^{-3}}\right)\left(\frac{1}{P_{CPR}}\right)\left(\frac{L_*}{L_\odot}\right)^{-1}
\end{equation}
where $P_{CPR}$ is defined in \citet[]{strubbe06}:
\begin{equation}
P_{CPR} \equiv Q_{rad} + Q_{SW}\frac{\dot M_* c^2}{L_*}
\end{equation}
which we can re-express as:
\begin{equation}
P_{CPR} =Q_{rad}\left(1+\frac{\tau_{PR}}{\tau_{SW}}\right)
\end{equation}

Drag forces predict a disk with constant vertical optical depth ($\equiv$ surface density).  In a steady state, the mass must cross an annulus is constant,
\begin{equation}
M_{\mbox{crossing}} = \mbox{constant} = 2\pi r\: \Sigma_r(r) \: v_r
\end{equation}
The average radial inward drift velocity of a grain goes as $v_r=r/\tau_{drag}\propto r^{-1}$.  Hence, $\Sigma_r(r)$ must also be constant as a function of orbital radius.  An infrared spectra consistent with a power-law follows directly from this.

While stellar wind drag and P-R drag forces can have similar magnitudes, the same is not true for the strength of the radial or ``blowout'' force unless $(\dot M_* / \dot M_\odot)(L_\odot/L_*)\sim3000$ \citep[]{plavchan05,gustafson94}.  Such high mass-loss rates are not supported by recent theoretical arguments \citep[]{holzwarth07}.  The additional factor of $\sim$1000 arises since the ratio of drag forces compared to the ratio of radial forces for P-R and stellar wind is enhanced by a factor of $(c/v_{sw})$ where $v_{sw}$ is the stellar wind velocity. \citet[]{strubbe06} writes the grain blowout radius from the combined effects of radiation and stellar wind as:
\begin{equation}
a_b = \frac{3}{8\pi}\frac{L_*P_{SWR}}{G M_* c \rho_g}
\end{equation}
where
\begin{equation}
P_{SWR} \equiv Q_{rad} + Q_{SW}\frac{\dot M_* c \: v_{sw}}{L_*}
\end{equation}

We estimate the minimum mass of the parent bodies -- asteroid and comet proto-planetary analogs -- by considering how quickly the dust mass must be replenished to maintain a steady-state as in \citet[]{chen05a}.  Grain-grain collision time-scales can be shorter than both P-R and stellar wind drag time-scales.  The shortest time-scale dictates which mechanism is responsible for removing dust, but that shortest time-scale will never be longer than the P-R drag time-scale for the orbital dust distances typically under consideration.  Hence, we use the P-R drag time-scale to estimate minimum total parent body mass, but note that the total parent body mass can be substantially larger if the grain-grain collision time-scale or the stellar wind drag time-scale are significantly shorter \citep[]{chen05a,plavchan05,minato06}.  The minimum total parent body mass is given by:

\begin{equation}
M_{PB} > 4 L_{IR} t_*/c^2
\end{equation}
as in \citet[]{chen01} where t$_*$ is the age of the star. 

Finally, throughout our analysis, we assume $\dot M_*$ to be the average mass-loss rate.  As noted in \citet[]{augereau06}, flaring activity can substantially enhance the stellar wind drag force relative to a quiescent state.  For a star that flares $\sim$10\% of the time with mass-loss rated enhanced by a factor of $\sim$100 during flared states with respect to quiescent activity, the average mass-loss rate can be $\sim$10 times greater than the quiescent mass-loss rate, and hence grain removal time-scales can be $\sim$10 times shorter.  The stellar activity cycle can be assumed to be much shorter than the $\sim$Myr drag-force removal time-scale, and hence individual grain lifetimes will be insensitive, on average, to the flared state of a star.  However, the methods in \citet[]{wood05} and \citet[]{schmidt07} to estimate mass-loss rates are sensitive to the flared state of a star, and a correction factor based on the activity level of the star should be applied to estimate the average mass-loss rate.

\subsubsection{The Transition Between Grain-Grain Collisions and Drag Forces in Disk Dynamical Evolution -- Theory}
From Equations A14 and A6, we derive:
\begin{equation}
\frac{\tau_{drag}}{\tau_{coll}} = \frac{1000}{\gamma}\left(\frac{r}{50\mbox{AU}}\right)^{1/2+\alpha}\left(\frac{Q_{coll}}{P_{CPR}}\right)\left(\frac{M_d}{10^{-3}M_\oplus}\right)^{1}\left(\frac{L_*}{L_\odot}\right)^{-1}\left(\frac{M_*}{M_\odot}\right)^{1/2}
\end{equation}
We consider the dust mass estimate from the fractional infrared excess \citep[$\S$4.4; ][]{chen05a,jura95}:
\begin{equation}
M_d \geq  \frac{16}{3}\pi \frac{L_{IR}}{L_*}\rho_g D^2_{\mbox{min}}\langle a\rangle
\end{equation}
If we set $r=D_{min}=D_{ring}$, and if we assume the dust is originating in a ring with relative width of 0.1, then Equation A21 approximates the local dust mass in the ring.  Furthermore, Equation A14 represents the total time for a dust grain to spiral into the star.  The average time for a dust grain to escape the ring by being dragged inwards before colliding with another grain is further reduced by a factor of $\sim\Delta D_{ring}/(2D_{ring})=0.05$.  If we assume $\langle a\rangle=1\mu$m, $\rho_g=2.5$ g cm$^{-3}$, and $\alpha=0$, we get:
\begin{equation}
\frac{\tau_{drag}}{\tau_{coll}}(\mbox{ring}) = \frac{L_{IR}/L_*}{9\times10^{-6}}  \left(\frac{D_{ring}}{50\mbox{AU}}\right)^{5/2}\left(\frac{Q_{coll}}{P_{CPR}}\right)\left(\frac{L_*}{L_\odot}\right)^{-1}\left(\frac{M_*}{M_\odot}\right)^{1/2}
\end{equation}
For a solar-type star with a young Kuiper-belt analog at 50 AU and $\dot M_{sw}= \dot M_\odot$ such that $Q_{coll}/P_{CPR}=0.75$, Equation A22 predicts that debris disks are drag-force dominated (and collision-force dominated otherwise) when 
\begin{equation}
\frac{L_{IR}}{L_*}<1.2\times10^{-5}
\end{equation}
Increased stellar winds increase the fractional infrared excess luminosity transition in Equation A23 by a factor of $\sim\frac{3}{4}(1+\frac{1}{3}(\dot M_{sw}/\dot M_\odot))$ for a solar-type star.  If $\dot M_{sw}=10\dot M_{\odot}$ as might be reasonable for young, low mass stars, the transition fractional infrared excess luminosity is 4$\times$10$^{-5}$. 

\subsubsection{Planet-Dust Interaction}
When discussing the role of different grain removal mechanisms in debris disks, the effect of planets on the dynamics cannot be neglected.  Dust trapped in orbital resonances with planets are invoked to explain clumps and non-azimuthally symmetric disk structures seen in resolved disks such as Vega and Eps Eri \citep[]{wyatt03,greaves05,krivov07a,krivov07b}.  Herein we consider the dynamically cold model of \citet[]{wyatt03}, with planetesimal (and dust) eccentricities of $<$1\%, to approximate when the presence of a planet is important for disk dynamical evolution relative to the processes already discussed.  

\citet[]{wyatt03} considers a migrating planet with migration rate $\dot a_p$ encountering planetesimals to estimate resonance capture probabilities.  Planetesimals will be captured in resonances with a probability Pr given by the functional form from \citet[]{wyatt03}:
\begin{equation}
\mbox{Pr} = \left[1+\left(\frac{\dot a_p}{\dot a_{0.5}}\right)^{\gamma}\right]^{-1}
\end{equation}
where $\dot a_{0.5}$ is the 50\% capture probability migration rate and $\gamma$ dictates how fast the probability turns over as a function of $\dot a_p$.  \citet[]{wyatt03} determined that $\dot a_{0.5}$ takes the form:
\begin{equation}
\dot a_{0.5} = \frac{1}{X}\left(\frac{M_p}{M_\oplus}\frac{M_\odot}{M_*}\right)^{u}\sqrt{\frac{M_*}{M_\odot}\frac{\mbox{AU}}{D_{pl}}}
\end{equation}
depending on the orbital resonance ($\dot a_{0.5}$ in units of AU/Myr; D$_{pl}$ is the planet orbital radius).  For the 2:1 orbital resonance, \citet[]{wyatt03} finds through computer simulations $X$=5.8 and $u$=1.4; for the 3:2 resonance, $X$=0.37 and $u$=1.4; for the 4:3 resonance X=0.23 and $u$=1.42; and higher order resonances have weaker capture probabilities (bigger X).  If $\dot a_p < \dot a_{0.5}$, planetesimals will have a $>$50\% of being caught in an orbital resonance and vice-versa.

We relate the planet--planetesimal interaction analysis of \citet[]{wyatt03} to planet--dust interaction as in \citet[]{krivov07b}.  First, we replace the planetesimals with dust grains.  Dust grains will orbit on slightly larger orbits than the Keplerian orbits of the planetesimals due to the effect of radiation pressure.  As a result, resonant orbital semi-major axes will be slightly larger and grain-size dependent.  The correction factor to a resonant semi-major axis is $(1-\beta)^{1/3}$, where $\beta$ is the ratio of the radiation pressure force to gravity and is dependent on grain properties.

Second, inward migrating dust under the action of a drag force is equivalent to an outward migrating planet.  We can replace $\dot a_p$  with the dust  inward drift velocity $v_r$  when $|v_r|>>|\dot a_p|$.   In the example of Vega in \citet[]{wyatt03}, a planet migration rate of $\dot a_p=0.45$AU/Myr is derived to model the resolved Vega sub-mm excess.  For comparison, from A.1.2, the dust migration rate in units of AU/Myr from the combined effects of P-R and stellar wind drag goes as:
\begin{equation}
v_r (\mbox{AU/Myr}) = \left(\frac{r}{AU}\right)\left(\frac{\mbox{Myr}}{\tau_{drag}}\right) =  25 \left[\left(\frac{r}{50\mbox{AU}}\right)\left(\frac{a}{\mu\mbox{m}}\right)\left(\frac{\rho_g}{\mbox{g cm}^{-3}}\right)\left(\frac{1}{P_{CPR}}\right)\left(\frac{L_*}{L_\odot}\right)^{-1}\right]^{-1}
\end{equation}
This is a factor of $\sim$50 larger than the inferred planet migration rate for Vega for the default values of grain size, density, orbital distance, etc.   Hence, replacing the planet migration rate with the dust inward drift velocity is appropriate.

With $|v_r|>>|\dot a_p|$ (and consequently $\tau_{coll}/\tau_{drag}>1$), the disk would be drag-force dominated excepting for the presence of a planet.  We expect the dust to spiral inwards towards the sublimation radius unless it is trapped in an orbital resonance, or scattered from the system in the ``resonance overlap region''.

Dust will be trapped into an orbital resonance with $>$50\% probability if
\begin{equation}
v_r < \dot a_{0.5} = \frac{1}{X}\left(\frac{M_p}{M_\oplus}\frac{M_\odot}{M_*}\right)^{u}\sqrt{\frac{M_*}{D_{pl}}}
\end{equation}
which we can re-express as:
\begin{equation}
\frac{M_p}{M_\oplus}> \left[177\:X\:P_{CPR}\:\left(\frac{M_*}{M_\odot}\right)^{u-0.5}\left(\frac{L_*}{L_\odot}\right)\left(\frac{D_{p+q:q}}{50\mbox{AU}}\right)^{-0.5}\left(\frac{a}{\mu\mbox{m}}\right)^{-1} \left(\frac{\rho_g}{\mbox{g cm}^{-3}}\right)^{-1} \right]^{1/u}
\end{equation}
If we set $u$=1.4, we can re-write this in terms of Jovian masses ($M_J$):
\begin{equation}
\frac{M_p}{M_J}> 0.13  \:X^{0.71}\:P_{CPR}^{0.71}\:\left(\frac{M_*}{M_\odot}\right)^{0.64}\left(\frac{L_*}{L_\odot}\right)^{0.71}\left(\frac{D_{p+q:q}}{50\mbox{AU}}\right)^{-0.64}\left(\frac{a}{\mu\mbox{m}}\right)^{-0.71} \left(\frac{\rho_g}{\mbox{g cm}^{-3}}\right)^{-0.71}
\end{equation}
In Equations A28 and A29, $D_{p+q:q}$ corresponds to the semi-major axis of the orbital resonance p+q:p, which can be related to the planet semi-major axis as in \citet[]{krivov07b}:
\begin{equation}
D_{p+q:q} = D_{pl}\left(1-\beta\right)^{1/3}\left(\frac{p+q}{p}\right)^{2/3}
\end{equation}

From Equation A29 we conclude that a proto-planet $>$0.05$M_J$ at 50 AU will trap a majority of 1$\mu$m dust (2.5 g cm$^{-3}$) around a solar type star in the 3:2 orbital resonance ($\dot M_{sw}/\dot M_{\odot}=1$ assumed).  At 10 AU, the minimum planet mass is 0.14 $M_J$, and at 5 AU the minimum planet mass is 0.22 $M_J$.  Since these planet masses are comparable to known extrasolar planet masses, we conclude that dust trapped in orbital resonances with Jovian planets should be a common phenomenon.  In our own Solar System, resonant trapping of a small fraction of inward migrating dust has been observed for the Earth \citep[]{dermott94,reach95}, and would be observable for the Jovian planets if we were not embedded in our own zodiacal cloud.

\citet[]{krivov07b} goes into more detail about what happens to the dust once it is trapped into resonance.  The resulting observability is determined by the survival time of grains in resonance relative to the collision and drag time-scales.  While in resonance, the dust eccentricities will be pumped up until the dust escapes resonance or collides with other dust.  If the dust escapes resonance, it can be scattered inward and continue spiraling into the star, or be scattered out of the system, and this is again dependent on the planet mass.   Additionally, the orbits of dust quickly become chaotic for planets of a few Jovian masses, providing an upper limit to the planet mass for observable resonant trapping in addition to the lower limit in Equation A29 \citep[]{kuchner03}.  We note that the minimum planet mass required to capture dust is proportional to $P_{CPR}^{0.7}$.  If the stellar wind mass-loss rate is substantially larger than solar, it affects the planet-dust interaction and disk dynamics overall.  Stars with high mass-loss rates are less likely to have observable dust in resonance with planets of a similar mass compared to stars with small mass-loss rates.

Dust in non-resonant orbits can enter the``resonance overlap region'' and be scattered out of the system on short time-scales:
\begin{equation}
\left|\frac{r}{D_{pl}}-1\right| < 1.3 \left(\frac{M_p}{M_*}\right)^{2/7}
\end{equation}
where $r$ and $D_{pl}$ are the dust and planet semi-major axes respectively, and $M_p$ and $M_*$ are the planet and stellar masses in the same units.   Again, if the planet is massive enough, the planet can effectively scatter a majority of the inspiraling dust out of the system, creating a disk with an inner hole.    This is the ``disk with inner hole'' model is widely used to explain infrared spectra with no short-wavelength ``hot excess'', but longer wavelength ``cold excess''.    Such a disk would have a constant surface density truncated at $R_{IN}=D_{pl}$ and $R_{OUT}$ equal to the orbital radius of the colliding parent bodies producing the dust.   Conversely, the presence of a warm excess where collisional time-scales are long relative to drag time-scales implies the lack of a planet of sufficient mass interior to where dust is being produced to halt the inward migration.


In summary, if  $\tau_{coll}/\tau_{drag}<1$, then the radial inward drift velocity of dust is effectively zero.  Since dust grains undergo destructive collisions before they can drift into a planet, dust will be confined to where the parent bodies are located, such as in a ring or in resonance with a planet (if the planet is migrating).  The presence of a planet can likely be inferred from the presence of clumpy disk structure.  If  $\tau_{coll}/\tau_{drag}>1$, the presence of Jovian planets halts or substantially arrests the inward migration of dust under drag forces.  If the planet is on the order of a Jovian mass, the dust will also likely be trapped in clumpy resonances that could dominate the fractional infrared excess.  The presence of non-azimuthal disk structures for disks with $L_{IR}/L_*$$<$10$^{-4}$ can be used to infer the presence of Jovian planets without directly detecting the planets themselves.  Conversely, a power-law infrared excess can be used to exclude the presence of Jovian planets at the orbital radii of the emitting dust.

\subsection{Spectral Type Disk Fraction Dependence}

Herein we present six factors that complicate the interpretation of the dependence of the 70 $\mu$m disk fraction on spectral-type.  Three of these factors apply particularly to our sample, and three are intrinsic to differences between spectral types.  We consider each in turn and its impact on the apparent discrepancy between the 70 $\mu$m disk fractions as a function of spectral type.     We conclude from these six factors that the paucity of bright 70$\mu$m excesses (F$_{70}$/F$_*$$>$$\sim$5--10) around late-K and M dwarfs is likely a real phenomenon rather than due to sensitivity or systematic bias. 

First, the average age of the T$_*$$<$5000 K sub-sample is 51 Myr (median 12.25 Myr), compared to the average age of 380 Myr (median 340 Myr) for the 5000 K $<$ T$_*$ $<$ 6000 K sub-sample.  Second, the average flux ratio upper-limit of the T$_*$$<$5000 K sub-sample is $F_{70}/F_*$=18.5 (median 6.1), and is  $F_{70}/F_*$=2.1 (median 1.8) for the 5000 K $<$ T$_*$ $<$ 6000 K sub-sample.   The first two factors cancel to first order, when accounting for the third factor -- the assumed age evolution of the fractional infrared excess.  We assume that L$_{IR}$/L$_*$ $\propto$ t$_*^{-1}$ \citep[e.g.,][]{rieke05}.    While we are $\sim$3--9 times less sensitive on average to 70 $\mu$m excesses around the cooler stars, a steady-state collisional evolution model predicts that they should be $\sim$10 times brighter at younger ages.  

Note that the first two factors do not cancel if we assume a different time-dependence for the evolution of L$_{IR}$/L$_*$. \citet[]{currie08b} posits that debris disks around solar type stars undergo a ``rise and fall'' of debris disk brightness, peaking at ages of $\sim$10--15 Myr.   Low mass K and M dwarfs could undergo a similar but delayed ``rise and fall'' of debris disk brightness, peaking at ages of 30--50 Myr \citep[e.g. NGC 2547,][]{forbrich08,kenyon08}.  Under such an alternate scenario, for the median age of the T$_*$$<$5000 K sub-sample, we would be catching the stars during the rise and before the peak in emission from debris disks.  

Fourth, we have a slightly higher fraction of excesses at 24 $\mu$m for stars with T$_*$ $<$ 5000 K (18.5$^{9.6}_{5.2}$\%) than for stars with 5000 K $<$ T$_*$ $<$ 6000 K (14.3$^{9.0}_{4.3}$\%).  This result implies that small 24$\mu$m excesses (e.g. $\sim$10--15\% flux density excess)  for young T$_*$ $<$ 5000 K stars are as common as they are around hotter stars, whereas larger fractional infrared excesses -- e.g. $>$100\% as probed with our 70 $\mu$m observations -- remain rare around K and M dwarfs relative to higher mass stars.  The lack of large fractional infrared excesses around K and M dwarfs still necessitates an explanation.

Fifth, at 70 $\mu$m the emitting dust orbits at smaller radii for cooler stars.    At small orbital radii ($\sim$3--10 AU, Table 4), we are equally sensitive to dust for all stars in our sample.  However, the hotter stars with strong 70 $\mu$m excesses have dust at orbital radii that we cannot probe with the 70 $\mu$m observations of the cooler stars.   The fifth factor implies that the lack of low mass debris disks we observe at 70 $\mu$m could depend on disk geometry.  If young Kuiper belt analogs (that produce the dust we are observing) form at radii that scale with the stellar luminosity, then we are indeed observing fewer massive, bright debris disks around the cooler K and M dwarfs compared to G dwarfs at comparable fractional infrared excess (e.g., F$_{70}$/F$_*$ $>$ 5--10).  Theoretical modeling indicates that the orbital distance for the formation of Jovian planets at the ``snow line'' has a weak dependence on stellar mass \citep[]{kennedy08,kennedy07,kennedy06}.   This implies that young Kuiper Belt analogs may form at smaller radii for lower mass stars, but this is not conclusive.   

If instead the young Kuiper Belt analogs form at comparable orbital radii independent of spectral type, our observations are not sensitive to detect these cold debris disks around late-K and M dwarfs.      Sub-millimeter observations by \citet[]{lestrade06,liu04} have revealed three cold debris disks around M dwarfs, albeit at low S/N for GJ 842.2.  The second sub-mm excess source, GJ 182, is not detected in excess in our sample, although an excess 5 times the photosphere at 70 $\mu$m would not have been detected.   The third source, AU Mic, is detected in excess of $\sim$9 times the photosphere at 70$\mu$m in our sample.  These initial sub-mm observations of low-mass stars imply cold sub-mm disks around late-K and M dwarfs could be as common as 70 $\mu$m excesses around older G dwarfs \citep[$\sim$15\%, ][]{bryden06}.  Observing more M dwarfs at sub-mm wavelengths will further address this possibility.

The dust removal mechanisms and time-scales are the sixth factor.  The dust lifetimes around K- and M-type stars due to P-R drag and collision time-scales are (sometimes significantly) longer and the dust blowout radii are smaller, scaling with either the stellar luminosity or the stellar mass.  For example, an M0 dwarf has a P-R drag time-scale $\sim$10 times longer than for a G2 star, a collision time-scale 40\% longer (for the same disk mass), and a blowout radius for dust grains $\sim$10 times smaller (these follow from ${\S}$A.1).   A collisional-cascade steady-state predicts the number density of grains scales as a$^{-3.5}$ where a is the grain size and the minimum grain size is set by the blowout radius.  The smaller blowout radius and longer grain lifetimes for K and M dwarfs imply larger dust abundances and more smaller grains relative to G-type stars for a similar amount of colliding parent bodies.  Such an excess of dust should be more easily detectable, and this is not observed at 70 $\mu$m.  A mechanism like stellar wind drag is necessary to account for this paucity of bright 70 $\mu$m debris disks around M dwarfs.

\begin{deluxetable}{lllrrrrrrrr}
\rotate
\tablecolumns{11}
\tabletypesize{\scriptsize}
\tablewidth{0pc}
\tablecaption{Stellar Properties}
\tablehead{
\colhead{Name}	&	\colhead{HD / }	&	\colhead{Spectral}	&	\colhead{D\tablenotemark{a}}	&	\colhead{Assoc}	&	\colhead{ROSAT} & \colhead{Lithium} & \colhead{log(R$^\prime_{HK}$)\tablenotemark{c}} &  \multicolumn{2}{c}{Stellar Rotation} & \colhead{References\tablenotemark{d}}\\
\colhead{}	&	\colhead{Name 2}	&	\colhead{Type\tablenotemark{a}}	&	\colhead{(pc)}	&	\colhead{-iation\tablenotemark{b}}	&	\colhead{(cnts/sec)}	&	\colhead{Abundance}	&	\colhead{}&	\colhead{v\textit{sin}i}&	\colhead{Period} &	\colhead{}\\
\colhead{}	&	\colhead{}	&	\colhead{}	&	\colhead{}	&	\colhead{}	&	\colhead{}	&	\colhead{log(N(Li))}	&	\colhead{}&	\colhead{(km/s)}&	\colhead{(days)}&	\colhead{}\\
\colhead{}	&	\colhead{}	&	\colhead{}	&	\colhead{}	&	\colhead{}	&	\colhead{}	&	\colhead{(H=12)}	&	\colhead{}&	\colhead{}&	\colhead{}&	\colhead{}
}
\startdata
\nodata	&	1835	\tablenotemark{e} &	G0 	&	20	&	HSC	&	2.89E-01	&	2.58$\pm$0.10	&	-4.46	&	3.6	&	7.81	&	1,1,2,1	\\
\nodata	&	7590	&	G0 	&	24	&	\nodata	&	1.91E-01	&	2.75$\pm$0.10	&	-4.46	&	6.4	&	5.67	&	1	\\
BB Scl\tablenotemark{f,g}	&	9770	&	G5 	&	24	&	\nodata	&	2.59E+00	&	\nodata	&	\nodata	&	2	&	\nodata	&	3	\\
\nodata	&	10008	&	G5 	&	24	&	LA	&	1.44E-01	&	1.96$\pm$0.12	&	-4.38	&	2.9	&	\nodata	&	1	\\
\nodata	&	11131	&	G0 	&	23	&	\nodata	&	3.86E-01	&	2.42$\pm$0.10	&	-4.47	&	3.4	&	8.92	&	1,1,4,1	\\
\nodata	&	15089\tablenotemark{f,g}	&	A5p 	&	43	&	\nodata	&	2.28E-01	&	\nodata	&	\nodata	&	49	&	1.74	&	5,6,7	\\
\nodata	&	16287\tablenotemark{h}	&	K0 	&	24	&	\nodata	&	\nodata	&	0.55	&	-4.32	&	4.9	&	\nodata	&	8	\\
\nodata	&	16555	&	A5 	&	44	&	\nodata	&	3.20E-03	&	\nodata	&	\nodata	&	315	&	\nodata	&	9,7	\\
\nodata	&	17240\tablenotemark{f}	&	F2 	&	47	&	\nodata	&	\nodata	&	\nodata	&	\nodata	&	45	&	\nodata	&	10,11,12	\\
DO Eri	&	24712	&	F0 	&	49	&	\nodata	&	\nodata	&	\nodata	&	\nodata	&	18	&	12.46	&	5,6,11,12	\\
v833 Tau\tablenotemark{e}	&	GJ 171.2	&	K3\tablenotemark{i} 	&	18	&	\nodata	&	2.69E+00	&	\nodata	&	-4.057	&	\nodata	&	\nodata	&	13	\\
\nodata	&	29697	&	K2 	&	13	&	UMG	&	2.36E+00	&	0.754	&	-4.1	&	9.5	&	4	&	14,13,15,15	\\
Gl 182	&	\nodata	&	M0\tablenotemark{j} 	&	27	&	LA	&	6.51E-01	&	1.8	&	\nodata	&	14	&	1.9	&	16,3,17	\\
\nodata	&	36435	&	G0 	&	20	&	\nodata	&	1.42E-01	&	1.60$\pm$0.15	&	-4.44	&	4.5	&	\nodata	&	1,1,18	\\
AB Dor\tablenotemark{e}	&	36705	&	G0	&	15	&	ABDMG	&	7.22E+00	&	3.1	&	\nodata	&	53	&	0.51	&	16,3,17	\\
\nodata	&	41824\tablenotemark{f}	&	G5	&	30	&	\nodata	&	1.90E+00	&	\nodata	&	-4.17	&	8	&	3.3	&	19,3,20	\\
\nodata	&	41593	&	K0 	&	15	&	UMG	&	2.57E-01	&	1.22$\pm$0.17	&	-4.36	&	5.0	&	7.97	&	1	\\
GJ 3400A\tablenotemark{f}	&	48189	&	G0	&	22	&	ABDMG	&	2.33E+00	&	3.3	&	-4.29	&	17.6	&	\nodata	&	16,19,3	\\
\nodata	&	51849\tablenotemark{f}	&	K0 	&	22	&	\nodata	&	5.03E-01	&	\nodata	&	\nodata	&	\nodata	&	\nodata	&	\nodata	\\
\nodata	&	52698	&	K0 	&	15	&	\nodata	&	1.99E-01	&	\nodata	&	-4.59	&	3.5	&	\nodata	&	1	\\
\nodata	&	59967	&	G5 	&	22	&	\nodata	&	3.20E-01	&	2.50$\pm$0.10	&	-4.44	&	3.4	&	\nodata	&	1	\\
\nodata	&	63433	&	G0 	&	22	&	UMG	&	2.24E-01	&	2.33$\pm$0.10	&	-4.34	&	6.1	&	6.46	&	1	\\
\nodata	&	72760	&	G5 	&	22	&	\nodata	&	9.65E-02	&	0.57$\pm$0.45	&	-4.39	&	4.2	&	\nodata	&	1	\\
\nodata	&	73350\tablenotemark{g}	&	G0 	&	24	&	HSC	&	1.30E-01	&	2.09$\pm$0.10	&	-4.49	&	3.8	&	6.14	&	1	\\
\nodata	&	74576	&	K0 	&	11	&	\nodata	&	4.60E-01	&	\nodata	&	-4.31	&	3.9	&	\nodata	&	19,3	\\
\nodata	&	92139\tablenotemark{f}	&	F2	&	27	&	\nodata	&	2.03E-01	&	\nodata	&	\nodata	&	15	&	\nodata	&	9,11,12	\\
\nodata	&	92945	&	K0	&	22	&	LA	&	1.18E-01	&	2.41	&	-4.39	&	5.1	&	\nodata	&	8,21	\\
\nodata	&	97334\tablenotemark{g}	&	G0 	&	22	&	LA	&	3.20E-01	&	2.64$\pm$0.09	&	-4.41	&	5.6	&	8.25	&	1	\\
TWA 14\tablenotemark{k}	&	\nodata	&	M0 	&	55	&	TW Hya	&	1.07E-01	&	\nodata	&	\nodata	&	\nodata	&	0.63	&	22	\\
TWA 12\tablenotemark{k}	&	\nodata	&	M2 	&	55	&	TW Hya	&	4.29E-01	&	\nodata	&	\nodata	&	15	&	3.303	&	3,23	\\
TWA 13A\tablenotemark{k}	&	\nodata	&	M1 	&	55	&	TW Hya	&	1.16E-01	&	\nodata	&	\nodata	&	12.3	&	5.56	&	3,22	\\
TWA 13B\tablenotemark{k}	&	\nodata	&	M1	&	55	&	TW Hya	&	1.16E-01	&	\nodata	&	\nodata	&	\nodata	&	5.35	&	22	\\
\nodata	&	103928\tablenotemark{e}	&	F0 	&	47	&	\nodata	&	5.00E-03	&	\nodata	&	\nodata	&	94	&	0.70	&	5,24,11,12	\\
\nodata	&	105963\tablenotemark{g}	&	K0	&	27	&	\nodata	&	1.50E-01	&	2.06	&	-4.22	&	6	&	7.44	&	8,8,8,20	\\
\nodata	&	109011	&	K0 	&	24	&	UMG	&	6.00E-02	&	0.81$\pm$0.15	&	-4.36	&	5.5	&	8.81	&	1	\\
TWA 15B\tablenotemark{k}	&	\nodata	&	M2	&	55	&	TW Hya	&	1.16E-01	&	\nodata	&	\nodata	&	\nodata	&	0.72	&	22	\\
TWA 15A\tablenotemark{k}	&	\nodata	&	M1.5 	&	55	&	TW Hya	&	1.16E-01	&	\nodata	&	\nodata	&	\nodata	&	0.65	&	22	\\
TWA 16\tablenotemark{k}	&	\nodata	&	M1.5 	&	55	&	TW Hya	&	1.07E-01	&	\nodata	&	\nodata	&	\nodata	&	\nodata	&	\nodata	\\
\nodata	&	112429	&	F0 	&	29	&	\nodata	&	\nodata	&	\nodata	&	\nodata	&	130	&	\nodata	&	9,11,12	\\
\nodata	&	113449	&	K0 	&	22	&	\nodata	&	1.70E-01	&	2.08$\pm$0.12	&	-4.34	&	5.8	&	6.54	&	1,13,1,1	\\
TWA 17\tablenotemark{k}	&	\nodata	&	K5 	&	55	&	TW Hya	&	7.21E-02	&	\nodata	&	\nodata	&	\nodata	&	0.69	&	22	\\
TWA 18\tablenotemark{k}	&	\nodata	&	M0.5 	&	55	&	TW Hya	&	6.22E-02	&	\nodata	&	\nodata	&	\nodata	&	1.11	&	22	\\
\nodata	&	116956	&	G5 	&	22	&	LA	&	2.01E-01	&	1.42$\pm$0.12	&	-4.447	&	5.6	&	7.8	&	1,13,1,1	\\
\nodata	&	123998	&	A2p 	&	43	&	\nodata	&	\nodata	&	\nodata	&	\nodata	&	15	&	\nodata	&	9,11,12	\\
\nodata	&	125158	&	A3 	&	48	&	\nodata	&	\nodata	&	\nodata	&	\nodata	&	\nodata	&	\nodata	&	11,12	\\
\nodata	&	128987	&	G5 	&	24	&	\nodata	&	9.81E-02	&	1.78$\pm$0.11	&	\nodata	&	3.2	&	9.35	&	1	\\
\nodata	&	128400	&	G0 	&	20	&	UMG	&	1.40E-01	&	\nodata	&	-4.56	&	4.4	&	\nodata	&	1,3	\\
\nodata	&	135599	&	K0 	&	16	&	UMG	&	9.93E-02	&	0.34$\pm$0.28	&	\nodata	&	4.6	&	5.97	&	1	\\
\nodata	&	141272	&	G5 	&	21	&	LA	&	1.87E-01	&	0.48$\pm$0.27	&	-4.452	&	4.0	&	14.01	&	1,13,1,1	\\
\nodata	&	144197	&	A3p 	&	38	&	\nodata	&	\nodata	&	\nodata	&	\nodata	&	7	&	\nodata	&	25,11,12	\\
\nodata	&	148367\tablenotemark{f}	&	A2 	&	37	&	\nodata	&	2.88E-01	&	\nodata	&	\nodata	&	18	&	\nodata	&	25,11,12	\\
\nodata	&	165185	&	G5 	&	17	&	UMG	&	5.85E-01	&	2.65$\pm$0.10	&	-4.49	&	7.2	&	\nodata	&	1	\\
RE1816+541\tablenotemark{l}	&	EY Dra	&	M1.5 	&	30	&	LA	&	2.92E-01	&	\nodata	&	\nodata	&	61	&	0.459	&	26,17	\\
\nodata	&	175742	&	K0 	&	21	&	UMG	&	1.88E+00	&	1.39	&	-3.90	&	14	&	2.907	&	8,8,8,23	\\
\nodata	&	177724	&	A0 	&	26	&	\nodata	&	1.41E-01	&	\nodata	&	\nodata	&	317	&	\nodata	&	5,11,12	\\
\nodata	&	180161	&	K0 	&	20	&	\nodata	&	1.55E-01	&	0.86$\pm$0.14	&	-4.44	&	3.1	&	9.7	&	1	\\
\nodata	&	186219	&	A3 	&	42	&	\nodata	&	\nodata	&	\nodata	&	\nodata	&	125	&	\nodata	&	9,7	\\
AT Mic\tablenotemark{f}	&	196982	&	M4.5\tablenotemark{j} 	&	10	&	BPMG	&	3.91E+00	&	\nodata	&	\nodata	&	10.1	&	\nodata	&	3	\\
eta Ind	&	197157	&	F0	&	24	&	\nodata	&	\nodata	&	\nodata	&	\nodata	&	\nodata	&	\nodata	&	11,12	\\
AU Mic\tablenotemark{e}	&	197481	&	M0\tablenotemark{j} 	&	10	&	BPMG	&	5.95E+00	&	\nodata	&	\nodata	&	9.3	&	4.9	&	3,17	\\
BO Mic	&	197890	&	K0 	&	44	&	LA	&	6.11E+00	&	\nodata	&	\nodata	&	120	&	0.38	&	27,28	\\
\nodata	&	358623\tablenotemark{m}	&	K6 	&	24	&	BPMG	&	2.36E-01	&	\nodata	&	\nodata	&	15.6	&	3.41	&	3,23	\\
SAO 145139\tablenotemark{f}	&	\nodata	&	K5\tablenotemark{j} 	&	26	&	\nodata	&	6.03E-01	&	\nodata	&	\nodata	&	\nodata	&	\nodata	&	\nodata	\\
\nodata	&	202730\tablenotemark{e}	&	A5 	&	30	&	\nodata	&	4.19E-01	&	\nodata	&	\nodata	&	135	&	\nodata	&	25,11,12	\\
\nodata	&	203244	&	G5 	&	20	&	\nodata	&	1.90E-01	&	1.64$\pm$0.15	&	-4.39	&	4	&	\nodata	&	1,1,3	\\
LO Peg 	&	HIP 106231	&	K3\tablenotemark{i} 	&	25	&	ABDMG	&	9.23E-01	&	1.624	&	-3.906	&	\nodata	&	0.42	&	14,13,23	\\
Gl 859AB\tablenotemark{f}	&	212697	&	G0 	&	16	&	\nodata	&	8.47E-01	&	2.9	&	\nodata	&	9.7	&	\nodata	&	29,18	\\
HKAqr	&	Gl 890	&	M0\tablenotemark{j} 	&	22	&	CMG	&	4.27E-01	&	\nodata	&	\nodata	&	\nodata	&	0.431	&	17	\\
\nodata	&	218738\tablenotemark{g}	&	G5 	&	25	&	\nodata	&	1.93E+00	&	2.13	&	-3.90	&	7	&	\nodata	&	8	\\
Gl 907.1\tablenotemark{f}	&	\nodata	&	K7\tablenotemark{j} 	&	27	&	\nodata	&	4.04E-01	&	\nodata	&	\nodata	&	3.4	&	\nodata	&	3	\\
\enddata
 \tablenotetext{a}{Spectral type from the HD Catalog unless otherwise noted \citep[]{nesterov95}.  Trigonometric parallax distance from Hipparcos unless otherwise noted.}
 \tablenotetext{b}{TW Hya = $\sim$ 8 Myr TW Hya association \citep[]{kastner97,webb99}; BPMG = $\sim$12 Myr $\beta$ Pic moving group \citep[]{zuckerman01b}; ABDMG = $\sim$50--100 Myr AB Dor moving group \citep[]{zuckerman04,luhman05}; LA = $\sim$20--100 Myr Local Association \citep[]{montes01a}; CMG = $\sim$200 Myr Castor moving group \citep[]{montes01a,barrado98}; UMG =  $\sim$300--500 Myr Ursa major group \citep[]{king03,soderblom93,soderblom93b}; and HSC = Hyades supercluster\citep[]{montes01a}}
\tablenotetext{c}{Calcium II R$^\prime_{HK}$ $\equiv$ log(L$_{HK}$/L$_{bol}$) corrected for photospheric contribution.}
\tablenotetext{d}{References listed in order of Association,Lithium Abundance,log(R$^\prime_{HK}$),Stellar Rotation, and young A star sources where appropriate.}
\tablenotetext{e}{Visual companion with separation $>$5$^{\prime\prime}$ not detected with \textit{Spitzer}.  The visual companion is of unknown age/distance, and is excluded from our sample. Projected separation is greater than 100 AU.}
\tablenotetext{f}{All photometry in this paper includes both A and B components (separation $<$5$^{\prime\prime}$), including co-adding photometry if the binary is resolved in some (but not all) bandpasses.  Projected separation is less than 100 AU.}
\tablenotetext{g}{Visual companion with separation $>$5$^{\prime\prime}$ detected with \textit{Spitzer} at 24 $\mu$m and not at 70 $\mu$m.  The visual companion is of unknown age/distance, and is determined to be photospheric at 24 $\mu$m from the K$_s$-[24] color.  Projected separation is greater than 100 AU.  We exclude this companion from our sample.}
\tablenotetext{h}{Einstein X-ray detection of 3.7$\times 10^{-13}$ ergs/cm$^2$/s \citep[]{johnson86}.}
\tablenotetext{i}{\citet[]{gray03}}
\tablenotetext{j}{\citet[]{hawley96}}
\tablenotetext{k}{Distance set to median distance of TWA 1--19 members with known trigonometric parallax as in \citep[]{low05}.  We note that \citet[]{lawson05} present evidence for a median distance of $\sim$90pc. Due to uncertainties in age and distance, we do not use photometric distances on a star by star basis.  Spectral types from \citet[]{barrado06}.}
\tablenotetext{l}{Distance estimated from absolute K magnitude distance modulus for a 115 Myr M1.5 dwarf.  Age inferred from possible Pleiades measurement and L$_X$/L$_*$ implied age of $<$300 Myr \citep[]{messina03}. Spectral type from \citet[]{jeffries94}.  The distance derived from the K$_s$ magnitude is not consistent with Pleiades membership; hence the age, distance, and estimated radius for this source are not well-constrained.}
\tablenotetext{m}{Distance estimated from absolute K magnitude distance modulus for a 12 Myr K7 dwarf.  Age inferred from membership in the $\beta$ Pic moving group.  Spectral type from \citet[]{torres06}.}
\tablerefs{1 -- \citet[]{gaidos00}, 2 -- \citet[]{valenti05}, 3 -- \citet[]{torres06}, 4 -- \citet[]{soderblom93}, 5 -- \citet[]{kudryavtsev03}, 6 -- \citet[]{royer02}, 7 -- \citet[]{stauffer00}, 8 -- \citet[]{strassmeier00}, 9 -- \citet[]{glebocki00}, 10 --  \citet[]{danziger72}, 11 -- \citet[]{song01}, 12 -- \citet[]{song00}, 13 -- \citet[]{gray03}, 14 -- \citet[]{fischer98}, 15 -- \citet[]{montes01b}, 16 -- \citet[]{favata98}, 17 -- \citet[]{messina03}, 18 -- \citet[]{pace03},  19 -- \citet[]{henry96}, 20 -- \citet[]{kazarovets06}, 21 -- \citet[]{gray06}, 22 -- \citet[]{lawson05}, 23 -- \citet[]{pojmanski05}, 24 -- \citet[]{koen02}, 25 -- \citet[]{erspamer03}, 26 -- \citet[]{jeffries94}, 27 -- \citet[]{bromage92}, 28 -- \citet[]{cutispoto97}, 29 -- \citet[]{pallavicini87}}

\end{deluxetable} 

\begin{deluxetable}{lrrrrrrrr}
\tablecolumns{9}
\tabletypesize{\scriptsize}
\tablewidth{0pc}
\tablecaption{MIPS 24 $\mu$m and 70 $\mu$m Fluxes (Not Color- Corrected) }
\tablehead{
\colhead{}	&	\colhead{Measured\tablenotemark{a}}	&	\colhead{Predicted}	& \colhead{}	& \colhead{Measured\tablenotemark{a}}	&	\colhead{Predicted} & \colhead{} & \colhead{} & \colhead{} \\
\colhead{}	&	\colhead{MIPS}	& \colhead{Photosphere\tablenotemark{b}}	&	\colhead{}	&	\colhead{MIPS}	&		\colhead{Photosphere\tablenotemark{b}}	&	\colhead{}&	\colhead{}& \colhead{New} \\
\colhead{}	&	\colhead{F$_{\nu}$(24 $\mu$m)}	 &	\colhead{F$_{\nu}$(24 $\mu$m)}	&	\colhead{}	&	\colhead{F$_{\nu}$(70 $\mu$m)}	&		\colhead{F$_{\nu}$(70 $\mu$m)}	&	\colhead{}&	\colhead{}& \colhead{Detection of} \\
\colhead{Name/HD}	&	\colhead{(mJy)}		 & \colhead{(mJy)}	&	\colhead{$\chi_{24}$}	&	\colhead{(mJy)}	&	\colhead{(mJy)}	&	\colhead{$\chi_{70}$}&	\colhead{References\tablenotemark{c}}& \colhead{Excess?} \\
}
\startdata
\multicolumn{9}{c}{Stars with MIPS-24 and MIPS-70 excesses\tablenotemark{d}}\\
\hline
\dataset[ADS/Sa.Spitzer#0004626944]{10008}	&	41.6$\pm$0.5	&	36	&	12.1	&	30$\pm$1\tablenotemark{e} 	&	4.1	&	19.8	&	1	& Y \\
\dataset[ADS/Sa.Spitzer#0004639744]{73350}	&	66.2$\pm$1.3	&	58	&	6.22	&	123$\pm$4	&	6.6	&	32.3	&	\nodata & Y 	\\
\dataset[ADS/Sa.Spitzer#0004627712]{112429}	&	132.8$\pm$1.6	&	121	&	7.11	&	58$\pm$4\tablenotemark{f}	&	14	&	10.3	&	1	& N \\
\dataset[ADS/Sa.Spitzer#0004629504]{135599}	&	82.5$\pm$0.9	&	74	&	9.15	&	106$\pm$10	&	8.5	&	9.4	&	\nodata	& Y \\
\hline
\multicolumn{9}{c}{Stars with MIPS-70 excesses\tablenotemark{d}}\\
\hline
\dataset[ADS/Sa.Spitzer#0004634624]{7590}	&	62.7$\pm$1.0	&	61	&	2.07	&	233$\pm$16	&	6.9	&	13.9	&	\nodata	& Y \\
\dataset[ADS/Sa.Spitzer#0004633856]{59967}	&	70.5$\pm$2.5	&	65	&	2.14	&	36$\pm$4	&	7.4	&	7.5	&	\nodata	& Y \\
\dataset[ADS/Sa.Spitzer#0004640256]{92945}	&	39.5$\pm$1.0	&	39	&	0.57	&	305$\pm$15	&	4.4	&	19.4	&	1	& N \\
\dataset[ADS/Sa.Spitzer#0004637440]{AU Mic}	&	155.2$\pm$3.2	&	150	&	1.65	&	223$\pm$26	&	17	&	8.0	&	1,2	& N \\
\hline
\multicolumn{9}{c}{Stars with MIPS-24 excesses\tablenotemark{d}}\\
\hline
\dataset[ADS/Sa.Spitzer#0004639232]{GJ 3400A}	&	120.1$\pm$1.4	&	109	&	8.25	&	16$\pm$5\tablenotemark{g}	&	12	&	0.7	&	1	& Y \\
\dataset[ADS/Sa.Spitzer#0004628480]{123998}	&	136.6$\pm$2.3	&	123	&	5.7	&	$<$20	&	14	&	\nodata	&	\nodata	& Y \\
\dataset[ADS/Sa.Spitzer#0004636928]{175742}	&	46.9$\pm$0.6	&	41	&	10.8	&	$<$13	&	4.7	&	\nodata	&	1	& Y \\
\dataset[ADS/Sa.Spitzer#0004637184]{AT Mic}	&	131.6$\pm$2.1\tablenotemark{f}	&	114	&	8.21	&	22$\pm$5\tablenotemark{f}	&	13	&	1.8	&	1,2	& Y \\
\dataset[ADS/Sa.Spitzer#0004637696]{BO Mic}	&	16.4$\pm$0.4	&	14	&	5.66	&	$<$12	&	1.6	&	\nodata	&	1	& Y \\
\dataset[ADS/Sa.Spitzer#0004643840]{358623}	&	14.2$\pm$0.2	&	13	&	7.28	&	$<$9	&	1.5	&	\nodata	&	1,2	& Y \\
\dataset[ADS/Sa.Spitzer#0004635904]{Gl 907.1}	&	26.7$\pm$0.4	&	24	&	5.44	&	$<$13	&	2.8	&	\nodata	&	1	& Y \\
\hline
\multicolumn{9}{c}{Stars with no MIPS excesses\tablenotemark{d}}\\
\hline
\dataset[ADS/Sa.Spitzer#0004631808]{1835}	&	86.0$\pm$1.0	&	81	&	4.95	&	18$\pm$4\tablenotemark{g}	&	9.3	&	2.3	&	1	& \nodata\\
\dataset[ADS/Sa.Spitzer#0004640768]{BB Scl}	&	87.6$\pm$1.4	&	97	&	-6.89	&	$<$13	&	11	&	\nodata	&	1	& \nodata \\
\dataset[ADS/Sa.Spitzer#0004627456]{11131}	&	64.2$\pm$2.3	&	62	&	0.83	&	$<$16	&	7.1	&	\nodata	&	1	& \nodata \\
\dataset[ADS/Sa.Spitzer#0004630528]{15089}	&	142.1$\pm$8.6	&	143	&	-0.08	&	28$\pm$3\tablenotemark{h}	&	16	&	3.7	&	\nodata	& \nodata\\
\dataset[ADS/Sa.Spitzer#0004636672]{16287}	&	30.8$\pm$0.8	&	30	&	0.89	&	$<$13	&	3.4	&	\nodata	&	\nodata	& \nodata\\
\dataset[ADS/Sa.Spitzer#0004630784]{16555}	&	108.5$\pm$1.7	&	111	&	-1.24	&	$<$12	&	13	&	\nodata	&	\nodata	& \nodata\\
\dataset[ADS/Sa.Spitzer#0004631040]{17240}	&	56.2$\pm$0.7	&	54	&	3.07	&	31$\pm$6\tablenotemark{i}	&	6.2	&	3.9	&	\nodata	& \nodata\\
\dataset[ADS/Sa.Spitzer#0004632832]{DO Eri}	&	56.6$\pm$1.0	&	56	&	0.53	&	$<$12	&	6.4	&	\nodata	&	\nodata	& \nodata \\
\dataset[ADS/Sa.Spitzer#0004641792]{v833 Tau}	&	63.5$\pm$2.5	&	62	&	0.77	&	$<$28	&	7.0	&	\nodata	&	\nodata	& \nodata \\
\dataset[ADS/Sa.Spitzer#0004638464]{29697}	&	69.7$\pm$0.7	&	68	&	2.53	&	$<$31	&	7.7	&	\nodata	&	1	& \nodata \\
\dataset[ADS/Sa.Spitzer#0004635392]{Gl 182}	&	28.7$\pm$0.7	&	27	&	2.06	&	$<$19	&	3.1	&	\nodata	&	1	& \nodata \\
\dataset[ADS/Sa.Spitzer#0004633088]{36435}	&	61.0$\pm$0.7	&	59	&	2.24	&	$<$12	&	6.8	&	\nodata	&	\nodata	& \nodata \\
\dataset[ADS/Sa.Spitzer#0004638720]{AB Dor}	&	106.0$\pm$8.8	&	95	&	1.22	&	$<$47	&	11	&	\nodata	&	1	& \nodata \\
\dataset[ADS/Sa.Spitzer#0004638976]{41824}	&	85.4$\pm$1.5	&	84	&	0.74	&	$<$11	&	9.6	&	\nodata	&	\nodata	& \nodata \\
\dataset[ADS/Sa.Spitzer#0004633344]{41593}	&	84.6$\pm$1.4	&	84	&	0.34	&	$<$22	&	9.6	&	\nodata	&	\nodata	& \nodata \\
\dataset[ADS/Sa.Spitzer#0004639488]{51849}	&	27.9$\pm$0.4	&	26	&	4.67	&	$<$10	&	3.0	&	\nodata	&	1	& \nodata \\
\dataset[ADS/Sa.Spitzer#0004633600]{52698}	&	102.1$\pm$2.0	&	100	&	1.15	&	$<$13	&	11	&	\nodata	&	\nodata	& \nodata \\
\dataset[ADS/Sa.Spitzer#0004634112]{63433}	&	58.6$\pm$0.6	&	56	&	3.64	&	$<$13	&	6.4	&	\nodata	&	\nodata	& \nodata \\
\dataset[ADS/Sa.Spitzer#0004634368]{72760}	&	48.8$\pm$0.8	&	48	&	0.61	&	$<$12	&	5.5	&	\nodata	&	\nodata	& \nodata \\
\dataset[ADS/Sa.Spitzer#0004640000]{74576}	&	130.2$\pm$1.7	&	129	&	0.71	&	$<$16	&	15	&	\nodata	&	1	& \nodata \\
\dataset[ADS/Sa.Spitzer#0004634880]{92139}	&	419.5$\pm$3.5	&	403	&	4.57	&	59$\pm$18	&	46	&	0.7	&	1	& \nodata \\
\dataset[ADS/Sa.Spitzer#0004640512]{97334}	&	73.3$\pm$1.3	&	74	&	-0.66	&	$<$15	&	8.5	&	\nodata	&	\nodata	& \nodata \\
\dataset[ADS/Sa.Spitzer#0004642560]{TWA 14}	&	4.21$\pm$0.20	&	3.8	&	2.24	&	$<$13	&	0.43	&	\nodata	&	3	& \nodata \\
\dataset[ADS/Sa.Spitzer#0004642048]{TWA 12}	&	6.04$\pm$0.33	&	5.7	&	1.17	&	$<$8	&	0.65	&	\nodata	&	3	& \nodata \\
\dataset[ADS/Sa.Spitzer#0004642304]{TWA 13A}	&	8.67$\pm$3.44	&	9.1	&	-0.11	&	$<$38	&	1.0	&	\nodata	&	3	& \nodata \\
\dataset[ADS/Sa.Spitzer#0004642304]{TWA 13B}	&	8.67$\pm$3.44	&	9.0	&	-0.09	&	$<$38	&	1.0	&	\nodata	&	3	& \nodata \\
\dataset[ADS/Sa.Spitzer#0004627200]{103928}	&	42.0$\pm$1.0	&	41	&	0.83	&	$<$11	&	4.7	&	\nodata	&	\nodata	& \nodata \\
\dataset[ADS/Sa.Spitzer#0004636160]{105963}	&	36.1$\pm$11.8	&	36	&	-0.03	&	$<$10	&	4.2	&	\nodata	&	1	& \nodata \\
\dataset[ADS/Sa.Spitzer#0004636416]{109011}	&	39.7$\pm$0.8	&	40	&	-0.13	&	$<$11	&	4.6	&	\nodata	&	\nodata	& \nodata \\
\dataset[ADS/Sa.Spitzer#0004642816]{TWA 15B}	&	1.60$\pm$0.62	&	1.4	&	0.31	&	$<$76	&	0.16	&	\nodata	&	3	& \nodata \\
\dataset[ADS/Sa.Spitzer#0004642816]{TWA 15A}	&	1.44$\pm$0.62	&	1.3	&	0.27	&	$<$76	&	0.15	&	\nodata	&	3	& \nodata \\
\dataset[ADS/Sa.Spitzer#0004643072]{TWA 16}	&	5.87$\pm$0.14	&	5.5	&	2.91	&	$<$17	&	0.62	&	\nodata	&	3	& \nodata \\
\dataset[ADS/Sa.Spitzer#0004627968]{113449}	&	48.5$\pm$0.8	&	45	&	4.34	&	$<$15	&	5.2	&	\nodata	&	\nodata	& \nodata \\
\dataset[ADS/Sa.Spitzer#0004643328]{TWA 17}	&	2.20$\pm$0.80	&	2.1	&	0.17	&	$<$18	&	0.24	&	\nodata	&	3	& \nodata \\
\dataset[ADS/Sa.Spitzer#0004643584]{TWA 18}	&	2.78$\pm$0.09	&	2.7	&	0.64	&	$<$9	&	0.31	&	\nodata	&	3	& \nodata \\
\dataset[ADS/Sa.Spitzer#0004628224]{116956}	&	50.5$\pm$1.7	&	49	&	0.9	&	14$\pm$2\tablenotemark{f}	&	5.6	&	3.6	&	\nodata	& \nodata \\
\dataset[ADS/Sa.Spitzer#0004628736]{125158}	&	104.3$\pm$9.8	&	97	&	0.77	&	$<$429	&	11	&	\nodata	&	\nodata	& \nodata \\
\dataset[ADS/Sa.Spitzer#0004629248]{128987}	&	44.3$\pm$0.5	&	44	&	1	&	$<$22	&	5.0	&	\nodata	&	\nodata	& \nodata \\
\dataset[ADS/Sa.Spitzer#0004628992]{128400}	&	67.0$\pm$1.2	&	67	&	0.15	&	$<$11	&	7.6	&	\nodata	&	\nodata	& \nodata \\
\dataset[ADS/Sa.Spitzer#0004629760]{141272}	&	43.3$\pm$0.8	&	45	&	-2.13	&	$<$14	&	5.1	&	\nodata	&	1	& \nodata \\
\dataset[ADS/Sa.Spitzer#0004630016]{144197}	&	131.8$\pm$1.4	&	140	&	-6.04	&	$<$26	&	16	&	\nodata	&	\nodata	& \nodata \\
\dataset[ADS/Sa.Spitzer#0004630272]{148367}	&	159.0$\pm$2.2	&	154	&	2.25	&	$<$23	&	18	&	\nodata	&	\nodata	& \nodata \\
\dataset[ADS/Sa.Spitzer#0004635136]{165185}	&	115.1$\pm$3.8	&	116	&	-0.35	&	$<$16	&	13	&	\nodata	&	\nodata	& \nodata \\
\dataset[ADS/Sa.Spitzer#0004641280]{RE1816+541}	&	7.52$\pm$0.13	&	7.4	&	0.72	&	$<$13	&	0.85	&	\nodata	&	1	& \nodata \\
\dataset[ADS/Sa.Spitzer#0004631296]{177724}	&	498.2$\pm$8.5	&	485	&	1.56	&	50$\pm$14\tablenotemark{f}	&	55	&	-0.4	&	1	& \nodata \\
\dataset[ADS/Sa.Spitzer#0004631552]{180161}	&	59.2$\pm$0.6	&	59	&	-0.14	&	$<$11	&	6.8	&	\nodata	&	1	& \nodata \\
\dataset[ADS/Sa.Spitzer#0004626688]{186219}	&	90.2$\pm$0.9	&	86	&	4.38	&	23$\pm$6	&	9.8	&	2.1	&	1	& \nodata \\
\dataset[ADS/Sa.Spitzer#0004632064]{eta Ind}	&	201.3$\pm$3.1	&	212	&	-3.37	&	31$\pm$6	&	24	&	1.2	&	1	& \nodata \\
\dataset[ADS/Sa.Spitzer#0004641536]{SAO 145139}	&	21.1$\pm$0.9	&	21	&	0.37	&	$<$15	&	2.4	&	\nodata	&	\nodata	& \nodata \\
\dataset[ADS/Sa.Spitzer#0004632320]{202730}	&	157.3$\pm$17.1	&	157	&	0.02	&	27$\pm$4\tablenotemark{g}	&	18	&	2.2	&	1,4	& \nodata \\
\dataset[ADS/Sa.Spitzer#0004632576]{203244}	&	61.2$\pm$1.1	&	59	&	2.47	&	$<$14	&	6.7	&	\nodata	&	\nodata	& \nodata \\
\dataset[ADS/Sa.Spitzer#0004641024]{LO Peg}	&	23.6$\pm$0.4	&	21	&	4.97	&	$<$8	&	2.4	&	\nodata	&	1	& \nodata \\
\dataset[ADS/Sa.Spitzer#0004637952]{Gl 859A}	&	165.3$\pm$2.9	&	163	&	0.7	&	$<$18	&	19	&	\nodata	&	1	& \nodata \\
\dataset[ADS/Sa.Spitzer#0004635648]{HK Aqr}	&	13.3$\pm$0.4	&	13	&	0.29	&	$<$16	&	1.5	&	\nodata	&	1	& \nodata \\
\dataset[ADS/Sa.Spitzer#0004638208]{218738}	&	44.1$\pm$10.4	&	39	&	0.5	&	$<$8	&	4.4	&	\nodata	&	1	& \nodata \\
\enddata
\tablenotetext{a}{Flux from DAT pipeline PSF fitting unless otherwise noted.  Detection confirmed with MOPEX pipeline PRF fitting unless otherwise noted.}
\tablenotetext{b}{From K$_s$-[24] empirical model derived in \citet[]{gautier08} and stellar effective temperature.  Stellar effective temperatures derived from PHOENIX NextGen model photospheres fit to optical and near-IR photometry.}
\tablenotetext{c}{Previous publications of \textit{Spitzer} MIPS photometry for these sources.}
\tablenotetext{d}{A 70 $\mu$m excess is defined as $\chi_{70}$$>$5.  A 24 $\mu$m excess is defined as $\chi_{24}$$>$5. $\chi_{24,70}\equiv (F_{24,70}-F_{*})/\sigma_{24,70}$ where the uncertainty $\sigma_{24,70}$ does not include the uncertainty in the model photosphere flux density, but does include the uncertainty in the detection S/N and the flux density calibration uncertainty.}
\tablenotetext{e}{Centroided position of 70 $\mu$m detection is offset 3.$^{\prime\prime}$7 in right ascension and 2.$^{\prime\prime}$9 in declination from position expected from 24 $\mu$m detection.}
\tablenotetext{f}{DAT aperture corrected flux density. Detection confirmed with MOPEX PRF fitting.}
\tablenotetext{g}{Flux from MOPEX pipeline PRF fitting. Detection not confirmed with DAT pipeline PSF fitting.}
\tablenotetext{h}{DAT pipeline detection not confirmed with MOPEX pipeline PRF fitting.}
\tablenotetext{i}{Centroided position of 70 $\mu$m detection is offset 7.$^{\prime\prime}$4 in right ascension and -1.$^{\prime\prime}$5 in declination from position expected from 24 $\mu$m detection.}

\tablerefs{1--\citet[]{chen05a}, 2--\citet[]{rebull08}, 3--\citet[]{low05},4--\citet[]{rieke05}}

\end{deluxetable}

\begin{deluxetable}{lrrrrrrrrrrrl}
\rotate
\tablecolumns{13}
\tabletypesize{\scriptsize}
\tablewidth{0pc}
\tablecaption{Optical and Near-IR Photo-electric Photometry and Derived Stellar Properties}
\tablehead{
\colhead{Name/HD}	&	\colhead{R$_*$\tablenotemark{a}} & \colhead{T$_*$\tablenotemark{a}} & \colhead{L$_*$\tablenotemark{a}}&	\colhead{U}	&\colhead{B}	 & \colhead{V}	&	\colhead{R$_c$}	&	\colhead{I$_c$}	&	\colhead{J}	&	\colhead{H}	&	\colhead{K$_s$} & \colhead{References} \\
\colhead{}	&	\colhead{(R$_\odot$)} & \colhead{(K)} & \colhead{(L$_\odot$)}&	\colhead{(mag)}	&\colhead{(mag)}	 & \colhead{(mag)}	&	\colhead{(mag)}	&	\colhead{(mag)}	&	\colhead{(mag)}	&	\colhead{(mag)}	&	\colhead{(mag)}
}
\startdata
1835	&	0.98	&	5840$\pm$20	&	1.00	&	7.16	&	7.04	&	6.38	&	6.01	&	5.68	&	5.253$\pm$0.021	&	5.035$\pm$0.034	&	4.861$\pm$0.016	&	1,2,2,3,4,5,5,5	\\
7590	&	0.97	&	5980$\pm$20	&	1.07	&	\nodata	&	7.17	&	6.59	&	6.25	&	5.93	&	5.515$\pm$0.018	&	5.258$\pm$0.029	&	5.177$\pm$0.016	&	-,2,2,3,4,5,5,5	\\
BB Scl	&	1.38	&	4820$\pm$20	&	0.91	&	8.66	&	8.09	&	7.16	&	6.62	&	6.09	&	5.34$\pm$0.023	&	4.973$\pm$0.076	&	4.69$\pm$0.018	&	1,2,2,6,6,5,5,5	\\
10008	&	0.80	&	5340$\pm$20	&	0.46	&	8.90	&	8.46	&	7.66	&	\nodata	&	6.82	&	6.225$\pm$0.024	&	5.899$\pm$0.036	&	5.753$\pm$0.018	&	1,2,2,-,4,5,5,5	\\
11131	&	0.98	&	5780$\pm$20	&	0.95	&	7.48	&	7.36	&	6.73	&	6.34	&	6.00	&	5.536$\pm$0.023	&	5.289$\pm$0.023	&	5.149$\pm$0.02	&	1,2,2,3,4,5,5,5	\\
15089	&	2.19	&	8740$\pm$20	&	25.0	&	4.69	&	4.64	&	4.48	&	4.37	&	4.31	&	3.981$\pm$0.428	&	4.29$\pm$0.036	&	4.248$\pm$0.031	&	1,2,2,7,7,5,5,5	\\
16287	&	0.78	&	5040$\pm$20	&	0.35	&	9.81	&	9.08	&	8.13	&	\nodata	&	7.14	&	6.518$\pm$0.024	&	6.035$\pm$0.033	&	5.938$\pm$0.026	&	1,2,2,-,4,5,5,5	\\
16555	&	2.19	&	7280$\pm$20	&	12.1	&	\nodata	&	5.58	&	5.29	&	\nodata	&	4.96	&	4.564$\pm$0.306\tablenotemark{b}	&	4.6$\pm$0.023	&	4.525$\pm$0.021	&	-,2,2,-,4,5,5,5	\\
17240	&	1.69	&	6860$\pm$20	&	5.63	&	6.61	&	6.61	&	6.24	&	\nodata	&	5.79	&	5.504$\pm$0.032	&	5.369$\pm$0.017	&	5.302$\pm$0.023	&	8,2,2,-,9,5,5,5	\\
DO Eri	&	1.73	&	7280$\pm$20	&	7.49	&	6.34	&	6.31	&	5.99	&	\nodata	&	5.62	&	5.432$\pm$0.018	&	5.314$\pm$0.024	&	5.262$\pm$0.02	&	10,2,2,-,4,5,5,5	\\
v833 Tau	&	0.84	&	4500$\pm$20	&	0.26	&	10.29	&	9.20	&	8.10	&	\nodata	&	6.92	&	5.945$\pm$0.023	&	5.4$\pm$0.018	&	5.24$\pm$0.023	&	11,2,2,-,4,5,5,5	\\
29697	&	0.67	&	4440$\pm$20	&	0.15	&	10.16	&	9.24	&	8.13	&	7.48	&	6.97	&	5.854$\pm$0.019	&	5.31$\pm$0.02	&	5.146$\pm$0.02	&	1,2,2,12,4,5,5,5	\\
Gl 182	&	0.89	&	3860$\pm$100	&	0.16	&	12.61	&	11.48	&	10.10	&	\nodata	&	8.26	&	7.117$\pm$0.02	&	6.45$\pm$0.031	&	6.261$\pm$0.017	&	1,2,2,-,9,5,5,5	\\
36435	&	0.84	&	5480$\pm$20	&	0.56	&	8.06	&	7.76	&	6.99	&	6.57	&	6.20	&	5.704$\pm$0.018	&	5.342$\pm$0.049	&	5.2$\pm$0.018	&	1,2,2,3,4,5,5,5	\\
AB Dor	&	0.84	&	5100$\pm$20	&	0.43	&	8.14	&	7.77	&	6.94	&	\nodata	&	6.00	&	5.316$\pm$0.019	&	4.845$\pm$0.033	&	4.686$\pm$0.016	&	1,2,2,-,4,5,5,5	\\
41824	&	1.51	&	5520$\pm$20	&	1.88	&	7.53	&	7.27	&	6.59	&	6.18	&	5.81	&	5.345$\pm$0.026	&	4.971$\pm$0.034	&	4.82$\pm$0.02	&	1,2,2,6,6,5,5,5	\\
41593	&	0.80	&	5320$\pm$20	&	0.46	&	8.00	&	7.58	&	6.75	&	6.28	&	5.90	&	5.317$\pm$0.018	&	4.942$\pm$0.038	&	4.822$\pm$0.017	&	1,2,2,3,4,5,5,5	\\
GJ 3400A	&	1.22	&	5740$\pm$20	&	1.45	&	6.86	&	6.76	&	6.14	&	5.78	&	5.45	&	5.079$\pm$0.272	&	4.747$\pm$0.092	&	4.544$\pm$0.026	&	1,2,2,6,9,5,5,5	\\
51849	&	0.68	&	4460$\pm$20	&	0.16	&	11.31	&	10.25	&	9.16	&	\nodata	&	\nodata	&	6.903$\pm$0.026	&	6.368$\pm$0.053	&	6.19$\pm$0.027	&	1,2,2,-,-,5,5,5	\\
52698	&	0.84	&	5120$\pm$20	&	0.43	&	8.21	&	7.60	&	6.70	&	6.22	&	5.79	&	5.152$\pm$0.017	&	4.845$\pm$0.047	&	4.636$\pm$0.015	&	1,2,2,3,4,5,5,5	\\
59967	&	0.95	&	5720$\pm$20	&	0.87	&	\nodata	&	7.29	&	6.65	&	6.28	&	5.94	&	5.527$\pm$0.026	&	5.253$\pm$0.023	&	5.099$\pm$0.021	&	-,2,2,3,4,5,5,5	\\
63433	&	0.90	&	5600$\pm$20	&	0.71	&	\nodata	&	7.59	&	6.91	&	6.52	&	6.17	&	5.624$\pm$0.043	&	5.359$\pm$0.026	&	5.258$\pm$0.016	&	-,2,2,3,4,5,5,5	\\
72760	&	0.86	&	5260$\pm$20	&	0.51	&	\nodata	&	8.13	&	7.31	&	6.89	&	6.49	&	5.917$\pm$0.026	&	5.551$\pm$0.036	&	5.423$\pm$0.02	&	-,2,2,3,4,5,5,5	\\
73350	&	0.96	&	5840$\pm$20	&	0.97	&	7.56	&	7.39	&	6.74	&	6.36	&	6.02	&	5.626$\pm$0.023	&	5.318$\pm$0.027	&	5.223$\pm$0.029	&	1,2,2,3,4,5,5,5	\\
74576	&	0.74	&	5040$\pm$20	&	0.31	&	8.16	&	7.52	&	6.58	&	6.06	&	5.61	&	4.937$\pm$0.037	&	4.441$\pm$0.212	&	4.358$\pm$0.02	&	1,2,2,3,4,5,5,5	\\
92139	&	2.49	&	7360$\pm$20	&	16.2	&	4.20	&	4.14	&	3.83	&	3.59	&	3.43	&	3.27	&	3.17$\pm$0.208	&	3.12	&	1,2,2,7,7,7,5,7	\\
92945	&	0.77	&	5180$\pm$20	&	0.38	&	9.17	&	8.61	&	7.71	&	7.21	&	6.79	&	6.176$\pm$0.024	&	5.77$\pm$0.046	&	5.66$\pm$0.018	&	1,2,2,13,4,5,5,5	\\
97334	&	0.99	&	5920$\pm$20	&	1.08	&	7.14	&	7.02	&	6.41	&	6.06	&	5.74	&	5.265$\pm$0.024	&	5.021$\pm$0.018	&	4.959$\pm$0.017	&	1,2,2,3,4,5,5,5	\\
TWA 14	&	0.72	&	3600$\pm$100	&	0.08	&	\nodata	&	\nodata	&	13.80	&	11.85	&	10.60	&	9.415$\pm$0.028	&	8.727$\pm$0.04	&	8.495$\pm$0.031	&	-,-,14,15,16,5,5,5	\\
TWA 12	&	0.88	&	3600$\pm$100	&	0.12	&	\nodata	&	14.38	&	12.85	&	\nodata	&	\nodata	&	8.999$\pm$0.034	&	8.334$\pm$0.033	&	8.053$\pm$0.029	&	-,17,17,-,-,5,5,5	\\
TWA 13A	&	1.08	&	3760$\pm$20	&	0.21	&	\nodata	&	12.88	&	11.46	&	\nodata	&	9.57	&	8.431$\pm$0.043	&	7.727$\pm$0.067	&	7.491$\pm$0.038	&	-,17,17,-,17,5,5,5	\\
TWA 13B	&	1.03	&	3900$\pm$20	&	0.22	&	\nodata	&	13.43	&	11.96	&	\nodata	&	9.88	&	8.429$\pm$0.037	&	7.684$\pm$0.055	&	7.46$\pm$0.027	&	-,17,17,-,17,5,5,5	\\
103928	&	1.45	&	7120$\pm$20	&	4.81	&	\nodata	&	6.74	&	6.42	&	\nodata	&	6.03	&	5.757$\pm$0.026	&	5.66$\pm$0.026	&	5.599$\pm$0.024	&	-,2,2,-,4,5,5,5	\\
105963	&	0.94	&	5000$\pm$20	&	0.49	&	9.49	&	8.93	&	8.00	&	\nodata	&	7.11	&	6.313$\pm$0.023	&	5.867$\pm$0.027	&	5.734$\pm$0.017	&	1,2,2,-,9,5,5,5	\\
109011	&	0.89	&	4820$\pm$20	&	0.38	&	9.69	&	9.05	&	8.11	&	7.32	&	7.18	&	6.324$\pm$0.052	&	5.814$\pm$0.027	&	5.662$\pm$0.02	&	1,2,2,7,4,5,5,5	\\
TWA 15B	&	0.44	&	3600$\pm$100	&	0.03	&	\nodata	&	\nodata	&	14.00	&	13.41	&	11.81	&	10.487$\pm$0.038	&	9.826$\pm$0.035	&	9.561$\pm$0.038	&	-,-,14,15,15,5,5,5	\\
TWA 15A	&	0.42	&	3600$\pm$100	&	0.03	&	\nodata	&	\nodata	&	14.10	&	13.51	&	11.94	&	10.562$\pm$0.026	&	9.935$\pm$0.023	&	9.673$\pm$0.023	&	-,-,14,15,15,5,5,5	\\
TWA 16	&	0.87	&	3600$\pm$100	&	0.11	&	\nodata	&	\nodata	&	12.30	&	11.64	&	10.17	&	8.994$\pm$0.026	&	8.332$\pm$0.038	&	8.09$\pm$0.023	&	-,-,14,15,15,5,5,5	\\
112429	&	1.50	&	7180$\pm$20	&	5.36	&	5.55	&	5.52	&	5.23	&	\nodata	&	4.88	&	4.803$\pm$0.230	&	4.604$\pm$0.204	&	4.425$\pm$0.016	&	1,2,2,-,4,5,5,5	\\
113449	&	0.87	&	4940$\pm$20	&	0.40	&	\nodata	&	8.57	&	7.70	&	7.28	&	6.81	&	6.053$\pm$0.021	&	5.674$\pm$0.038	&	5.509$\pm$0.023	&	-,2,2,3,4,5,5,5	\\
TWA 17	&	0.49	&	4080$\pm$20	&	0.06	&	\nodata	&	\nodata	&	12.70	&	11.69	&	10.78	&	9.806$\pm$0.022	&	9.187$\pm$0.023	&	9.013$\pm$0.019	&	-,-,14,15,15,5,5,5	\\
TWA 18	&	0.61	&	3600$\pm$100	&	0.06	&	\nodata	&	\nodata	&	12.90	&	\nodata	&	10.92	&	9.744$\pm$0.023	&	9.076$\pm$0.023	&	8.846$\pm$0.019	&	-,-,14,-,15,5,5,5	\\
116956	&	0.87	&	5280$\pm$20	&	0.52	&	\nodata	&	8.11	&	7.29	&	6.85	&	6.46	&	5.812$\pm$0.023	&	5.481$\pm$0.021	&	5.411$\pm$0.024	&	-,2,2,3,4,5,5,5	\\
123998	&	2.13	&	7860$\pm$20	&	15.5	&	5.24	&	5.14	&	4.89	&	\nodata	&	4.65	&	4.873$\pm$0.288	&	4.624$\pm$0.248	&	4.406$\pm$0.024	&	1,2,2,-,4,5,5,5	\\
125158	&	2.16	&	7700$\pm$20	&	14.7	&	5.64	&	5.50	&	5.21	&	4.97	&	4.93	&	5.059$\pm$0.224\tablenotemark{b}	&	4.788$\pm$0.047	&	4.67$\pm$0.023	&	1,2,2,7,4,5,5,5	\\
128987	&	0.86	&	5520$\pm$20	&	0.62	&	\nodata	&	7.97	&	7.23	&	6.84	&	6.47	&	5.947$\pm$0.021	&	5.629$\pm$0.018	&	5.531$\pm$0.018	&	-,2,2,3,4,5,5,5	\\
128400	&	0.91	&	5600$\pm$20	&	0.73	&	\nodata	&	7.44	&	6.73	&	\nodata	&	5.97	&	5.525$\pm$0.021	&	5.214$\pm$0.044	&	5.072$\pm$0.018	&	-,2,2,-,4,5,5,5	\\
135599	&	0.77	&	5180$\pm$20	&	0.38	&	\nodata	&	7.76	&	6.93	&	6.47	&	6.06	&	5.484$\pm$0.018	&	5.115$\pm$0.027	&	4.958$\pm$0.017	&	-,2,2,13,4,5,5,5	\\
141272	&	0.82	&	5240$\pm$20	&	0.45	&	\nodata	&	8.25	&	7.44	&	7.01	&	6.60	&	5.991$\pm$0.021	&	5.61$\pm$0.027	&	5.501$\pm$0.018	&	-,2,2,13,4,5,5,5	\\
144197	&	2.00	&	7920$\pm$20	&	14.1	&	5.11	&	4.96	&	4.71	&	4.51	&	4.51	&	4.519$\pm$0.300\tablenotemark{b}	&	4.285$\pm$0.268	&	4.267$\pm$0.316	&	1,2,2,7,4,5,5,5	\\
148367	&	2.06	&	8040$\pm$20	&	15.8	&	4.91	&	4.82	&	4.63	&	4.43	&	4.34	&	4.271$\pm$0.244	&	4.155$\pm$0.2	&	4.165$\pm$0.036	&	1,2,2,7,7,5,5,5	\\
165185	&	0.99	&	5940$\pm$20	&	1.09	&	6.60	&	6.53	&	5.94	&	5.59	&	5.26	&	4.835$\pm$0.037	&	4.614$\pm$0.016	&	4.469$\pm$0.016	&	1,2,2,3,9,5,5,5	\\
RE1816+541	&	0.54	&	3620$\pm$100	&	0.04	&	\nodata	&	13.43	&	11.83	&	\nodata	&	9.73	&	8.616$\pm$0.021	&	7.96$\pm$0.031	&	7.75$\pm$0.02	&	-,18,18,-,19,5,5,5	\\
175742	&	0.81	&	4780$\pm$20	&	0.31	&	9.68	&	9.13	&	8.19	&	\nodata	&	7.24	&	6.243$\pm$0.019	&	5.762$\pm$0.018	&	5.637$\pm$0.02	&	1,2,2,-,9,5,5,5	\\
177724	&	2.27	&	9620$\pm$20	&	39.4	&	3.00	&	2.99	&	2.95	&	2.98	&	2.96	&	2.93	&	3.048$\pm$0.28	&	2.92	&	1,2,2,7,4,7,5,7	\\
180161	&	0.86	&	5400$\pm$20	&	0.56	&	8.28	&	7.84	&	7.03	&	\nodata	&	6.19	&	5.651$\pm$0.03	&	5.297$\pm$0.036	&	5.202$\pm$0.02	&	1,2,2,-,4,5,5,5	\\
186219	&	1.76	&	7600$\pm$20	&	9.26	&	5.74	&	5.64	&	5.39	&	\nodata	&	5.13	&	5.01$\pm$0.288	&	4.888$\pm$0.018	&	4.797$\pm$0.017	&	1,2,2,-,4,5,5,5	\\
AT Mic	&	0.81	&	3200$\pm$100	&	0.06	&	12.69	&	11.80	&	10.24	&	8.98	&	7.31	&	5.807$\pm$0.026	&	5.201$\pm$0.046	&	4.944$\pm$0.042	&	1,12,12,12,12,5,5,5	\\
eta Ind	&	1.64	&	7400$\pm$20	&	7.18	&	4.86	&	4.78	&	4.50	&	\nodata	&	4.20	&	3.909$\pm$0.278	&	3.692$\pm$0.28	&	3.82$\pm$0.268	&	1,2,2,-,4,5,5,5	\\
AU Mic	&	0.84	&	3500$\pm$100	&	0.09	&	11.16	&	10.21	&	8.76	&	7.81	&	6.70	&	5.436$\pm$0.017	&	4.831$\pm$0.016	&	4.529$\pm$0.02	&	1,2,2,20,20,5,5,5	\\
BO Mic	&	1.00	&	4680$\pm$20	&	0.43	&	10.93	&	10.39	&	9.45	&	8.89	&	8.33	&	7.513$\pm$0.021	&	6.93$\pm$0.029	&	6.794$\pm$0.026	&	1,2,2,6,17,5,5,5	\\
358623	&	0.54	&	4060$\pm$20	&	0.07	&	\nodata	&	11.60	&	10.62	&	\nodata	&	9.13	&	7.849$\pm$0.021	&	7.249$\pm$0.031	&	7.039$\pm$0.02	&	-,2,2,-,17,5,5,5	\\
SAO 145139	&	0.72	&	4460$\pm$20	&	0.18	&	11.60	&	10.61	&	9.42	&	\nodata	&	\nodata	&	7.146$\pm$0.024	&	6.589$\pm$0.036	&	6.428$\pm$0.018	&	11,2,2,-,-,5,5,5	\\
202730	&	1.61	&	8360$\pm$20	&	11.3	&	4.76	&	4.66	&	4.48	&	\nodata	&	4.27	&	3.426$\pm$0.434	&	4.224$\pm$0.076	&	4.145$\pm$0.026	&	1,2,2,-,9,5,5,5	\\
203244	&	0.87	&	5440$\pm$20	&	0.59	&	\nodata	&	7.71	&	6.98	&	6.58	&	6.21	&	5.651$\pm$0.019	&	5.297$\pm$0.024	&	5.216$\pm$0.018	&	-,2,2,3,4,5,5,5	\\
LO Peg	&	0.70	&	4520$\pm$20	&	0.18	&	11.04	&	10.24	&	9.24	&	\nodata	&	8.14	&	7.075$\pm$0.021	&	6.524$\pm$0.04	&	6.382$\pm$0.026	&	1,2,2,-,9,5,5,5	\\
Gl 859A	&	1.11	&	5960$\pm$20	&	1.39	&	6.22	&	6.14	&	5.52	&	5.17	&	4.84	&	4.129$\pm$0.206	&	4.199$\pm$0.076	&	4.102$\pm$0.016	&	1,2,2,3,9,5,5,5	\\
HK Aqr	&	0.53	&	3660$\pm$20	&	0.05	&	13.85	&	12.85	&	10.71	&	\nodata	&	\nodata	&	7.979$\pm$0.021	&	7.301$\pm$0.029	&	7.114$\pm$0.031	&	1,2,2,21,21,5,5,5	\\
218738	&	0.92	&	5040$\pm$20	&	0.48	&	9.41	&	8.84	&	7.91	&	\nodata	&	\nodata	&	6.225$\pm$0.02	&	5.788$\pm$0.026	&	5.659$\pm$0.026	&	1,2,2,-,-,5,5,5	\\
Gl 907.1	&	0.81	&	4240$\pm$20	&	0.19	&	12.04	&	10.91	&	9.55	&	\nodata	&	8.00	&	7.045$\pm$0.023	&	6.485$\pm$0.024	&	6.294$\pm$0.018	&	1,2,2,-,9,5,5,5	\\
\enddata
 \tablenotetext{a}{Radii are fit  to the nearest 0.5\% and the uncertainty is dominated by the uncertainty in the trigonometric parallax.  For unresolved binaries listed in Table 1, the derived radius, effective temperature, and luminosity listed correspond to the equivalent single star (e.g., L$_*$=L$_A$+L$_B$; R$_*$$\sim$$\sqrt{2}$ R$_{A}$ if T$_*$$\sim$T$_A$$\sim$T$_B$ and R$_A$$\sim$R$_B$).  For sources with non-trigonometric distances in Table 1, the derived radius and luminosity are dependent on the distance assumed.}
 \tablenotetext{b}{Not included in $\chi^2$ fit to model photospheres.}
\tablerefs{1 -- \citet[]{mermilliod91}, 2-- Tycho-2 using the color transformations of \citet[]{mamajek02,mamajek06}, 3 -- \citet[]{taylor05,taylor03}, 4 -- UCAC2 Bright Star Supplement \citep[]{ucac2bss}, 5 -- 2MASS \citep[]{skrutskie06}, 6 -- \citet[]{cutispoto99}, 7 -- \citet[]{morel78}, 8 -- \citet[]{oja91}, 9 -- the OSACA database \citet[]{bobylev06}, 10 -- \citet[]{ducati02}, 11 -- Sky 2000 Catalog \citep[]{sky00}, 12 -- \citet[]{reid02}, 13 -- \citet[]{kotoneva02}, 14 -- \citet[]{low05}, 15 -- \citet[]{zuckerman01c}, 16 -- N-band in Guide Star Catalogue 2.3.2 \citet[]{gsc232}, 17 -- \citet[]{torres06}, 18 -- \citet[]{mason95}, 19 -- \citet[]{droege06}, 20 -- \citet[]{cousins80}, 21 -- \citet[]{koen02b}}
\end{deluxetable}

\begin{deluxetable}{lrrrrrrrrrrr}
\rotate
\tablecolumns{12}
\tabletypesize{\scriptsize}
\tablewidth{0pc}
\tablecaption{Derived Stellar and Thin Dust Disk Model Dust Properties}
\tablehead{
\colhead{Name/HD}	&	\colhead{F$_{70}$\tablenotemark{a}} & \colhead{T$_d$} & \colhead{R$_d$} & \colhead{log(L$_{IR}$/L$_*$)\tablenotemark{b}} & \colhead{M$_D$} &\colhead{M$_{PB}$} &	\colhead{log(L$_X$/L$_*$)}	&\colhead{$\dot M_{sw}$}	 & \colhead{Age}	&	\colhead{Age}	&	\colhead{M$_*$}	\\
\colhead{}	& \colhead{(mJy)} &	\colhead{(K)} & \colhead{(AU)} & \colhead{} &  \colhead{(M$_{\mbox{moon}}$)} &\colhead{(M$_{\mbox{moon}}$)} &	\colhead{}	&\colhead{($\dot M_\odot$)\tablenotemark{c}}	 & \colhead{(Myr)}	&	\colhead{Method}	&	\colhead{(M$_\odot$)}
}
\startdata
\multicolumn{12}{c}{Stars with MIPS-24 and MIPS-70 excesses}\\
\hline
10008	&	33$\pm$1	&	$<$78	&	$>$8.7	&	-4.09	&	2.4E-05	&	$>$0.11	&	-4.44	&	\nodata	&	275$\pm$25	&	1,3	&	0.94	\\
73350	&	136$\pm$4	&	$<$63	&	$>$19	&	-3.84	&	3.7E-04	&	$>$0.6	&	-4.80	&	\nodata	&	513$\pm$136	&	1,2,3	&	1.09	\\
112429	&	63$\pm$5	&	$<$86	&	$>$24	&	-4.69	&	3.2E-04	&	$>$0.2	&	$<$-5.62	&	\nodata	&	250$\pm$200	&	5	&	1.56	\\
135599	&	117$\pm$12	&	$<$62	&	$>$12	&	-3.88	&	6.6E-05	&	$>$0.05	&	-5.11	&	14	&	214$\pm$17	&	3,4	&	0.92	\\
\hline
\multicolumn{12}{c}{Stars with MIPS-70 excesses}\\
\hline
7590	&	259$\pm$18	&	$<$40	&	$>$51	&	-3.57	&	5.2E-03	&	$>$1.1	&	-4.69	&	\nodata	&	462$\pm$35	&	1,2,3	&	1.15	\\
59967	&	39$\pm$4	&	$<$71	&	$>$14	&	-4.44	&	4.6E-05	&	$>$0.12	&	-4.46	&	\nodata	&	353$\pm$68	&	1,2	&	1.09	\\
92945	&	341$\pm$17	&	$<$46	&	$>$23	&	-3.12	&	1.2E-03	&	$>$1.0	&	-4.47	&	\nodata	&	294$\pm$23	&	1,2	&	0.93	\\
AU Mic	&	247$\pm$28	&	$<$59	&	$>$6.8	&	-3.36	&	2.0E-03	&	$>$0.004	&	-2.82	&	\nodata	&	12$\pm$2	&	4	&	0.31	\\\hline
\multicolumn{12}{c}{Stars with MIPS-24 excesses}\\
\hline
GJ 3400A	&	16$\pm$5	&	$<$200	&	$>$2.2	&	-4.86	&	7.6E-07	&	$>$0.006	&	-3.76	&	\nodata	&	75$\pm$25	&	4	&	1.03	\\
123998	&	\nodata	&	$<$100	&	$>$31	&	-5.22	&	3.9E-04	&	$>$0.35	&	$<$-5.74	&	\nodata	&	250$\pm$200	&	5	&	1.77	\\
175742	&	\nodata	&	$<$110	&	$>$3.5	&	-4.46,$<$-4.27\tablenotemark{d}	&	1.3E-06	&	$>$0.002	&	-3.18	&	\nodata	&	51$\pm$3	&	3	&	0.77	\\
AT Mic	&	23$\pm$5	&	$<$120	&	$>$1.0	&	-4.24	&	3.0E-07	&	$>$0.0003	&	-2.82	&	\nodata	&	12$\pm$2	&	4	&	0.15	\\
BO Mic	&	\nodata	&	$<$81	&	$>$7.8	&	-4.43,$<$-3.90\tablenotemark{d}	&	8.5E-06	&	$>$0.003	&	-2.16	&	\nodata	&	25.5$\pm$8.5	&	7	&	0.90	\\
358623	&	\nodata	&	$<$88	&	$>$2.7	&	-4.32,$<$-3.70\tablenotemark{d}	&	2.8E-07	&	$>$0.0003	&	-3.36	&	\nodata	&	12$\pm$2	&	4	&	0.67	\\
Gl 907.1	&	\nodata	&	$<$93	&	$>$3.9	&	-4.48,$<$-3.91\tablenotemark{d}	&	1.3E-06	&	$>$0.001	&	-3.47	&	\nodata	&	12.5$\pm$5.5	&	7	&	0.60	\\
\hline
\multicolumn{12}{c}{Stars with no MIPS excesses}\\
\hline
1835	&	19$\pm$4	&	$<$86	&	$>$11	&	$<$-4.97	&	$<$8.5E-6	&	\nodata	&	-4.60	&	\nodata	&	466$\pm$55	&	1,2,3	&	1.09	\\
BB Scl	&	\nodata	&	$<$260	&	$>$1.1	&	$<$-4.69	&	$<$1.4E-7	&	\nodata	&	-3.45	&	\nodata	&	180$\pm$120	&	1,7	&	1.17	\\
11131	&	\nodata	&	$<$91	&	$>$9.2	&	$<$-4.83	&	$<$8.8E-6	&	\nodata	&	-4.39	&	\nodata	&	506$\pm$255	&	1,2,3	&	1.07	\\
15089	&	30$\pm$3	&	$<$100	&	$>$38	&	$<$-5.47	&	$<$5.1E-4	&	\nodata	&	-5.42	&	\nodata	&	200$\pm$150	&	5	&	1.87	\\
16287	&	\nodata	&	$<$76	&	$>$7.9	&	$<$-4.41	&	$<$7.5E-6	&	\nodata	&	-4.71	&	41	&	340$\pm$252	&	2	&	0.88	\\
16555	&	\nodata	&	$<$270	&	$>$3.7	&	$<$-5.21	&	$<$5.3E-6	&	\nodata	&	-6.87	&	\nodata	&	200$\pm$150	&	5	&	1.56	\\
17240	&	34$\pm$7	&	$<$62	&	$>$48	&	$<$-4.68	&	$<$1.5E-3	&	\nodata	&	$<$-5.22	&	\nodata	&	1100$\pm$550	&	6	&	1.37	\\
DO Eri	&	\nodata	&	$<$98	&	$>$22	&	$<$-5.21	&	$<$1.1E-4	&	\nodata	&	$<$-5.31	&	\nodata	&	160$\pm$110	&	5	&	1.58	\\
v833 Tau	&	\nodata	&	$<$73	&	$>$7.4	&	$<$-4.20	&	$<$8.6E-6	&	\nodata	&	-3.09	&	\nodata	&	15$\pm$3	&	3	&	0.81	\\
29697	&	\nodata	&	$<$77	&	$>$5.1	&	$<$-4.15	&	$<$3.0E-6	&	\nodata	&	-3.21	&	\nodata	&	49$\pm$37	&	2,3	&	0.75	\\
Gl 182	&	\nodata	&	$<$69	&	$>$6.5	&	$<$-3.78	&	$<$1.9E-5	&	\nodata	&	-3.18	&	\nodata	&	9$\pm$4	&	7	&	0.45	\\
36435	&	\nodata	&	$<$100	&	$>$5.6	&	$<$-4.87	&	$<$1.9E-6	&	\nodata	&	-4.79	&	\nodata	&	504$\pm$145	&	1,2	&	0.98	\\
AB Dor	&	\nodata	&	$<$80	&	$>$8.0	&	$<$-4.31	&	$<$1.2E-5	&	\nodata	&	-3.04	&	\nodata	&	75$\pm$25	&	4	&	0.91	\\
41824	&	\nodata	&	$<$300	&	$>$1.0	&	$<$-4.71	&	$<$2.0E-7	&	\nodata	&	-3.61	&	\nodata	&	109$\pm$9	&	3	&	1.38	\\
41593	&	\nodata	&	$<$90	&	$>$6.5	&	$<$-4.72	&	$<$3.1E-6	&	\nodata	&	-4.59	&	\nodata	&	326$\pm$93	&	1,2,3	&	0.93	\\
51849	&	\nodata	&	$<$94	&	$>$3.6	&	$<$-4.16	&	$<$1.6E-6	&	\nodata	&	-3.48	&	\nodata	&	65$\pm$25	&	7	&	0.71	\\
52698	&	\nodata	&	$<$280	&	$>$1.0	&	$<$-4.74	&	$<$3.1E-8	&	\nodata	&	-4.73	&	48	&	809$\pm$379	&	1,2	&	0.89	\\
63433	&	\nodata	&	$<$30	&	$>$73	&	$<$-4.75	&	$<$5.1E-4	&	\nodata	&	-4.55	&	\nodata	&	299$\pm$96	&	1,2,3	&	1.03	\\
72760	&	\nodata	&	$<$90	&	$>$6.8	&	$<$-4.70	&	$<$3.9E-6	&	\nodata	&	-4.74	&	57	&	415$\pm$194	&	1,2	&	0.94	\\
74576	&	\nodata	&	$<$300	&	$>$1.0	&	$<$-3.75	&	$<$9.6E-10	&	\nodata	&	-4.43	&	\nodata	&	215$\pm$94	&	1,3	&	0.84	\\
92139	&	62$\pm$19	&	$<$120	&	$>$21	&	$<$-5.50	&	$<$1.1E-4	&	\nodata	&	-5.76	&	\nodata	&	100$\pm$50	&	5	&	1.58	\\
97334	&	\nodata	&	$<$100	&	$>$8.0	&	$<$-4.96	&	$<$5.3E-6	&	\nodata	&	-4.53	&	\nodata	&	451$\pm$199	&	1,2,3	&	1.12	\\
TWA 14	&	\nodata	&	$<$50	&	$>$8.7	&	$<$-3.01	&	$<$1.7E-4	&	\nodata	&	-3.00	&	\nodata	&	8$\pm$2	&	4	&	0.27	\\
TWA 12	&	\nodata	&	$<$60	&	$>$7.5	&	$<$-3.41	&	$<$7.3E-5	&	\nodata	&	-2.57	&	\nodata	&	8$\pm$2	&	4	&	0.27	\\
TWA 13A	&	\nodata	&	$<$50	&	$>$14	&	$<$-2.97	&	$<$1.1E-3	&	\nodata	&	-3.26	&	\nodata	&	8$\pm$2	&	4	&	0.32	\\
TWA 13B	&	\nodata	&	$<$50	&	$>$15	&	$<$-2.99	&	$<$1.0E-3	&	\nodata	&	-3.29	&	\nodata	&	8$\pm$2	&	4	&	0.38	\\
103928	&	\nodata	&	$<$67	&	$>$38	&	$<$-5.18	&	$<$2.4E-4	&	\nodata	&	-6.23	&	\nodata	&	440$\pm$390	&	5	&	1.53	\\
105963	&	\nodata	&	$<$90	&	$>$6.7	&	$<$-4.63	&	$<$4.7E-6	&	\nodata	&	-4.39	&	\nodata	&	266$\pm$11	&	1,3	&	0.87	\\
109011	&	\nodata	&	$<$87	&	$>$6.4	&	$<$-4.54	&	$<$4.1E-6	&	\nodata	&	-4.65	&	50	&	348$\pm$120	&	3	&	0.84	\\
TWA 15B	&	\nodata	&	$<$37	&	$>$9.7	&	$<$-1.71	&	$<$6.4E-4	&	\nodata	&	-2.41	&	\nodata	&	8$\pm$2	&	4	&	0.66	\\
TWA 15A	&	\nodata	&	$<$35	&	$>$10	&	$<$-1.63	&	$<$1.4E-3	&	\nodata	&	-2.37	&	\nodata	&	8$\pm$2	&	4	&	0.38	\\
TWA 16	&	\nodata	&	$<$53	&	$>$9.4	&	$<$-3.07	&	$<$1.8E-4	&	\nodata	&	-3.17	&	\nodata	&	8$\pm$2	&	4	&	0.38	\\
113449	&	\nodata	&	$<$76	&	$>$8.6	&	$<$-4.51	&	$<$8.4E-6	&	\nodata	&	-4.26	&	\nodata	&	212$\pm$33	&	2,3	&	0.86	\\
TWA 17	&	\nodata	&	$<$51	&	$>$7.1	&	$<$-2.74	&	$<$5.3E-5	&	\nodata	&	-3.02	&	\nodata	&	8$\pm$2	&	4	&	0.81	\\
TWA 18	&	\nodata	&	$<$52	&	$>$6.8	&	$<$-3.01	&	$<$5.3E-5	&	\nodata	&	-2.98	&	\nodata	&	8$\pm$2	&	4	&	0.38	\\
116956	&	15$\pm$3	&	$<$86	&	$>$7.7	&	$<$-4.64	&	$<$5.6E-6	&	\nodata	&	-4.34	&	\nodata	&	334$\pm$96	&	1,2,3	&	0.98	\\
125158	&	\nodata	&	$<$53	&	$>$110	&	$<$-3.87	&	$<$1.0E-1	&	\nodata	&	$<$-5.61	&	\nodata	&	125$\pm$75	&	5	&	1.70	\\
128987	&	\nodata	&	$<$73	&	$>$11	&	$<$-4.44	&	$<$2.3E-5	&	\nodata	&	-4.85	&	54	&	623$\pm$74	&	1,3	&	1.01	\\
128400	&	\nodata	&	$<$130	&	$>$4.1	&	$<$-5.03	&	$<$8.8E-7	&	\nodata	&	-4.86	&	62	&	795$\pm$143	&	1,2	&	1.03	\\
141272	&	\nodata	&	$<$83	&	$>$7.5	&	$<$-4.59	&	$<$5.4E-6	&	\nodata	&	-4.40	&	\nodata	&	567$\pm$382	&	1,2,3	&	0.93	\\
144197	&	\nodata	&	$<$110	&	$>$26	&	$<$-5.37	&	$<$1.9E-4	&	\nodata	&	$<$-5.81	&	\nodata	&	62.5$\pm$12.5	&	5	&	1.75	\\
148367	&	\nodata	&	$<$120	&	$>$21	&	$<$-5.58	&	$<$8.2E-5	&	\nodata	&	-5.22	&	\nodata	&	290$\pm$240	&	5	&	1.77	\\
165185	&	\nodata	&	$<$160	&	$>$3.2	&	$<$-5.14	&	$<$5.6E-7	&	\nodata	&	-4.46	&	\nodata	&	437$\pm$186	&	1,2	&	1.13	\\
RE1816+541	&	\nodata	&	$<$57	&	$>$5.1	&	$<$-3.30	&	$<$2.0E-5	&	\nodata	&	-2.86	&	\nodata	&	115$\pm$100	&	1,4	&	0.23	\\
177724	&	52$\pm$14	&	$<$230	&	$>$9.4	&	$<$-5.75	&	$<$2.1E-5	&	\nodata	&	-6.26	&	\nodata	&	100$\pm$50	&	5	&	2.30	\\
180161	&	\nodata	&	$<$110	&	$>$4.7	&	$<$-4.92	&	$<$1.2E-6	&	\nodata	&	-4.68	&	\nodata	&	468$\pm$61	&	1,2,3	&	0.96	\\
186219	&	25$\pm$7	&	$<$89	&	$>$30	&	$<$-5.11	&	$<$3.1E-4	&	\nodata	&	$<$-5.55	&	\nodata	&	200$\pm$150	&	5	&	1.68	\\
eta Ind	&	32$\pm$6	&	$<$140	&	$>$11	&	$<$-5.41	&	$<$1.6E-5	&	\nodata	&	$<$-5.91	&	\nodata	&	250$\pm$200	&	5	&	1.62	\\
SAO 145139	&	\nodata	&	$<$64	&	$>$8.1	&	$<$-3.98	&	$<$1.3E-5	&	\nodata	&	-3.25	&	\nodata	&	35$\pm$7	&	7	&	0.79	\\
202730	&	30$\pm$4	&	$<$59	&	$>$76	&	$<$-5.77	&	$<$4.6E-4	&	\nodata	&	-5.11	&	\nodata	&	150$\pm$100	&	5	&	1.87	\\
203244	&	\nodata	&	$<$53	&	$>$21	&	$<$-4.99	&	$<$2.2E-5	&	\nodata	&	-4.58	&	\nodata	&	334$\pm$79	&	1,2	&	0.99	\\
LO Peg	&	\nodata	&	$<$96	&	$>$3.6	&	$<$-4.15	&	$<$1.8E-6	&	\nodata	&	-3.13	&	\nodata	&	75$\pm$25	&	4	&	0.71	\\
Gl 859A	&	\nodata	&	$<$270	&	$>$1.3	&	$<$-4.96	&	$<$2.0E-7	&	\nodata	&	-4.48	&	\nodata	&	314$\pm$63	&	1	&	0.99	\\
HK Aqr	&	\nodata	&	$<$61	&	$>$4.5	&	$<$-3.48	&	$<$8.2E-6	&	\nodata	&	-3.02	&	\nodata	&	25.5$\pm$12.5	&	7	&	0.29	\\
218738	&	\nodata	&	$<$120	&	$>$3.5	&	$<$-4.61	&	$<$1.3E-6	&	\nodata	&	-3.22	&	\nodata	&	40$\pm$8	&	7	&	0.85	\\
\enddata
 \tablenotetext{a}{Color-corrected to the dust temperature.}
 \tablenotetext{b}{Unless otherwise noted, the fractional infrared luminosity is determined from a single temperature blackbody fit to the 24 and 70 $\mu$m photometry (or upper limit).  70 $\mu$m photometry are color-corrected to the maximum blackbody temperature as described in ${\S}$4.4.}
  \tablenotetext{c}{$\dot M_\odot \sim 2 \times 10^{-14} M_\odot$/yr}
 \tablenotetext{d}{The first fractional infrared excess uses the approximation $L_{IR}\sim\nu F_\nu$ at 24 $\mu$m after subtracting off the expected stellar flux density and ignoring the 70 $\mu$m flux density upper-limit.  The fractional infrared excess upper limit is derived from the 70 $\mu$m upper limit using the single-temperature blackbody fit as described in table note (b).}
 \tablerefs{1 -- X-ray--age correlation \citep[${\S}$4.3, ][]{mamajek08}, 2 -- Calcium H\&K emission--age correlation \citep[${\S}$4.3, ]{mamajek08} , 3 -- Rotation--age correlation \citep[${\S}$4.3, ]{mamajek08},  4 -- cluster/group membership, 5 -- Stromgren photometry \citep[]{stauffer00,song01,song00}, 6 -- \citet[]{nordstrom04}, 7 -- isochrone-fitting \citep[${\S4.3}$, ][]{seiss00}}
  \end{deluxetable}

\begin{deluxetable}{lrrrrrrrrrrr}
\rotate
\tablecolumns{12}
\tabletypesize{\scriptsize}
\tablewidth{0pc}
\tablecaption{Sample Statistics}
\tablehead{
\colhead{Category}	&  \multicolumn{2}{c}{24 $\mu$m} &  &  \multicolumn{2}{c}{70 $\mu$m} & & \multicolumn{2}{c}{24 or 70 $\mu$m}& & \colhead{Median} & \colhead{Literature}\\
\cline{2-3}\cline{5-6} \cline{8-9}
\colhead{}	&  \colhead{\#} &  \colhead{Disk} & &  \colhead{\#} &   \colhead{Disk} &&  \colhead{\#} &   \colhead{Disk} &\colhead{\#} & \colhead{F$_70$/F$_*$} & \colhead{Comparison}\\
\colhead{}	&  \colhead{excesses} &   \colhead{Fraction}  & &\colhead{excesses} &  \colhead{Fraction}  & & \colhead{excesses} &   \colhead{Fraction} & \colhead{total} & \colhead{Sensitivity\tablenotemark{a}} &\colhead{}
}
\startdata
ALL 						& 11  & 15.7$^{+5.4}_{-3.4}$\%  & 	& 8  & 11.4$^{+4.9}_{-2.8}$\%  &		& 15  &  21.4$^{+5.6}_{-4.1}$\%	& 70	& 2.48 & 15\%\tablenotemark{b}\\
T$_*$ $<$ 5000 K  			& 5  &   18.5$^{+9.6}_{-5.2}$\%  & 	& 1  &  3.7$^{+7.6}_{-1.1}$\%   &		&  6   &  22.2$^{+9.8}_{-5.9}$\% 	& 27 & 6.32 &  \nodata\\
5000 K $<$ T$_*$ $<$ 6000 K  & 4  &    14.3$^{+9.0}_{-4.3}$\%  & 	& 6  &  21.4$^{+9.5}_{-5.7}$\% &		&  7  &  25.0$^{+9.6}_{-6.3}$\% 	& 28	& 2.20 &  8.5--19\%,10--20\%\tablenotemark{c}\\
T$_*$ $>$ 6000 K  			& 2 &   13.3$^{+13.3}_{-4.6}$\% & 	& 1 &  6.7$^{+12.5}_{-2.2}$\% &		&  2  &  13.3$^{+13.3}_{-4.6}$\% 	& 15 & 1.74 &   33\%\tablenotemark{d}\\
single star\tablenotemark{e} 	& 7  &   15.9$^{+7.0}_{-4.0}$\%  & 	&6  & 13.6$^{+6.8}_{-3.6}$\%  &		&  10 & 22.7$^{+3.0}_{-5.0}$\% 	& 44 & 2.64 & 15\%\tablenotemark{a} \\
all binaries  				& 4  &   15.4$^{+9.6}_{-4.6}$\%  & 	& 2  & 7.7$^{+8.6}_{-2.5}$\% &			&  5   &  19.2$^{+9.8}_{-5.4}$\% 	& 26 & 2.29 &   23\%\tablenotemark{f}\\
\enddata
\tablenotetext{a}{The median 24 $\mu$m S/N is 57 when including calibration uncertainties.  The minimum excess reported is F$_{24}$/F$_*$=1.09, and consequently the disk fractions listed are for F$_{24}$/F$_*$$\gtrsim$ 1.1.  At 70 $\mu$m, the median sensitivity is determined from the 3-$\sigma$ excess upper limit for detections and non-detections.}
 \tablenotetext{b}{\citet[]{lagrange00}}
 \tablenotetext{c}{Fractions are at 24 and 70 $\mu$m respectively from \citet[]{meyer08,meyer05}}
 \tablenotetext{d}{\citet[]{su06}}
 \tablenotetext{e}{Binarity as defined in Table 1; single star -- no known companion; spectroscopic or unresolved binary, footnote (f) in Table 1, projected separations of 4--78 AU; wide or visual binary -- footnotes (e),(g) in Table 1, projected separations of 137--4320 AU.}
\tablenotetext{f}{The $>$3 AU binary separation 70 $\mu$m excess fraction (11/48) inferred from Table 4 in \cite[]{trilling07}.  The one 24 $\mu$m excess source identified in \citet[]{trilling07} at these separations is also in excess at 70 $\mu$m.}
  \end{deluxetable}


\begin{figure}
\plotone{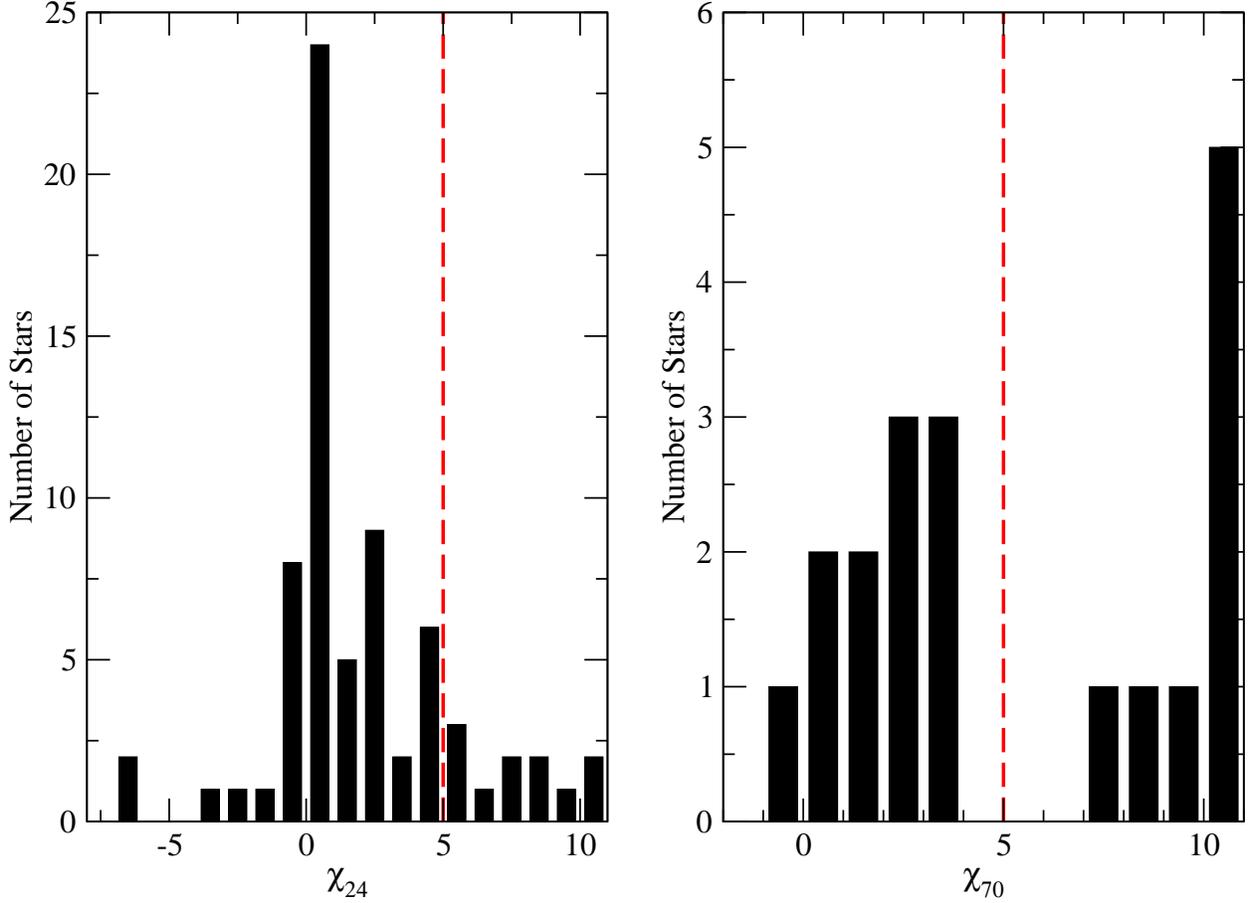}
\caption{Distribution of detected 24 and 70 $\mu$m photometry relative to the expected photospheric values for all objects reported in this paper.  $\chi_{70}$ is skewed to positive values since we are not sensitive to detect the photospheres for our entire sample, and non-detections are not included in the distribution.  As a result detections are more likely to be in excess than not.  The rightmost bin corresponds to the total number of sources with $\chi$$>$10.  The red vertical dashed line represents the threshold we set for the detection of excess.}
\end{figure}
\clearpage
\begin{figure}
\plotone{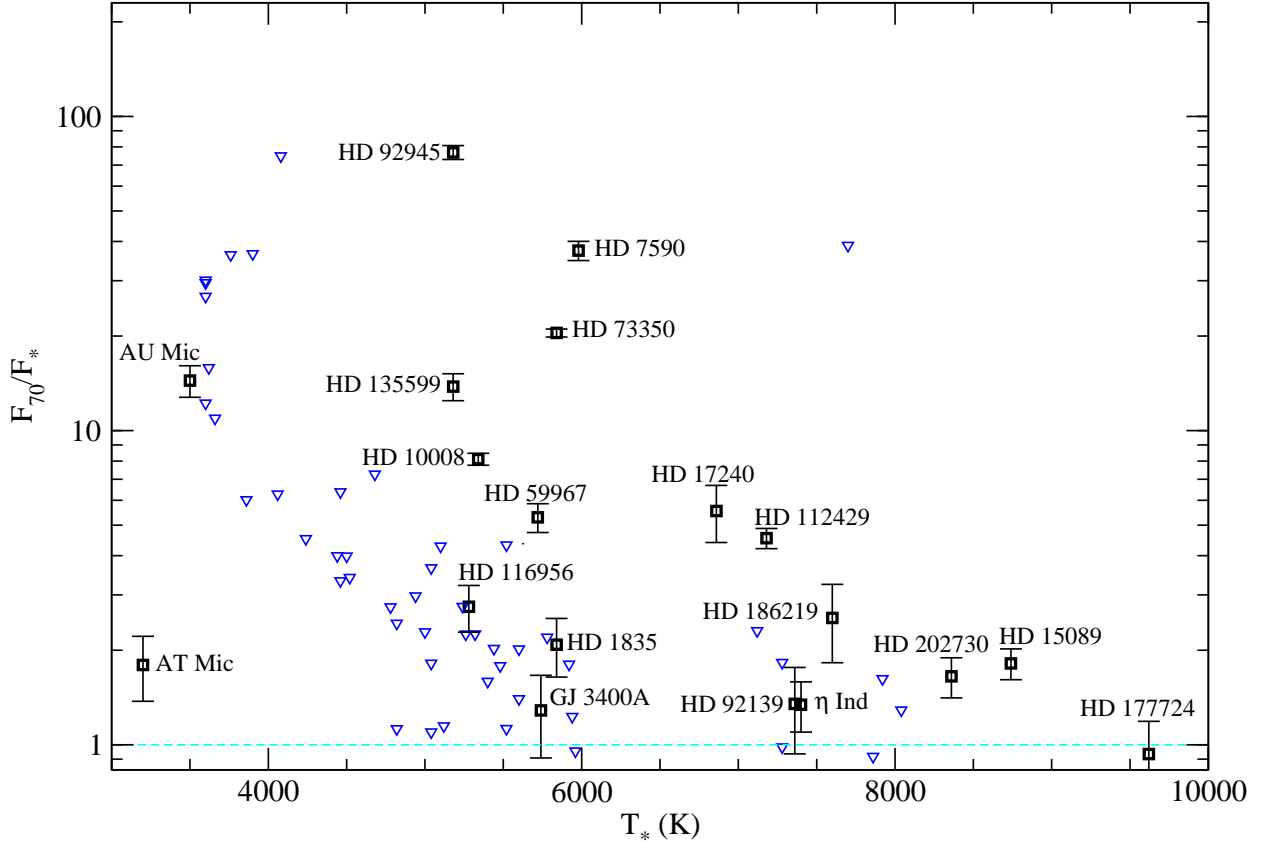}
\caption{The observed 70 $\mu$m MIPS flux density (not color-corrected) divided by the expected photospheric flux density as a function of effective stellar temperature for our sample.  3-$\sigma$ upper limits from non-detections at 70 $\mu$m are plotted with blue triangles, and detections are plotted with black squares and name labels.  The MIPS observations for some sources have been previously published (Table 2, Column 8).  However, all photometry plotted are derived in this work and represent improved absolute flux calibration uncertainties and data reduction methods.  We are less sensitive to 70 $\mu$m excess for cooler stars, introducing a bias in our sample.}
\end{figure}
\clearpage
\begin{figure}
\plotone{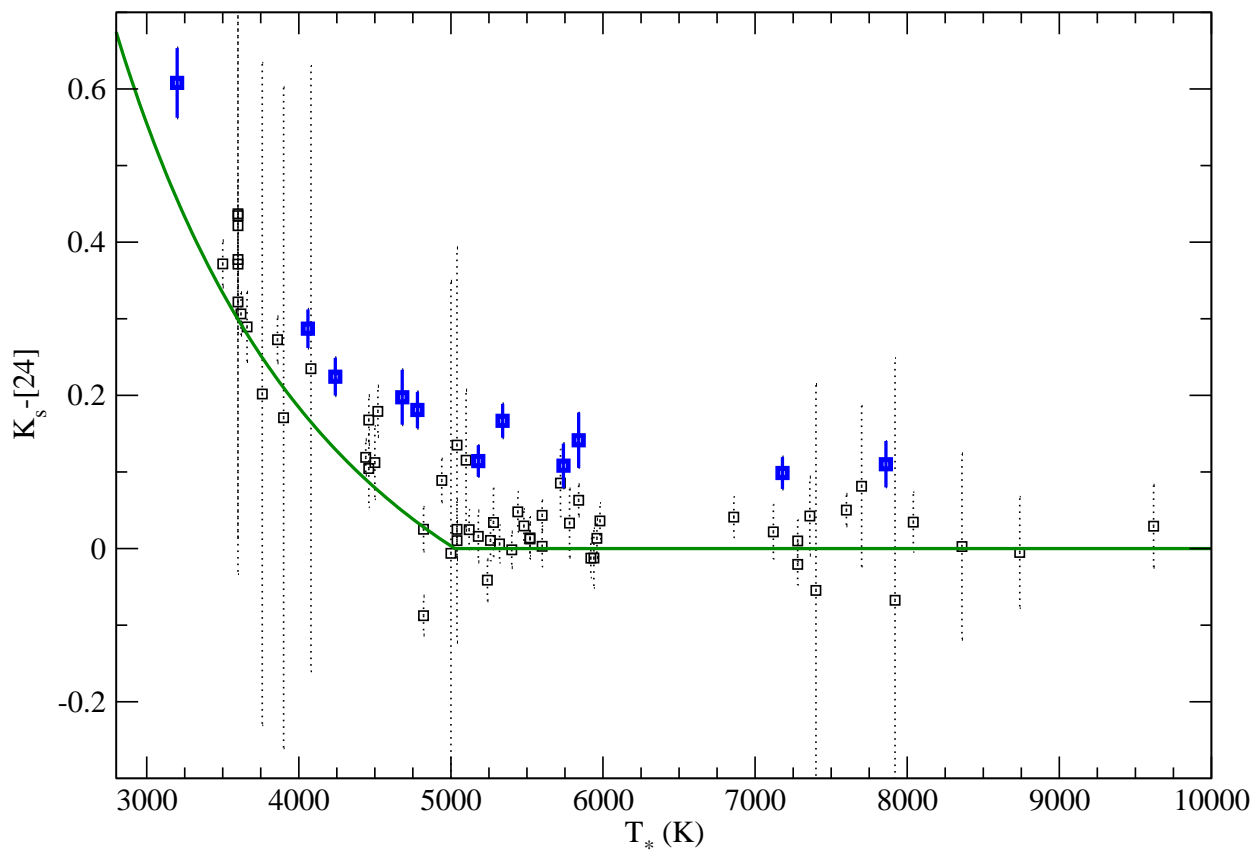}
\caption{The observed K$_s$-[24] color as a function of stellar effective temperature for our sample.   All colors plotted are re-derived in this work as in Figure 2.  The large ($>$0.2mag) uncertainties for several sources are either due to the saturation of the K$_s$ 2MASS images, or blending from companions at 24 $\mu$m.  In green we plot the empirical model of the intrinsic stellar K$_s$-[24] colors for $<$5pc main sequence dwarfs derived in \citet[]{gautier08}.  Sources with $\chi_{24}$$>$5 are plotted in blue, indicating possible warm excesses.}

\end{figure}

\clearpage
\begin{figure}
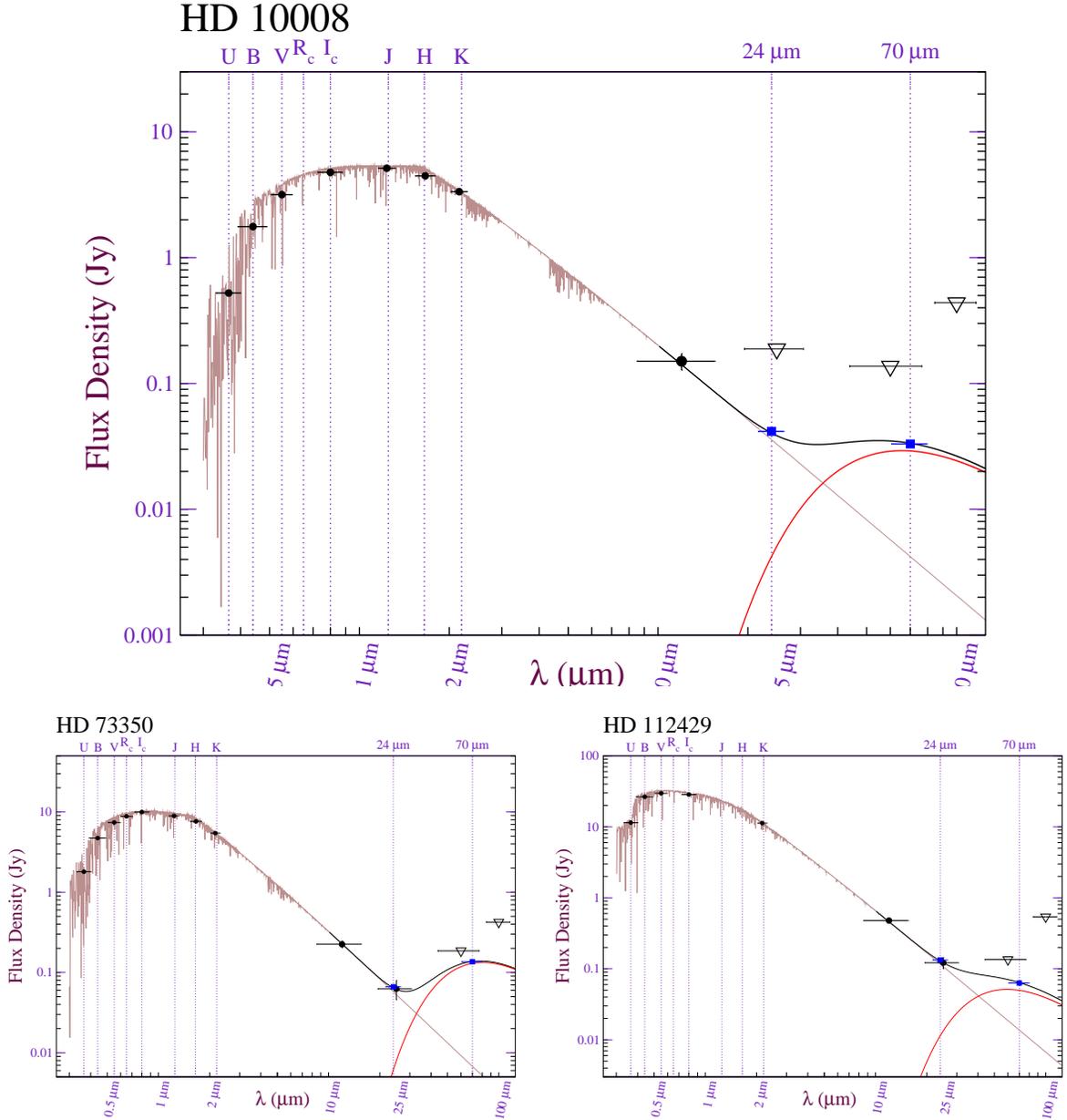

\epsscale{0.8}
\plotone{f4a}\\
\epsscale{1}
\plottwo{f4b}{f4c}
\caption{Spectral energy distributions (SEDs) for stars in our sample with 70 $\mu$m MIPS detections.  Stars are ordered from top-left to bottom-right as they are in Table 2.  First, 4 stars with 24 and 70 $\mu$m excesses are shown, followed by 4 stars with 70 $\mu$m excesses only, followed by two stars with 24 $\mu$m excesses only and 70 $\mu$m detections, and finally stars with photospheric 24 and 70 $\mu$m detections.  Optical and near-infrared photometry are shown in black.  MIPS color-corrected photometry are shown in blue, and IRAS color-corrected photometry are shown in black with 1-$\sigma$ uncertainties shown for detections.  Open triangles are shown for 3-$\sigma$ upper-limits.   Horizontal error bars on photometry represent effective bandwidths.  In light brown are the temperature-interpolated Phoenix NextGen sythetic spectra fit to the optical and near-IR photometry.  In red is the dust component SED, and in black is the summation of the synthetic stellar SED and dust component SED depicting the total model flux at wavelengths longward of 10$\mu$m.}
\end{figure}
\begin{figure}
\plottwo{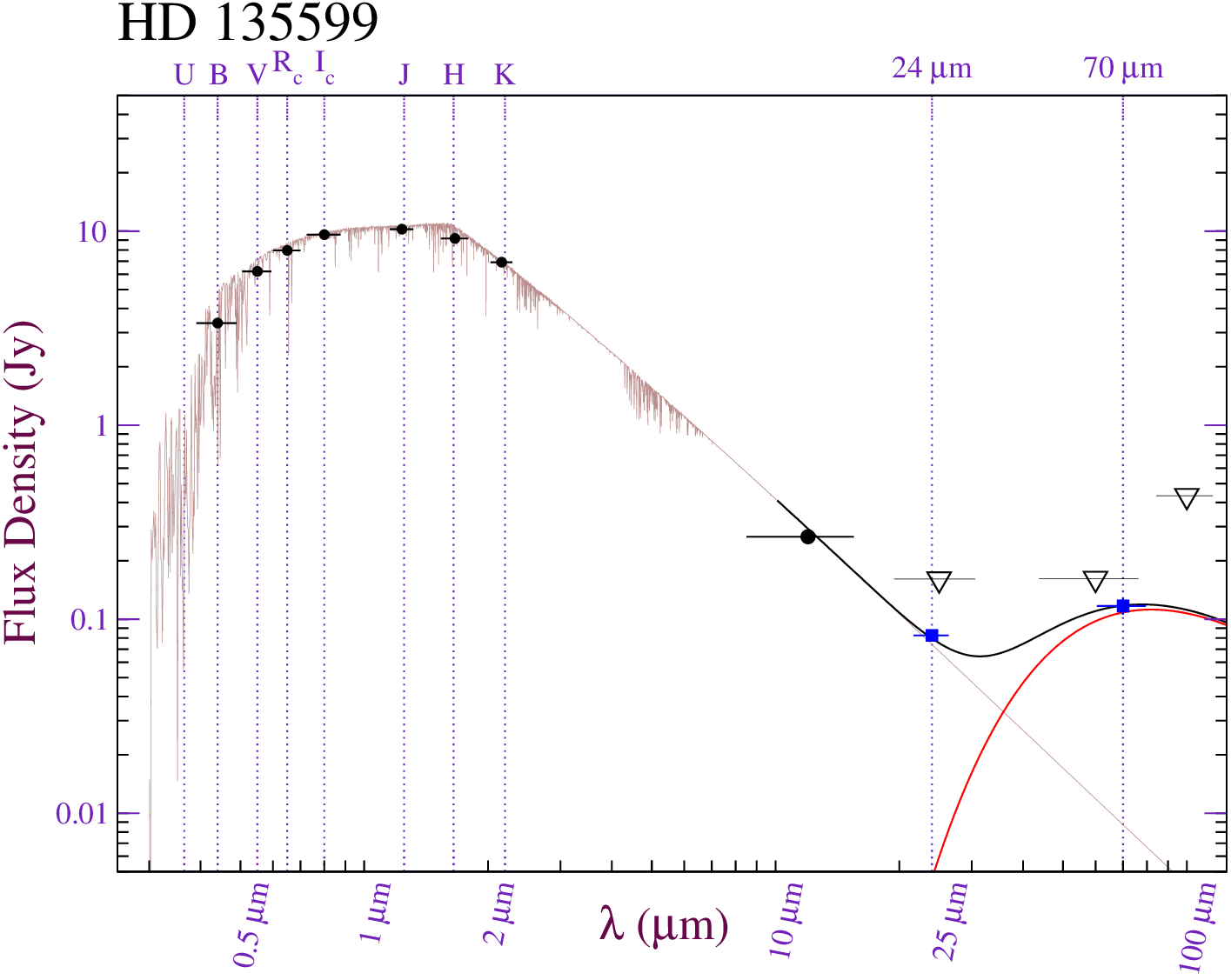}{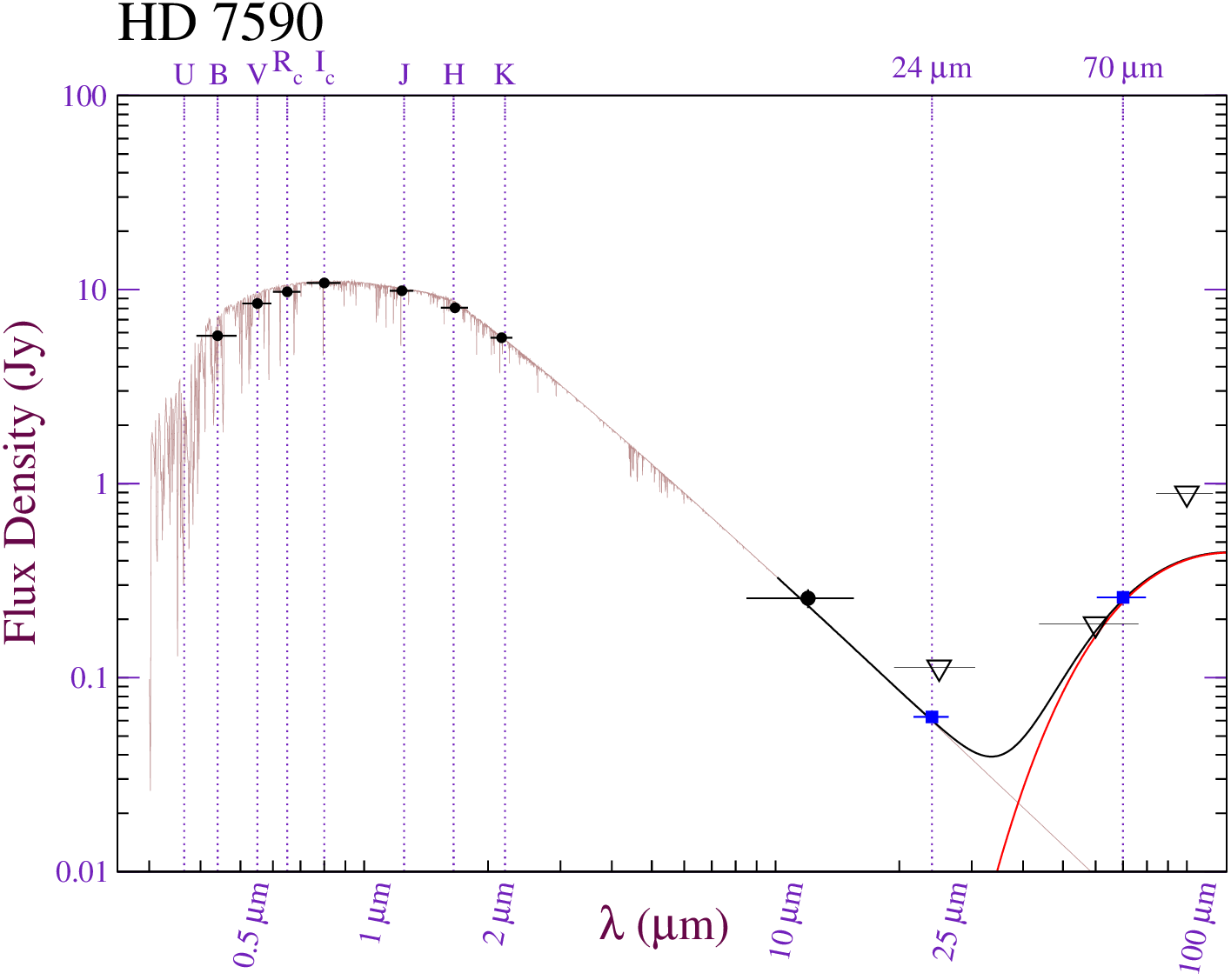}\\
\plottwo{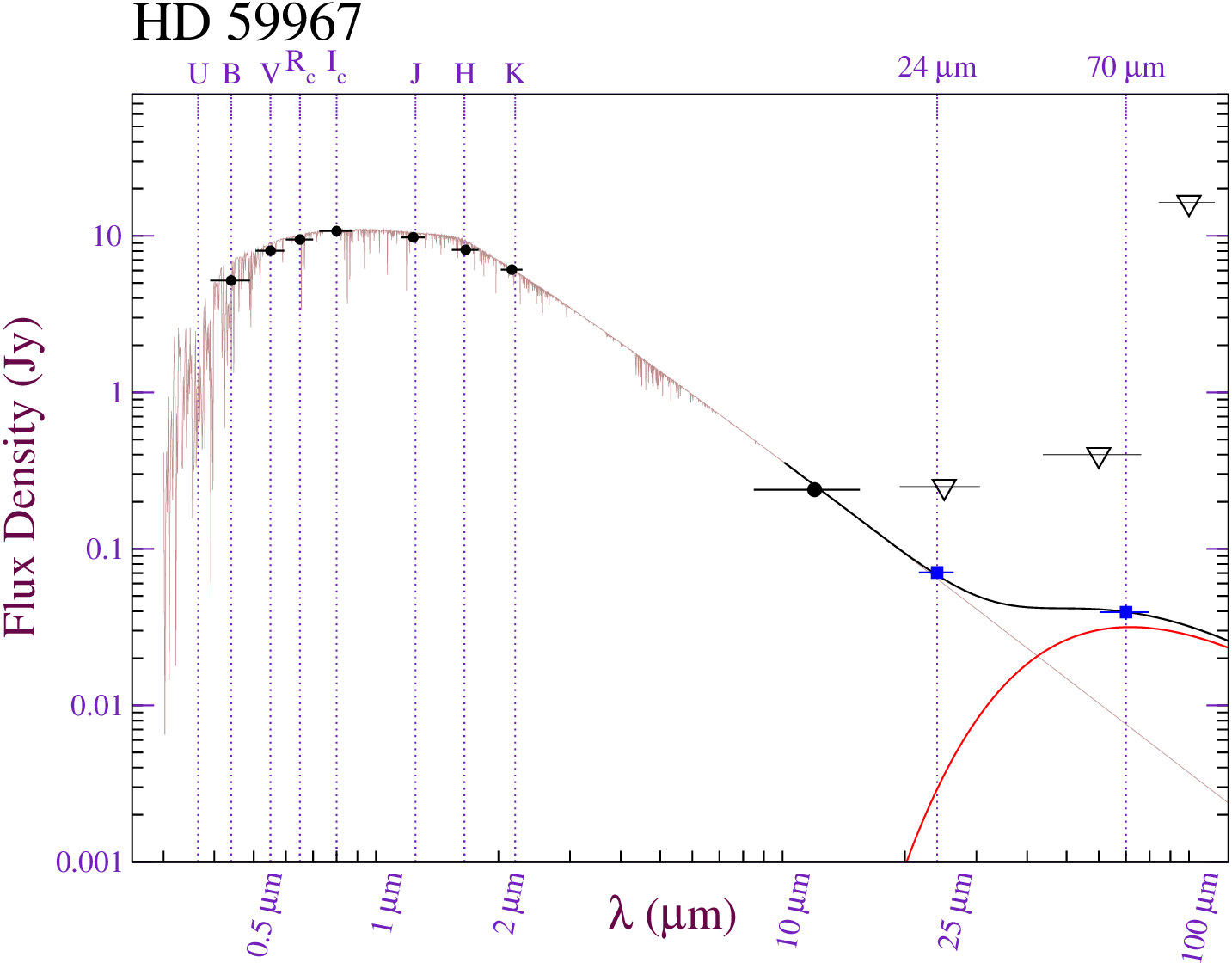}{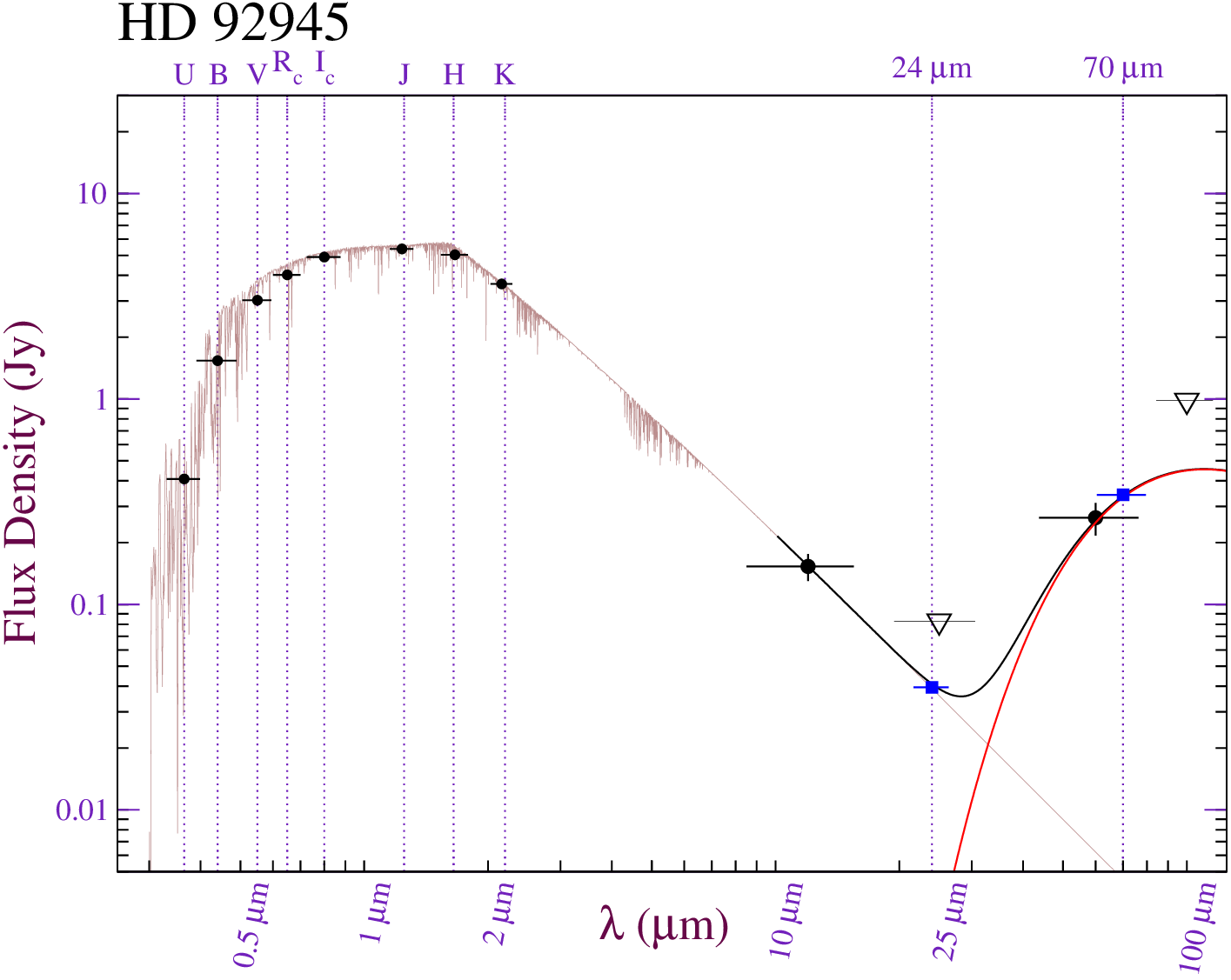}\\
\plottwo{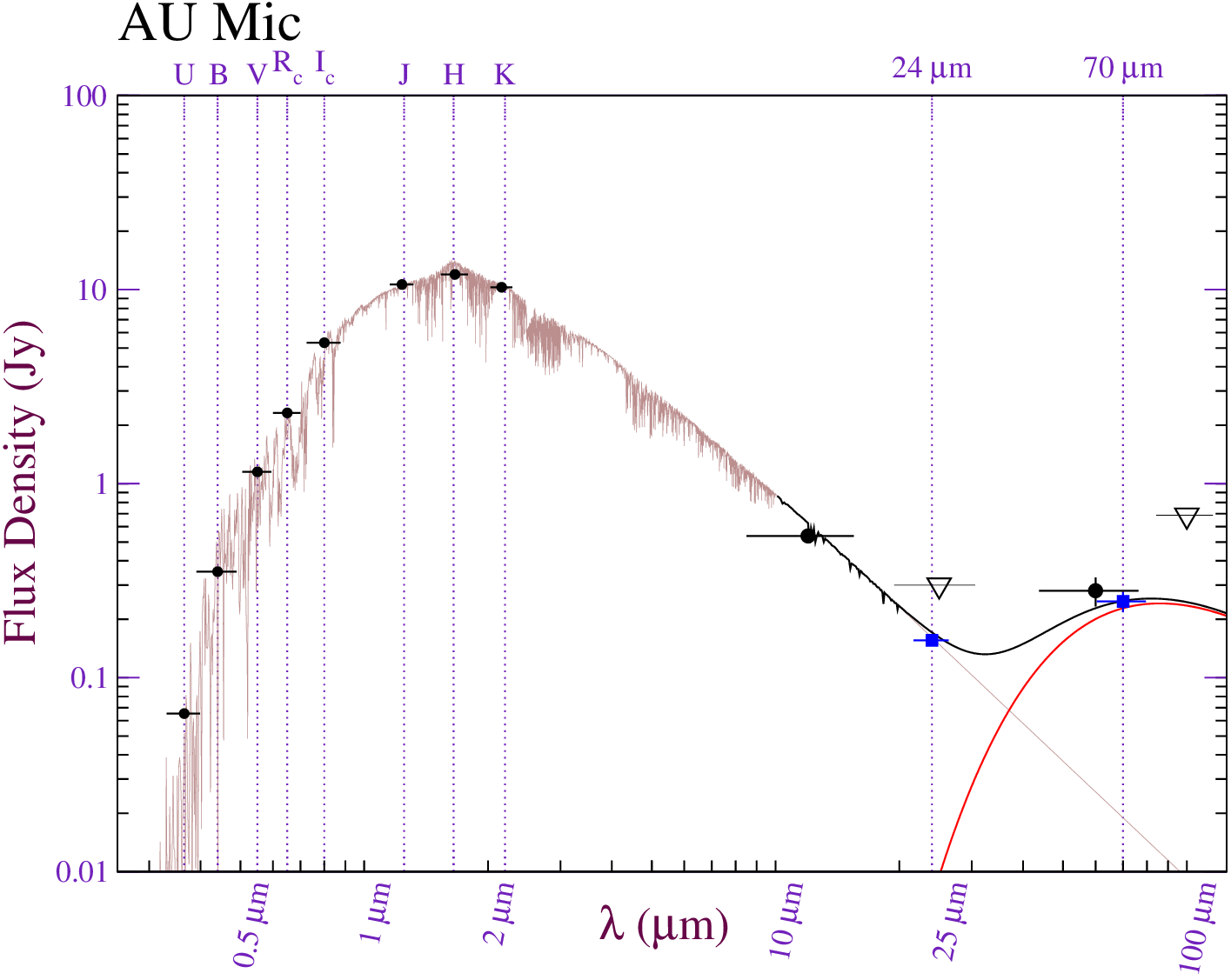}{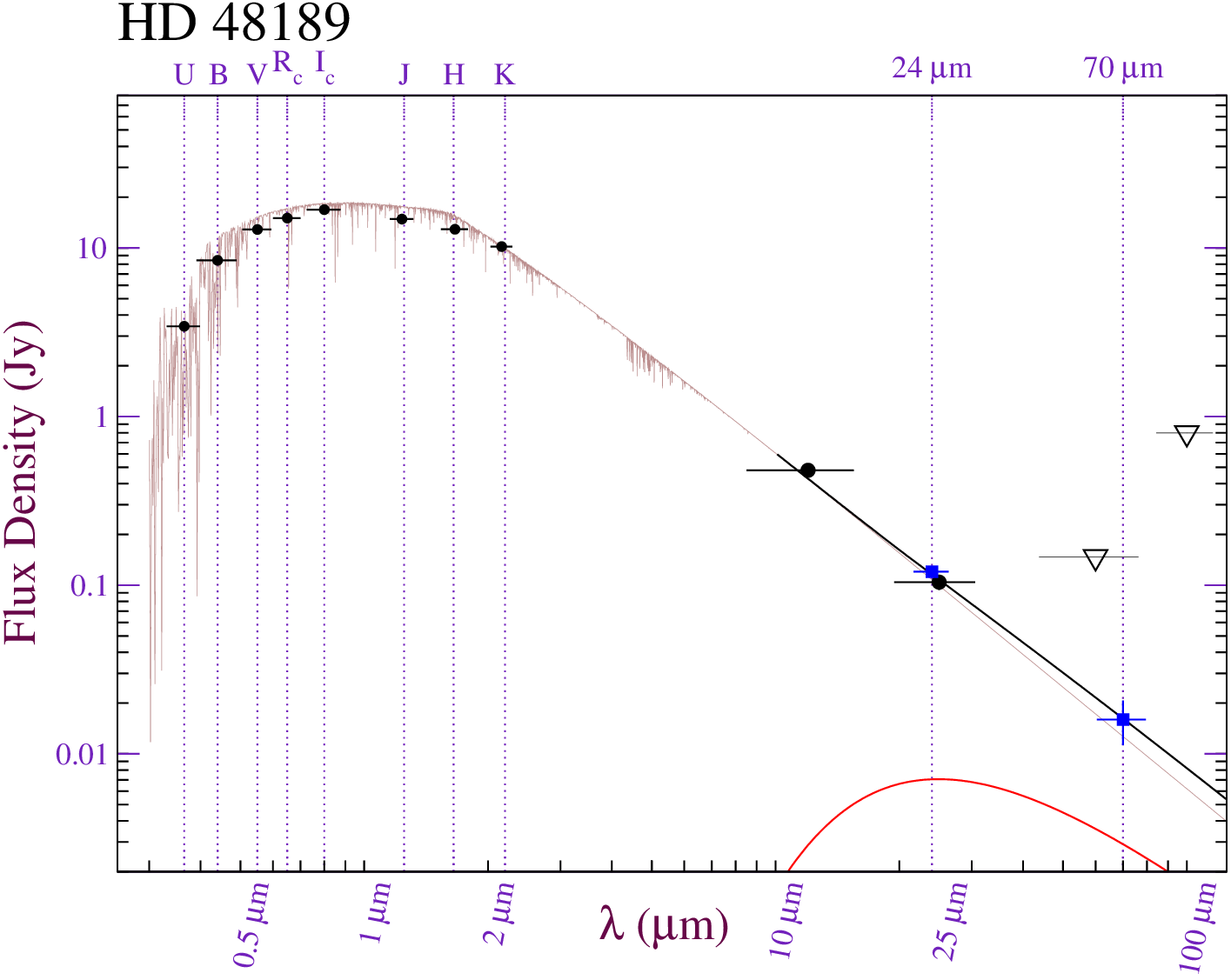}
\end{figure}
\begin{figure}
\plottwo{f4j}{f4k}\\
\plottwo{f4l}{f4m}\\
\plottwo{f4n}{f4o}
\end{figure}
\begin{figure}
\plottwo{f4p}{f4q}\\
\plottwo{f4r}{f4s}\\
\end{figure}

\clearpage
\begin{figure}
\plotone{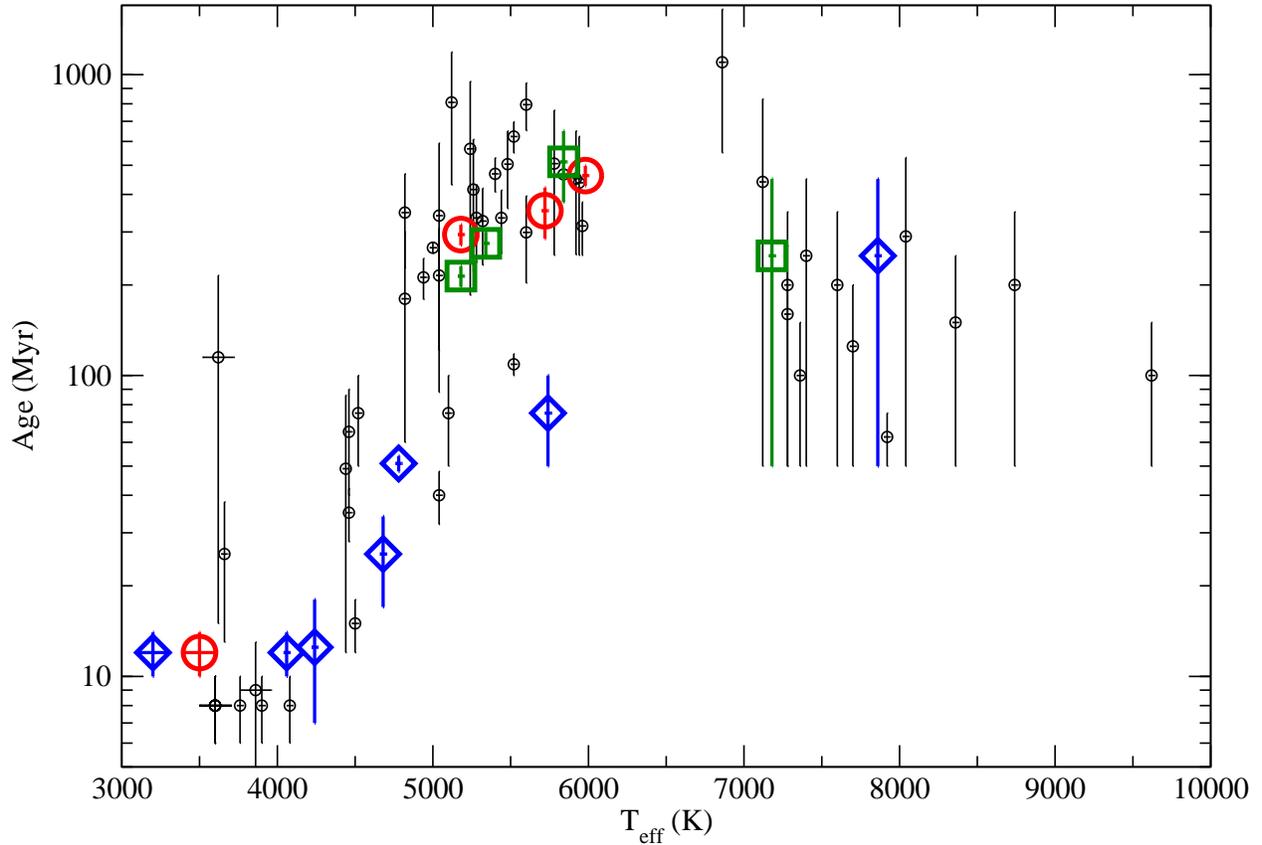}
\caption{We plot the distribution of our sources as a function of stellar effective temperature and estimated age.  Open black circles are stars without MIPS excess; open green squares with both 24 and 70 $\mu$m MIPS excesses; open red circles are stars with 70 $\mu$m MIPS excesses; open blue diamonds are stars with 24 $\mu$m excesses.  Uncertainties in age and effective temperature are shown as lines centered on each symbol.  See ${\S}$4.3 for discussion.}
\end{figure}

\clearpage
\begin{figure}
\plotone{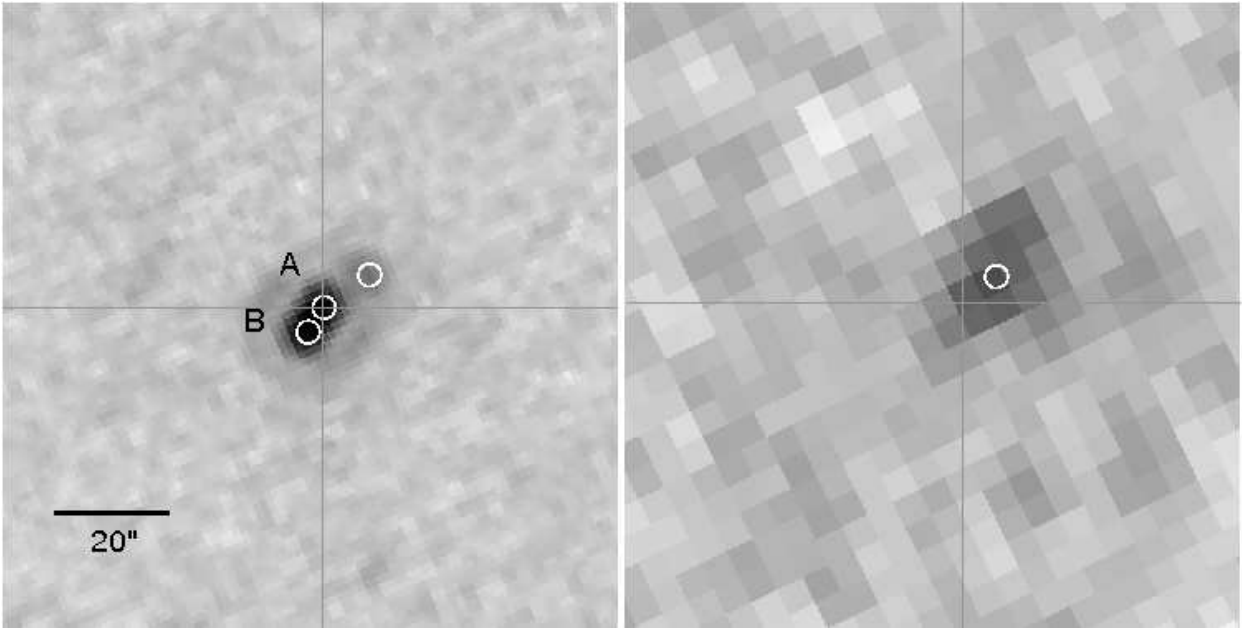}
\caption{Left: 24 $\mu$m image of TWA 13, with the A and B components labeled in black and circled in white. A third background galaxy is also circled.  The gray lines divide the image into quadrants, and the image is centered on TWA 13A.  Right: 70 $\mu$m image at the expected position of TWA 13.  The 70 $\mu$m emission is coincident with the background galaxy and not TWA 13A as previously published in \citet[]{chen05a,low05}. The same gray lines are used as in the 24 $\mu$m image, centered on the expected position of TWA 13A.  Both the left and right images are to scale, with north up and east to the left (a 20$^{\prime\prime}$ line is shown in black).}
\end{figure}

\clearpage
\begin{figure}
\plotone{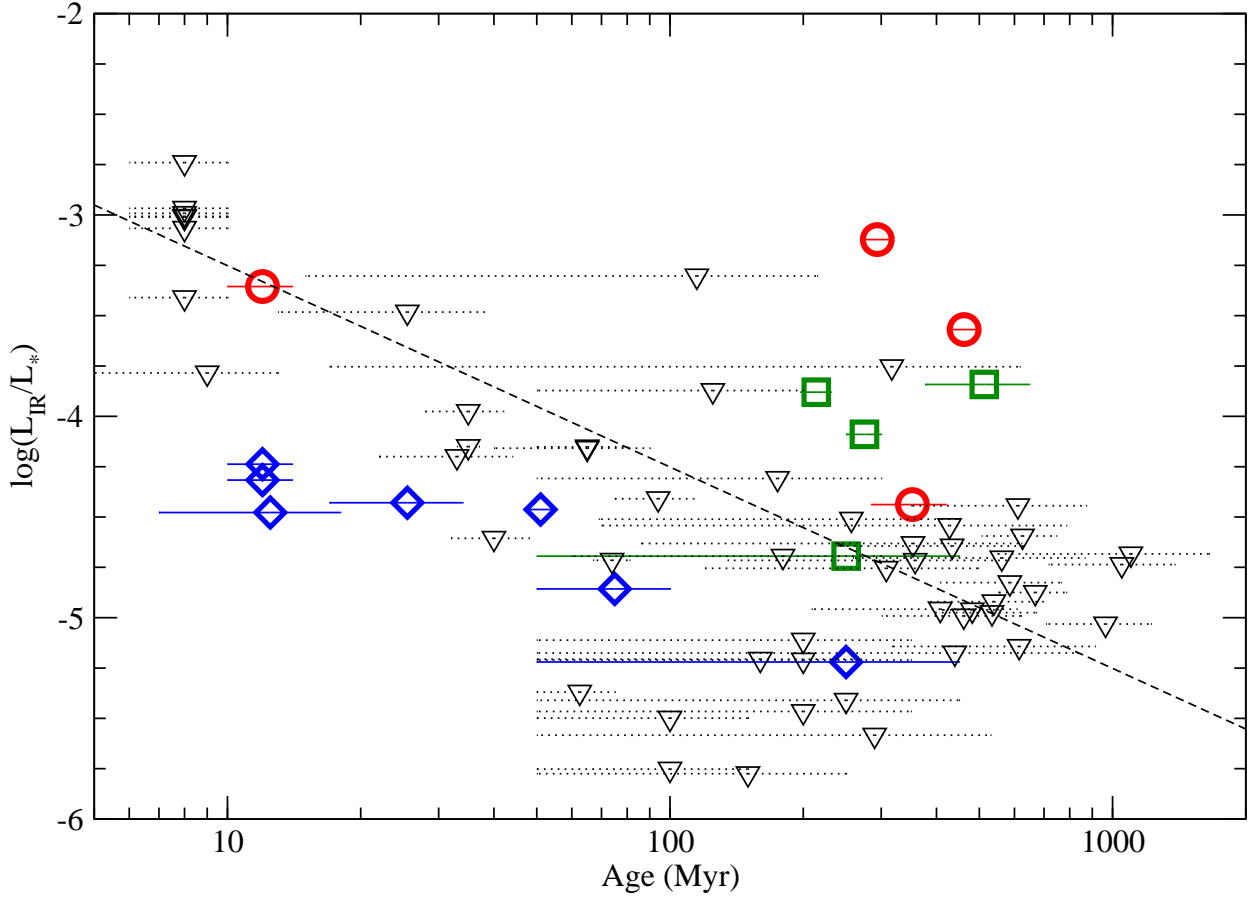}
\caption{Fractional infrared excess as a function of estimated age for all stars in our sample.  Open green squares are stars with both 24 and 70 $\mu$m MIPS excesses; open red circles are stars with 70 $\mu$m MIPS excesses; open blue diamonds are stars with 24 $\mu$m excesses; open downward triangles are upper limits for stars with no excess derived from the 70 $\mu$m flux upper limit.   Uncertainties in age are shown as horizontal lines. A t$_*^{-1}$ power law decay consistent with collisional disk evolution is shown as a black dotted line.  Due to 70 $\mu$m sensitivities, this plot is incomplete for log(L$_{IR}$/L$_*$) $<$ -4.5 and ages less than $\sim$50--100 Myr.}
\end{figure}

\clearpage
\begin{figure}
\plotone{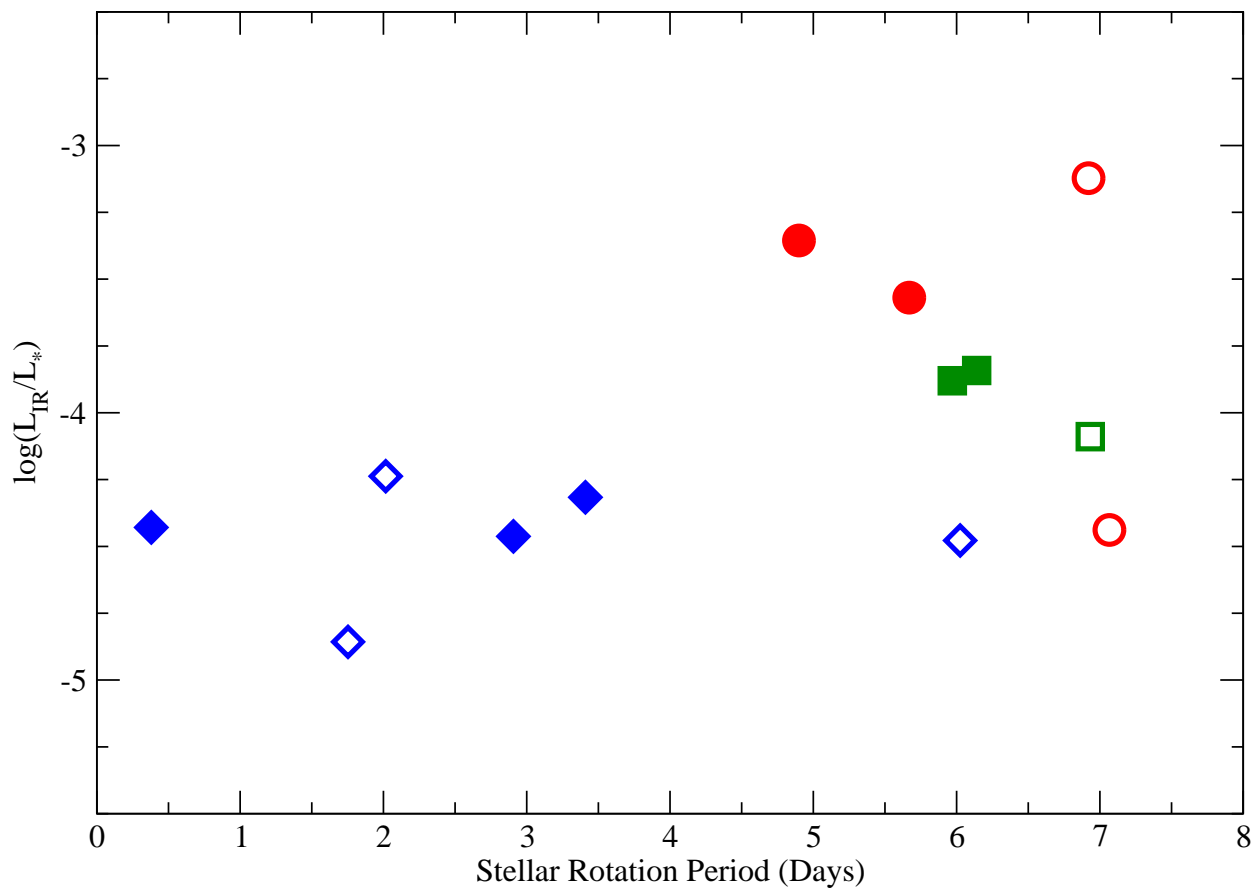}
\caption{Fractional infrared excess as a function of measured or estimated stellar rotation periods for stars in our sample with T$_*$$<$6000 K.   We choose 6000 K to be consistent with our previous subgrouping of our sample and to select stars with only convective atmospheres (there are no stars in our sample between 6000 and 6800 K).   Green squares are stars with 24 and 70 $\mu$m MIPS excesses; red circles are stars with 70 $\mu$m MIPS excesses; blue diamonds are stars with 24 $\mu$m excesses.  Filled symbols indicate a measured period; open symbols indicate a period estimated from v sin i and the radius from the SED fit.}
\end{figure}

\clearpage
\begin{figure}
\plotone{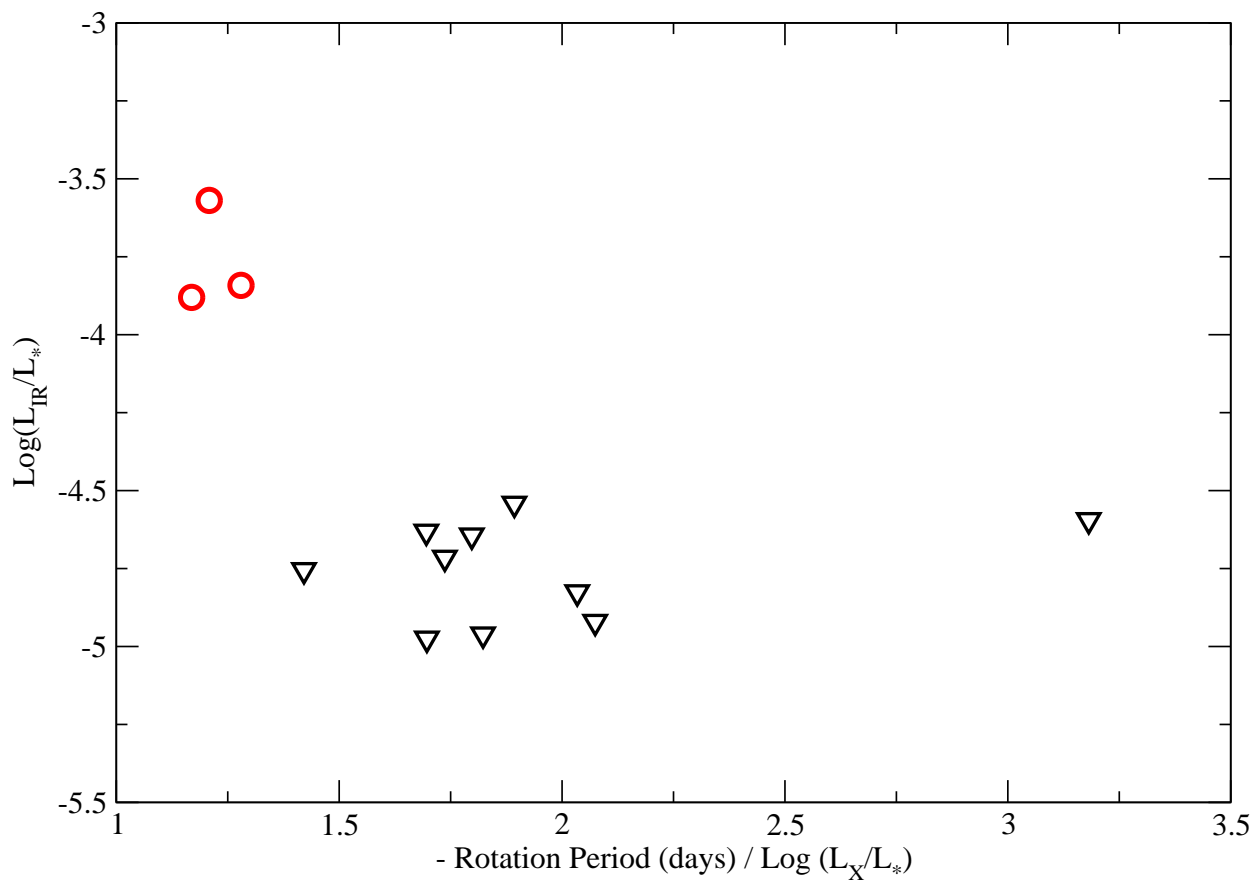}
\caption{Fractional infrared excess plotted as a function of the negative ratio of the measured rotation period to the logarithm of the fractional X-ray luminosity.  X-ray activity increases to the right, and period increases to the right.  Red circles are stars with 70 $\mu$m MIPS excesses (HD 135599, HD 7590 and HD 73350); upper limits for stars without 24 or 70 $\mu$m excesses are shown with open black triangles.  Only stars with measured rotation periods and X-ray fluxes with -5 $<$ Log(L$_X$/L$_*$) $<$ -4.3 are included (t$_*$ $>$ $\sim$300 Myr).}
\end{figure}

\end{document}